\documentclass[12pt]{article}
\usepackage{jheppub}

\pdfoutput=1

\usepackage{amsmath,bbm,array,amsfonts,graphicx,wrapfig,lscape,float,mathtools,multirow,longtable}
\usepackage[dvipsnames,table]{xcolor}
\usepackage{array}
\usepackage{color}
\usepackage{breqn}

\newcommand{\be}{\begin{equation}}
\newcommand{\ee}{\end{equation}}
\newcommand{\beq}{\begin{equation}}
\newcommand{\beql}[1]{\begin{equation}\label{#1}}
\newcommand{\eeq}{\end{equation}}
\newcommand{\ba}{\begin{array}}
\newcommand{\ea}{\end{array}}
\newcommand{\bea}{\begin{eqnarray}}
\newcommand{\beal}[1]{\begin{eqnarray}\label{#1}}
\newcommand{\eea}{\end{eqnarray}}
\newcommand{\ben}{\begin{enumerate}}
\newcommand{\een}{\end{enumerate}}
\newcommand{\bean}{\begin{eqnarray*}}
\newcommand{\eean}{\end{eqnarray*}}
\newcommand{\eref}[1]{(\ref{#1})}
\newcommand{\sref}[1]{\S\ref{#1}}
\newcommand{\tref}[1]{Table~\ref{#1}}
\newcommand{\nn}{\nonumber}

\newcommand{\fref}[1]{Figure \ref{#1}}
\newcommand{\btab}[1]{\begin{tabular}{#1}}
\newcommand{\etab}{\end{tabular}}

\def \Black {\pdfsetcolor{0 0 0 1}}

\newcommand{\comment}[1]{}

\newcommand{\ud}{\mathrm{d}}

\newcommand{\qed}{\nobreak \ifvmode \relax \else
      \ifdim\lastskip<1.5em \hskip-\lastskip
      \hskip1.5em plus0em minus0.5em \fi \nobreak
      \vrule height0.75em width0.5em depth0.25em\fi}

\newcommand{\Tr}{\text{Tr}}

\definecolor{darkspringgreen}{rgb}{0.09, 0.45, 0.27}
\definecolor{forestgreen}{rgb}{0.13, 0.55, 0.13}

\title{Hilbert Series of Bipartite Field Theories} 

\author[a]{Minsung Kho} 
\author[a,b]{and Rak-Kyeong Seong}

\affiliation[a]{
Department of Mathematical Sciences, and 
${}^{b}$Department of Physics,\\ 
Ulsan National Institute of Science and Technology,\\
50 UNIST-gil, Ulsan 44919, South Korea
}
\emailAdd{minsung@unist.ac.kr}	
\emailAdd{seong@unist.ac.kr}

\preprint{
\begin{flushright}
\end{flushright}
}

\abstract{
We study the algebraic structure of the mesonic moduli spaces of bipartite field theories by computing the Hilbert series.
Bipartite field theories form a large family of $4d$ $\mathcal{N}=1$ supersymmetric gauge theories that are defined by bipartite graphs on Riemann surfaces with boundaries. 
By calculating the Hilbert series, we are able to identify the generators and defining generator relations of the mesonic moduli spaces of these theories. 
Moreover, we show that certain bipartite field theories exhibit enhanced global symmetries which can be identified through the computation of the corresponding refined Hilbert series. 
As part of our study, we introduce two one-parameter families of bipartite field theories defined on cylinders whose mesonic moduli spaces are all complete intersection toric Calabi-Yau 3-folds.
}

\preprint{UNIST-MTH-24-RS-03}

\begin{document}

\maketitle

\section{Introduction}

One of the largest families of $4d$ $\mathcal{N}=1$ supersymmetric gauge theories is realized in terms of periodic bipartite graphs on a 2-torus.
These bipartite graphs are known as brane tilings \cite{Franco:2005rj,Franco:2005sm,Hanany:2005ve,Hanany:2012hi} and dimer models \cite{2003math.....10326K}.
The corresponding $4d$ $\mathcal{N}=1$ supersymmetric gauge theories are worldvolume theories of D3-branes probing a toric Calabi-Yau 3-fold.
The brane tilings represent a Type IIB brane configuration, which consists of D5-branes suspended between a NS5-brane that wraps a holomorphic curve $\Sigma$ defined by the Newton polynomial of the associated toric Calabi-Yau 3-fold.
Via T-duality, this Type IIB brane configuration is related to the D3-branes probing the toric Calabi-Yau 3-fold. 

The underlying bipartite graph of a brane tiling has powerful geometrical, combinatorial and representation-theoretic properties. 
For example, the bipartite graph is dual to a directed graph on the 2-torus known as the periodic quiver.
Such quiver diagrams can undergo local mutations known as `spider moves' or `urban renewal' \cite{ciucu1998complementation,kenyon1999trees}.
These local mutations are known to identify two distinct brane tilings that correspond to two $4d$ $\mathcal{N}=1$ supersymmetric gauge theories, which are related by Seiberg duality \cite{Seiberg:1994pq,Feng:2001xr,Feng:2001bn,Beasley:2001zp,Feng:2002zw}. 
These quiver mutations are also known to be associated to underlying cluster algebras \cite{fomin2002cluster,fomin2002cluster2}.
Another powerful combinatorial property that brane tilings exhibit are perfect matchings of the bipartite graph.
These are collections of edges in the graph corresponding to bifundamental chiral fields of the $4d$ $\mathcal{N}=1$ supersymmetric gauge theory, that connect to every white and black node of the graph uniquely once. 
These combinations of bifundamental chiral fields correspond to GLSM fields \cite{Witten:1993yc} that parameterize the moduli spaces of the associated $4d$ $\mathcal{N}=1$ supersymmetric gauge theories. 

The moduli spaces of the $4d$ $\mathcal{N}=1$ supersymmetric gauge theories realized in terms of brane tilings have rich algebro-geometric properties.
There are two moduli spaces that are of significant interest when studying brane tilings.
One of these moduli spaces is known as the mesonic moduli space $\mathcal{M}^{mes}$ \cite{Feng:2000mi,Feng:2001xr,Feng:2002fv,Feng:2002zw}, which is defined as the space of gauge invariant operators under the $F$- and $D$-terms of the $4d$ $\mathcal{N}=1$ supersymmetric gauge theory. 
When the $4d$ $\mathcal{N}=1$ theory is abelian with $U(1)$ gauge groups, the mesonic moduli space $\mathcal{M}^{mes}$ is exactly the probed toric Calabi-Yau 3-fold. 
Brane tilings also possess the master space $\mathcal{F}^\flat$ \cite{Forcella:2008eh,Forcella:2008bb,Forcella:2008ng,Forcella:2009bv,Hanany:2010zz,Zaffaroni:2008zz} as a moduli space, which is the freely generated space of chiral bifundamental fields of the $4d$ $\mathcal{N}=1$ theory subject to the $F$-terms and the non-abelian $SU(N)$ part of the $U(N)$ gauge symmetries.  
The remaining $U(1)$ symmetries become the baryonic part of the global symmetry of the master space $\mathcal{F}^\flat$.

The study of brane tilings and their moduli spaces led to the introduction of new brane configurations that are related to toric Calabi-Yau $n$-folds and realize supersymmetric gauge theories in diverse spacetime dimensions.
These include the study of worldvolume theories of M2-branes probing toric Calabi-Yau 4-folds with the introduction of brane crystals \cite{Lee:2006hw,Lee:2007kv} and brane tilings with Chern-Simons level assignments \cite{Hanany:2008cd,Hanany:2008fj,Franco:2008um,Hanany:2008gx,Davey:2009sr}.
More recently, the introduction of brane brick models which realize an infinite family of $2d$ $(0,2)$ supersymmetric gauge theories that are worldvolume theories of D1-brane probing toric Calabi-Yau $4$-folds, led to a variety of new discoveries. These include the brane realization \cite{Franco:2016nwv} and the mirror symmetry interpretation \cite{Franco:2016qxh,Franco:2016tcm} of Gadde-Gukov-Putrov triality \cite{Gadde:2013lxa} for $2d$ $(0,2)$ theories. 
These achievements further led to the construction of so-called brane hyper-brick models \cite{Franco:2016tcm} that realize $0d$ $\mathcal{N}=1$ worldvolume theories of D$(-1)$-branes that probe toric Calabi-Yau 5-folds.

In this work, we concentrate on a particular family of $4d$ $\mathcal{N}=1$ supersymmetric gauge theories known as bipartite field theories (BFT) \cite{Franco:2012mm,Hanany:2012vc,Xie:2012mr,Franco:2012wv,Heckman:2012jh,Cremonesi:2013aba,Franco:2013ana,Franco:2014nca}.
These theories are defined by bipartite graphs embedded on Riemann surfaces that can have an arbitrarily large genus and any number of boundaries. 
These boundaries on the Riemann surface introduce flavor symmetries in the BFT quiver.
Accordingly, one can think of BFTs as generalizations of brane tilings, which are defined on a torus with no boundaries. 
Many of the features of the moduli spaces associated to the BFTs depend on the structure of the bipartite graphs and the properties of the Riemann surfaces.
For instance, the dimensions of the moduli spaces are directly related to the genus of the Riemann surface and the number of boundaries.

In order to study some of the algebro-geometric features of the moduli spaces of BFTs,
we introduce a mathematical tool known as the Hilbert series \cite{Benvenuti:2006qr,Feng:2007ur,Butti:2006au,Butti:2007jv,hanany2007counting}.
The Hilbert series is a generating function that counts gauge invariant operators on the moduli space of a supersymmetric gauge theory.
By focusing on the mesonic moduli space of abelian BFTs, the Hilbert series that we calculate in this work count gauge invariant operators under the $F$- and $D$-terms of the corresponding $4d$ $\mathcal{N}=1$ supersymmetric gauge theory.
Our computation of the mesonic Hilbert series in this work enables us to identify the full algebraic structure of the mesonic moduli space of abelian BFTs.
Using plethystics \cite{Benvenuti:2006qr,Feng:2007ur,Butti:2006au,Butti:2007jv,hanany2007counting}, we are able to extract information about the generators and the defining relations amongst them that characterize algebraically the mesonic moduli spaces of the abelian BFTs that we study in this work.
Additional information that we can obtain from the Hilbert series includes whether the mesonic moduli space of the BFT is Calabi-Yau or a complete intersection.
Moreover, by rearranging fugacities associated to flavor charges in the refined Hilbert series into characters of representations of hidden enhanced global symmetries, we can potentially discover global symmetry enhancements in the mesonic moduli spaces of certain BFTs. 

The outline of our work is as follows. Section \sref{sbft} introduces BFTs and how they define the corresponding $4d$ $\mathcal{N}=1$ supersymmetric gauge theories as introduced in \cite{Franco:2012mm}. 
Sections \sref{smod} and \sref{salg} give respectively a brief introduction to the moduli spaces of BFTs and the forward algorithm for BFTs that is used to construct the moduli spaces using GLSM fields. 
Based on the forward algorithm, section \sref{shs} introduces the computation of Hilbert series and how plethystics can be used to identify the algebro-geometric structure of BFT moduli spaces.
In our work, we discuss examples of BFTs defined on a disk in section \sref{sec_e1}, including an example where we identify global symmetry enhancement of the mesonic moduli space through the Hilbert series. 
We also study infinite families of BFTs living on cylinders in sections \sref{sec_f01} and \sref{sec_f02} whose mesonic moduli spaces are identified as complete intersection toric Calabi-Yau 3-folds.
\\

\section{Background}

In the following section, we review the construction of bipartite field theories (BFT) as introduced in \cite{Franco:2012mm} and their master and mesonic moduli spaces \cite{Feng:2000mi,Feng:2001xr,Feng:2002fv,Feng:2002zw,Forcella:2008eh,Forcella:2008bb,Forcella:2008ng,Forcella:2009bv,Hanany:2010zz,Zaffaroni:2008zz}.
We then outline the forward algorithm \cite{Franco:2005rj,Feng:2000mi} for bipartite field theories and the computation of Hilbert series \cite{Benvenuti:2006qr,Feng:2007ur,Butti:2006au,Butti:2007jv,hanany2007counting} for moduli spaces of BFTs, which is going to be the central topic of discussion in this work. 

\subsection{Bipartite Field Theories \label{sbft}}

Bipartite field theories (BFT) \cite{Franco:2012mm,Hanany:2012vc,Xie:2012mr,Franco:2012wv,Heckman:2012jh,Cremonesi:2013aba,Franco:2013ana,Franco:2014nca} form a family of $4d$ $\mathcal{N}=1$ supersymmetric gauge theories that are defined by bipartite graphs $G$ embedded on Riemann surfaces $\Sigma$ with genus $g$.
These Riemann surfaces can have boundaries and we express the number of distinct boundaries in terms of $B$.
We consider BFTs as natural generalizations of brane tilings \cite{Franco:2005rj,Hanany:2005ve,Hanany:2012hi} defined on a torus $T^2$ with $B=0$. 

The bipartite graph $G$ is made of white and black nodes and edges that connect nodes of opposite color. 
Each component of the bipartite graph has an interpretation in terms of the corresponding $4d$ $\mathcal{N}=1$ supersymmetric gauge theory.
In the following paragraph, we summarize the dictionary on components of the bipartite graph $G$ on $\Sigma$ that define the corresponding $4d$ $\mathcal{N}=1$ supersymmetric gauge theory.

\begin{figure} [h]
\centering
\includegraphics[width=.95\textwidth]{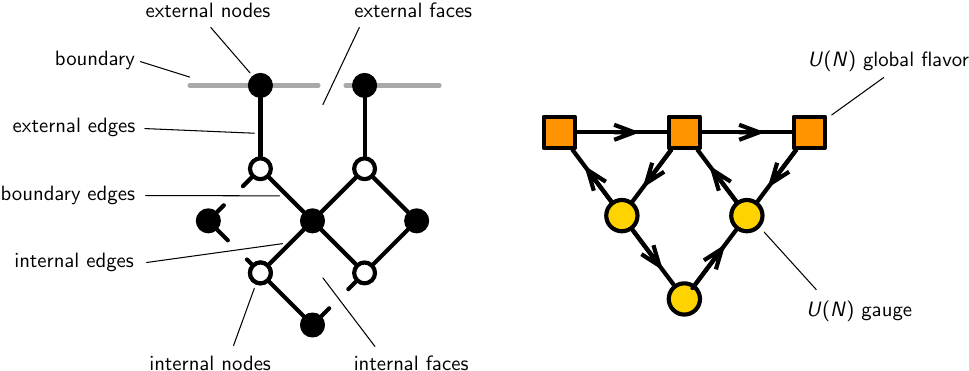}
\caption{
A bipartite graph on a bounded Riemann surface and the corresponding quiver diagram.
}
\label{fig_00}
\end{figure}

\paragraph{Dictionary.}
The following components of a bipartite graph on $\Sigma$ with $B\geq 0$ number of boundaries define the corresponding $4d$ $\mathcal{N}=1$ supersymmetric gauge theory, as introduced in \cite{Franco:2012mm}:

\begin{itemize}
\item \textbf{Faces.} Faces of the bipartite graph are bounded by edges and are considered to be all internal when $B=0$. When $B>0$, the faces which are adjacent to one of the boundaries of the bipartite graph are called external and are not fully encircled by edges of the bipartite graph. 

\noindent \textbf{Internal Faces.} Internal faces correspond to $U(N)$ gauge groups of the corresponding $4d$ $\mathcal{N}=1$ supersymmetric gauge theory. We note that the $U(1)$ part decouples from the gauge symmetry and contributes as baryonic directions of the \textit{master space} of the bipartite field theory, as we will elaborate on as part of the discussion on moduli spaces of bipartite field theories in the following section. 

\noindent \textbf{External Faces.} External faces correspond to $U(N)$ global (flavor) symmetry of the $4d$ $\mathcal{N}=1$ supersymmetric gauge theory. 

\item \textbf{Edges.} Edges correspond to chiral fields in the corresponding $4d$ $\mathcal{N}=1$ supersymmetric gauge theory.
 These fields transform in the bifundamental representation of the $U(N)$ symmetries identified with the neighboring two faces of the edge. 
The orientation on an edge adopted from the orientation associated to the two nodes where the edge ends identifies under which $U(N)$ symmetry the chiral field transforms as a fundamental and under which $U(N)$ symmetry the chiral field transforms as an anti-fundamental.
Since faces can be internal or external, we can classify edges into 3 different types as follows.

\noindent \textbf{Internal Edges.}
Internal edges correspond to chiral fields that transform in the bifundamental representation of two $U(N)$ gauge groups. These gauge groups correspond to the two neighboring internal faces of the bipartite graph. 

\noindent \textbf{Boundary Edges.}
Boundary edges correspond to chiral fields that transform either in the fundamental or anti-fundamental of a $U(N)$ gauge group and in the conjugate representation of a neighboring $U(N)$ flavor symmetry. These symmetries correspond respectively to the internal face and the external face adjacent to the boundary edge. 

\noindent \textbf{External Edges.}
External edges correspond to chiral fields that transforms in the bifundamental representation of two $U(N)$ flavor symmetries.

\item \textbf{Nodes.} Nodes in the bipartite graph are either white or black and have respectively clockwise and anti-clockwise orientation. 
This orientation is adopted by the edges connected to the nodes, determining the bifundamental representation of the chiral fields representation by the edges. 
Given that bipartite graphs can in general have $B>0$ boundaries, nodes can be attached to a boundary of the bipartite graph. We call these nodes external, whereas all other nodes are considered to be internal.

\noindent \textbf{Internal Nodes.} White internal nodes correspond to positive terms and black internal nodes correspond to negative terms in the superpotential $W$ of the $4d$ $\mathcal{N}=1$ supersymmetric gauge theory.
The set of chiral fields that form a particular superpotential term is formed by the edges connected to the corresponding internal node, where the order of fields in the term is given by the orientation around the node.

\noindent \textbf{External Nodes.} External nodes on boundaries of the bipartite graph have orientation according to their coloring, but do not correspond to terms in the superpotential $W$ of the $4d$ $\mathcal{N}=1$ theory.

\end{itemize}
\fref{fig_00} summarizes the above dictionary for a bipartite graph with $B$ boundaries. 
In general, we consider the rank of all gauge and flavor $U(N)$ symmetries to be the same $N$. 
For the remainder of this work, we concentrate on the case when $N=1$ such that the BFTs that we consider are all abelian. 
\\

\subsection{The Mesonic Moduli Space and the Master Space \label{smod}}

Bipartite field theories as $4d$ $\mathcal{N}=1$ supersymmetric gauge theories have two defining moduli spaces that are known as the mesonic moduli space $\mathcal{M}^{mes}$ \cite{Feng:2000mi,Feng:2001xr,Feng:2002fv,Feng:2002zw} and the master space $\mathcal{F}^\flat$ \cite{Forcella:2008eh,Forcella:2008bb,Forcella:2008ng,Forcella:2009bv,Hanany:2010zz,Zaffaroni:2008zz}. 
In the following work, we concentrate on moduli spaces of abelian BFTs
with $U(1)$ gauge symmetries and $U(1)$ global flavor symmetries. 

\paragraph{Master Space $\mathcal{F}^\flat$.}
The \textbf{master space $\mathcal{F}^\flat$} for a bipartite field theory with $U(1)$ gauge symmetries and $U(1)$ global flavor symmetries takes the following algebraic form, 
\beal{es05a1}
\mathcal{F}^\flat = \text{Spec} ~\mathbb{C}^{E} [X_{1}, \dots, X_{E}] / \langle \partial_{X_{i}} W = 0 \rangle ~,~
\eea
where $E$ is the number of chiral fields in the BFT corresponding to edges in the bipartite graph. 
The $F$-terms $\partial_{X_{i}} W = 0$ of the superpotential $W$ form a quotienting ideal in $\mathcal{F}^\flat$.

We note that the master space $\mathcal{F}^\flat$ for BFTs is analogous to master spaces defined for brane tilings \cite{Forcella:2008eh,Forcella:2008bb,Forcella:2008ng,Forcella:2009bv,Hanany:2010zz,Zaffaroni:2008zz} and more recently for brane brick models \cite{Franco:2015tna,Kho:2023dcm}.
In comparison to these master spaces, the master space $\mathcal{F}^\flat$ for BFTs has rich and unique algebro-geometric properties, which we summarize as follows:
\begin{itemize}

\item 
We restrict ourselves to bipartite graphs whose number of \textit{internal} white and internal black nodes is the same. 
This results in the $F$-terms $\partial_{X_{i}} W = 0$ of the superpotential $W$ to be binomial relations in terms of chiral fields.  
As a result, 
the resulting master space $\mathcal{F}^\flat$ in \eref{es05a1} is in terms of a \textbf{toric variety} \cite{fulton,cox1995homogeneous,sturmfels1996grobner}. 

\item
The master space $\mathcal{F}^\flat$ of a BFT is reducible into irreducible components under \textbf{primary decomposition} \cite{M2} of the  ideal $\langle \partial_{X_{i}} W = 0 \rangle$. 
The top-dimensional irreducible component is what we call the \textbf{coherent component} of the master space of the BFT, which we call ${}^{\text{Irr}}\mathcal{F}^\flat$.
In the following work, we will 
concentrate on the coherent component ${}^{\text{Irr}}\mathcal{F}^\flat$ and refer to it when we refer to master spaces. 

\item 
The master space  ${}^{\text{Irr}}\mathcal{F}^\flat$ of an abelian BFT exhibits the following global symmetries:

\noindent\textbf{Mesonic Part of the Global Symmetry.} 
Given that the BFT is defined by a bipartite graph on a Riemann surface $\Sigma$ with genus $g$,
there are $2g$ fundamental cycles on $\Sigma$, where each fundamental cycle contributes a $U(1)$ factor to the mesonic symmetry of the associated BFT \cite{Franco:2012mm,Franco:2012wv}. 
Accordingly, we identify the mesonic symmetry of a BFT defined on a Riemann surface $\Sigma$ with genus $g$ to be $U(1)_{x}^{2g}$ or an enhancement with a total rank of $2g$. 

\noindent\textbf{Baryonic Part of the Global Symmetry.} 
We note that each internal face of the bipartite graph on $\Sigma$ that defines a BFT corresponds to a $U(1)$ gauge symmetry. 
When $\Sigma$ has no boundaries, $B=0$, 
charges under one of the $U(1)$ gauge symmetries can be expressed in terms of charges under the other $U(1)$ gauge symmetries. 
This is because due to the bipartite nature of the graph on $\Sigma$. 
The internal faces are all even-sided and the number of fundamental and anti-fundamental chiral fields charged under a $U(1)$ gauge group is the same, making one of the $U(1)$ gauge symmetries dependent on the other $U(1)$ gauge symmetries. 
In the case when $\Sigma$ has boundaries, $B>0$, all $U(1)$ gauge symmetries are independent of each other..

The $U(1)$ gauge symmetries each contribute baryonic directions to the global symmetry of the master space ${}^{\text{Irr}}\mathcal{F}^\flat$ of a BFT.
Accordingly, the baryonic part of the global symmetry of the BFT master space ${}^{\text{Irr}}\mathcal{F}^\flat$ is written as 
$U(1)_{b}^{F_I-1}$, when $\Sigma$ has no boundaries with $F_I$ being the number of internal faces of the bipartite graph.
The baryonic part of the global symmetry becomes $U(1)_{b}^{F_I}$ when $\Sigma$ has boundaries. 

\noindent\textbf{Global Flavor Symmetries.} 
Every external face of the bipartite graph on $\Sigma$ with $B>0$ boundaries corresponds to a $U(1)_{f}$ global flavor symmetry of the BFT. 
Combined with the baryonic $U(1)_b$ of the master space global symmetry, the global flavor $U(1)_{f}$ are not all independent of each other when $B>0$.
Combined with the baryonic $U(1)_b$, the charges coming from one of the $U(1)_{f}$ of the global flavor symmetry of the BFT can be expressed in terms of the other $U(1)$ charges from the baryonic $U(1)_b$ and global flavor $U(1)_f$.
As a result, given in total $F_E$ external faces in the bipartite graph on $\Sigma$, the global flavor symmetry contribution to the master space ${}^{\text{Irr}}\mathcal{F}^\flat$ is given by $U(1)_{f}^{F_E - 1}$. 
We note that for $B=0$, there are no flavor symmetries for the master space ${}^{\text{Irr}}\mathcal{F}^\flat$.

\noindent\textbf{Global Boundary Symmetries.} 
For a BFT defined by a bipartite graph on $\Sigma$ with genus $g$ and $B>0$ boundaries, there are in addition to the $2g$ fundamental cycles on $\Sigma$, $B-1$ paths that connect the distinct $B$ boundaries of $\Sigma$. 
Each of these paths gives rise to an additional $U(1)_h$ global symmetry in the master space of the BFT, which we call here the \textbf{boundary global symmetry}. 

\item We can summarize the full global symmetry for BFTs defined on Riemann surfaces $\Sigma$ with $B=0$ and $B>0$ boundaries as follows, 
\beal{es05a10}
B= 0 &~:~&  U(1)_x^{2g} \times U(1)_b^{F_I - 1} \times U(1)_{R} ~,~\nn\\
B>0 &~:~&  U(1)_x^{2g} \times U(1)_b^{F_I} \times U(1)_f^{F_E - 1} \times U(1)_{h}^{B-1} \times U(1)_{R} ~,~
\eea
where $g$ is the genus of $\Sigma$, $F_I$ is the number of internal faces and $F_E$ is the number of external faces of the bipartite graph of the BFT, and $B$ is the number of boundaries of $\Sigma$.

\noindent\textbf{Dimension of the Master Space.} 
The dimension of the master space ${}^{\text{Irr}}\mathcal{F}^\flat$ of a BFT defined on $\Sigma$ with genus $g$ and $B$ boundaries can be obtained from the total rank of the corresponding global symmetry of ${}^{\text{Irr}}\mathcal{F}^\flat$ as follows:
\beal{es05a10}
B= 0 &~:~&  2g + F_I  ~,~\nn\\
B>0 &~:~&  2g + F_I + F_E + B - 1 ~.~
\eea

\end{itemize}

\paragraph{Mesonic Moduli Space $\mathcal{M}^{mes}$.}
The mesonic moduli space $\mathcal{M}^{mes}$ of a bipartite field theory is the gauge-invariant subspace of the master space ${}^{\text{Irr}}\mathcal{F}^\flat$. It is given as follows,
\beal{es05a20}
B= 0 &~:~& \mathcal{M}^{mes} = {}^{\text{Irr}}\mathcal{F}^\flat // U(1)_b^{F_{I} - 1} ~,~\nn\\
B> 0 &~:~& \mathcal{M}^{mes} = {}^{\text{Irr}}\mathcal{F}^\flat // U(1)_b^{F_{I}} ~,~
\eea
where $U(1)_b$ are the gauge symmetries coming from the $F_I$ internal faces of the BFT, which constitute the baryonic directions of the global symmetry of the master space ${}^{\text{Irr}}\mathcal{F}^\flat$.

The mesonic moduli space $\mathcal{M}^{mes}$ of a BFT therefore exhibits all global symmetries of its corresponding master space ${}^{\text{Irr}}\mathcal{F}^\flat$ except the baryonic directions given by $U(1)_b$. The remaining global symmetry associated to the BFT mesonic moduli space $\mathcal{M}^{mes}$ can be summarized as follows,
\beal{es05a25}
B= 0 &~:~&  U(1)_x^{2g}  \times U(1)_{R} ~,~\nn\\
B>0 &~:~&  U(1)_x^{2g}  \times U(1)_f^{F_E-1} \times U(1)_{h}^{B-1} \times U(1)_{R} ~.~
\eea
Accordingly, the dimension of the mesonic moduli space $\mathcal{M}^{mes}$ is given by,
\beal{es05a26}
B= 0 &~:~&  2g + 1  ~,~\nn\\
B>0 &~:~&  2g  + F_E + B - 1 ~.~
\eea

\paragraph{$U(1)_R$ Symmetry and Charges.} We expect the family of BFTs to flow to IR fixed points, as it is the case for brane tilings \cite{Franco:2005rj,Hanany:2005ve,Hanany:2012hi,Hanany:2015tgh} and similar theories proposed in \cite{Xie:2012mr,Heckman:2012jh}.
As discussed above, the $U(1)_R$ symmetry is part of the global symmetries of BFTs and charges under the $U(1)_R$ symmetries on chiral fields of the BFT can be determined using a process known as $a$-maximization \cite{Intriligator:2003jj,Martelli:2005tp,Butti:2005ps,Heckman:2012jh}.
This process involves the maximization of a trial $a$-function, which takes the form,
\beal{es11a01}
a_{\text{trial}}(R) = \frac{3}{32} \left(3 \Tr R^3 - \Tr R \right)
~.~
\eea
For BFTs, the trial $a$-function is determined by two constraints.
By denoting the $U(1)_R$ charge of a BFT chiral field $X_{ij}$ as $R(X_{ij})$,
the first constraint comes from the condition that the superpotential $W$ of a BFT has an overall $U(1)_R$ charge of $2$.
As a result, for every internal node of the bipartite graph corresponding to a superpotential term, we have
\beal{es11a02}
\text{for each internal node}~V : \sum_{X_{ij} \in V} R(X_{ij}) = 2 ~,~
\eea
where $V$ refers to the set of edges connected to the specific internal node of the BFT. 
The remaining constraint arises from the vanishing of the NSVZ beta function for each internal face of the BFT corresponding to a $U(N)$ gauge group of the corresponding $4d$ $\mathcal{N}=1$ theory.
These constraints take the following form, 
\beal{es11a03}
\text{for each internal face}~F : \sum_{X_{ij} \in F} (1-R(X_{ij})) = 2 ~,~
\eea
where $F$ refers to the set of edges forming the boundary of a specific face of the BFT.  
Combining these constraints, the trial $a$-function in \eref{es11a01} can be written as follows
\beal{es11a04}
a_{\text{trial}}(R) = \frac{3N^2}{32} \sum_{X_{ij}}{[3(R(X_{ij})-1)^3 - (R(X_{ij})-1)]}
~,~
\eea
where under maximization of the trial function above, the $U(1)_R$ charges $R(X_{ij})$ of chiral fields $X_{ij}$ are determined in the BFT. 
\\
 
\subsection{The Forward Algorithm and GLSM Fields \label{salg}}

The forward algorithm was first introduced in \cite{Feng:2000mi,Franco:2005rj} for brane tilings and corresponding $4d$ $\mathcal{N}=1$ supersymmetric gauge theories.
Applied to bipartite field theories, it allows us to construct the mesonic moduli space $\mathcal{M}^{mes}$ and the master space ${}^{\text{Irr}}\mathcal{F}^\flat$ of a BFT in terms of GLSM fields \cite{Witten:1993yc}.
In the following section, we review the forward algorithm for BFTs and illustrate how GLSM fields are used to parameterize the BFT moduli spaces. 

\paragraph{Forward Algorithm for BFTs.}
A BFT is defined on a bipartite graph $G$ on $\Sigma$.
The internal white and black nodes of the bipartite graph are each connected to an edge that represents a chiral field $X_{m}$ of the BFT. 
This implies that every chiral field $X_m$ appears precisely once in a positive and once in a negative term of the superpotential $W$ of the BFT. 
Due to this, the $F$-terms take the following binomial form,
\beal{es06a10}
\frac{\partial W}{\partial X_m} = F_m^{+} - F_{m}^{-} ~,~
\eea
where $F^{\pm}_m$ are monomial products of the chiral fields $X_m$.
Under these binomial relations, we note that not all chiral fields are independent to each other. 
This allows us to introduce a set of independent fields $v_k$, which we can express in terms of the original chiral fields $X_m$ as follows, 
\beal{es06a11}
X_m = \prod_{k=1}^{d_{\mathcal{F}^\flat}} v_k^{K_{mk}} ~,~
\eea
where $m=1, \dots, n^\chi$ is the label for the chiral fields $X_m$ with $n^\chi$ being the number of chiral fields in the BFT. 
The number of independent fields $v_k$ is given by the dimension of the BFT master space, $d_{\mathcal{F}^\flat}$, which is given in \eref{es05a10}.
As a result, we can identify the $K$-matrix to be $(n^\chi \times d_{\mathcal{F}^\flat})$-matrix, denoted as $K_{n^\chi \times d_{\mathcal{F}^\flat}}$.

The set of all $F$-terms form a binomial ideal
and the affine variety formed under this ideal is therefore toric \cite{fulton,cox1995homogeneous,sturmfels1996grobner}.
From the perspective of toric geometry, the row vectors $\vec{K}_m \in \mathbb{Z}^{d_{\mathcal{F}^\flat}}$ of the $K$-matrix in \eref{es06a11} form a cone $\mathbb{M}^+$.
This cone has a dual $\mathbb{N}^{+}$ whose defining vectors can be combined to form a matrix known as the $T$-matrix. 
The relationship between the $K$-matrix and the $T$-matrix is given by,
\beal{es06a12}
\vec{K} \cdot \vec{T} \geq 0 ~,~
\eea
where now the independent fields $v_k$ can be expressed in terms of a set of new fields $p_a$ via the $T$-matrix as follows,
\beal{es06a15}
v_k = \prod_{a=1}^{c} p_a^{T_{ka}}
\eea
Here, the new fields $p_a$ are identified as GLSM fields in the toric description of the BFT master space ${}^{\text{Irr}}\mathcal{F}^\flat$ and mesonic moduli space $\mathcal{M}^{mes}$. 
The number of GLSM fields is given by $c$, which makes the $T$-matrix a $(d_{\mathcal{F}^\flat} \times c)$-matrix denoted as $T_{d_{\mathcal{F}^\flat} \times c}$.

Under \eref{es06a11} and \eref{es06a12}, all chiral fields $X_m$ in the BFT can be expressed as products of GLSM fields $p_a$ as follows, 
\beal{es06a16}
X_m = \prod_{a=1}^{c} p_a^{P_{ma}} ~,~
\eea
where the GLSM field matrix $P$ is $(n^\chi \times c)$-dimensional and is defined by, 
\beal{es06a17}
P_{n^\chi \times c} = K_{n^\chi \times d_{\mathcal{F}^\flat}} \cdot T_{d_{\mathcal{F}^\flat} \times c} ~.~
\eea
Here, the labels on the matrices indicate their dimensions. 

The strict positivity of the entries of the $P$-matrix allows us to give a combinatorial meaning of the GLSM fields $p_a$ for BFTs, as first introduced in \cite{Franco:2012mm}.
As in the case of brane tilings \cite{Franco:2005rj,Hanany:2005ve,Hanany:2012hi} and brane brick models \cite{Franco:2015tna,Franco:2015tya},
the collection of chiral fields $X_m$ associated to a GLSM field $p_a$ defined by the $P$-matrix in \eref{es06a17}
correspond to a subset of edges in the bipartite graph of the BFT that connect exact once to all internal nodes of the bipartite graph.
While all internal nodes of the bipartite graph are connected uniquely once to an edge belonging to a chiral field in the GLSM field, not all external nodes are connected to the collection of edges.
Accordingly, we identify some GLSM fields of a BFT with \textit{almost} \textbf{perfect matchings} \cite{Franco:2012mm} of the corresponding bipartite graph on $\Sigma$. 
\\

The GLSM fields form a new basis that can be used to parameterize the $F$- and $D$-terms of the BFT. 
Under this change of basis, the collection of $F$- and $D$-terms of the BFT can be expressed in terms of $U(1)$ charges on the GLSM fields summarized by charge matrices called respectively the $Q_F$-matrix and the $Q_D$-matrix.
These charge matrices are defined as follows:
\begin{itemize}
\item \textbf{$Q_F$ matrix.}
The $F$-term relations of a BFT can be expressed in terms of a collection of $U(1)$ charges on the GLSM fields. 
These charges are summarized in a $((c-d_{\mathcal{F}^\flat}) \times c)$-matrix which is called the $Q_F$-matrix for BFTs.
It is given by the kernel of the $P$-matrix and takes the following form, 
\beal{es06a20}
( Q_{F} )_{(c-d_{\mathcal{F}^\flat}) \times c} = \text{ker} (P) ~.~
\eea

\item \textbf{$Q_D$ matrix.}
The $U(1)$ gauge charges of an abelian BFT are given by the $(F_I \times n^\chi)$-dimensional quiver incidence matrix $d$, where $F_I$ corresponds to the number of $U(1)_b$ gauge groups in the BFT.
When $B=0$, we note that all chiral fields in the BFT are bifundamental or adjoint, the incidence matrix satisfies $\sum_{i=1}^{F_I} d_{im} = 0$, where $m=1, \dots, n^\chi$. 
Accordingly, the quiver incidence matrix can be reduced to give a $((F_I-1) \times n^\chi)$-dimensional reduced quiver incidence matrix $\overline{d}$ when $B=0$. 
The $P$-matrix acts as a translation matrix between the chiral fields and GLSM fields and can be used to obtain the $Q_D$-matrix as follows, 
\beal{es06a25}
B= 0 &~:~&  \overline{d}_{(F_I-1) \times n^\chi} = (Q_{D})_{(F_I-1) \times c}  \cdot P^{t}_{c\times n^\chi}~,~\nn\\
B>0 &~:~&  d_{F_I \times n^\chi} = (Q_{D})_{F_I \times c}  \cdot P^{t}_{c\times n^\chi} ~.~
\eea
\end{itemize}
We can combine the $F$- and $D$-term charge matrices to form a total charge matrix as follows, 
\beal{es06a26}
B= 0 &~:~&  (Q_t)_{(c-d_{\mathcal{F}^\flat}+F_I - 1) \times c} = \Big( ( Q_{F} )_{(c-d_{\mathcal{F}^\flat}) \times c} ~,~ (Q_{D})_{(F_I-1) \times c} \Big)~,~\nn\\
B>0 &~:~&  (Q_t)_{(c-d_{\mathcal{F}^\flat}+F_I ) \times c} = \Big( ( Q_{F} )_{(c-d_{\mathcal{F}^\flat}) \times c} ~,~ (Q_{D})_{F_I \times c} \Big) ~.~
\eea
Accordingly, using the $F$- and $D$-term charge matrices and the GLSM fields encoded in the $P$-matrix, we can express the master space ${}^{\text{Irr}}\mathcal{F}^\flat$ and the mesonic moduli space $\mathcal{M}^{mes}$ of a BFT as follows, 
\beal{es06a26b}
{}^{\text{Irr}}\mathcal{F}^\flat &=& 
\text{Spec} ~\mathbb{C}[p_1, \dots, p_c] // Q_F
~,~
\nn\\
\mathcal{M}^{mes} &=& \text{Spec}~ \Big(\mathbb{C}[p_1, \dots, p_c] // Q_F\Big)// Q_D ~.~
\eea

The kernel of the total charge matrix $Q_t$ takes the following form, 
\beal{es06a27}
B= 0 &~:~&  (G_t)_{(d_{\mathcal{F}^\flat}-F_I + 1) \times c} = \text{ker}(Q_t)~,~\nn\\
B>0 &~:~&  (G_t)_{(d_{\mathcal{F}^\flat}-F_I) \times c} = \text{ker}(Q_t)~,~
\eea
and it encodes the \textbf{toric diagram} \cite{fulton,cox1995homogeneous,sturmfels1996grobner} of the mesonic moduli space $\mathcal{M}^{mes}$ of the BFT. 
We can see that the dimension $d_{\mathcal{M}^{mes}}$ of the mesonic moduli space can be identified as the number of rows of the toric diagram matrix $G_t$ as follows, 
\beal{es06a28}
B= 0 &~:~&  d_{\mathcal{M}^{mes}} = d_{\mathcal{F}^\flat}-F_I + 1 ~,~\nn\\
B>0 &~:~&  d_{\mathcal{M}^{mes}} = d_{\mathcal{F}^\flat}-F_I  ~,~
\eea
which matches the dimension formulas in \eref{es05a26}.
Each column of the $G_t$-matrix refers to a vertex in the toric diagram of the mesonic moduli space of the BFT.
As we note from the derivation of the $G_t$-matrices above, each such vertex in the toric diagram is associated to at least one of the GLSM fields in the corresponding BFT.
\\

\subsection{Hilbert Series and Plethystics \label{shs}}

The Hilbert series \cite{Benvenuti:2006qr,Butti:2006au,Butti:2007jv,hanany2007counting} captures the geometric structure of an algebraic variety, allowing us to construct using plethystics \cite{Feng:2007ur} the generators and the defining relations of the moduli spaces of a bipartite field theory. 
\\

\paragraph{Hilbert Series for BFTs.} Given an affine variety $Y$ over $\mathbb{C}^k[x_1, \dots, x_k]$, which in the context of the mesonic moduli space $\mathcal{M}^{mes}$ and the master space ${}^{\text{Irr}}\mathcal{F}^\flat$ of a BFT is a toric variety, the Hilbert series is defined as the generating function for the dimension of the graded pieces of the coordinate ring of $Y$.
When we refer to the mesonic moduli space $\mathcal{M}^{mes}$ and the master space ${}^{\text{Irr}}\mathcal{F}^\flat$ of a BFT, we are referring to the cone $\mathcal{X}$ over the affine variety Y in $\mathbb{C}^k$.
The general form of the coordinate ring of $Y$ is as follows, 
\beal{es10a09}
\mathbb{C}^k[x_1, \dots, x_k] / \mathcal{I} ~,~
\eea
where for the case of BFTs, $\mathcal{I}$ is the ideal in terms of $x_1, \dots, x_k$ of the variety $Y$.

The Hilbert series takes the form, 
\beal{es10a10}
g(t; \mathcal{X}) = \sum_{i=0}^{\infty} \text{dim}_{\mathbb{C}} (Y_i) t^i ~,~
\eea
where $\text{dim}_{\mathbb{C}} (Y_i)$ is the dimension of the $i$-th graded piece $Y_i$, which is the number of algebraically independent degree $i$ polynomials on the variety $Y$.
Here, $t$ is the fugacity that counts the degree of the polynomials. 
We can also introduce multiple fugacities $t_1, \dots, t_k$ when the coordinate ring in \eref{es10a09} is multi-graded with pieces $Y_{\vec{i}}$ and grading $\vec{i}=(i_1, \dots, i_k)$. 
The resulting refined Hilbert series takes the following form,
\beal{es10a11}
g(t_1, \dots, t_k; \mathcal{X}) = \sum_{\vec{i}=0}^{\infty} \text{dim}_{\mathbb{C}} (Y_{\vec{i}}) ~t_1^{i_1} \cdots  t_k^{i_k} ~.~
\eea
Here, the number of fugacities $t_i$ can be chosen to be as large as the dimension of the ambient space $\mathbb{C}^k$ or as few as the dimension of $\mathcal{X}$.
\\

\paragraph{Hilbert Series for the BFT Master Space.}
Focusing on the master space ${}^{\text{Irr}}\mathcal{F}^\flat$ of an abelian BFT, the Hilbert series becomes the generating function of the graded pieces of the coordinate ring of a toric variety $Y$, where the coordinate ring takes the form
\beal{es10a20}
\mathbb{C}^{E}[X_1, \dots, X_{n^\chi}] / \mathcal{I}_{\partial W}^{\text{Irr}} ~.~
\eea
Here, $X_{1}, \dots, X_{n^\chi}$ are the chiral fields of the BFT and $\mathcal{I}_{\partial W}^{\text{Irr}}$ is the top-dimensional irreducible component of the binomial ideal formed by the F-terms $\langle \partial_{X_i} W\rangle$ of the BFT. 
The coherent component $\mathcal{I}_{\partial W}^{\text{Irr}}$ can be obtained from $\langle \partial_{X_i} W\rangle$ by primary decomposition \cite{M2}.

The refined Hilbert series of the master space ${}^{\text{Irr}}\mathcal{F}^\flat$ then takes the following form, 
\beal{es10a25}
g(y_a, t; {}^{\text{Irr}}\mathcal{F}^\flat) = \sum_{\vec{a}=0}^{\infty}
\text{dim}_{\mathbb{C}}(Y_{\vec{q}}) ~ y_1^{q_1} \cdots y_{d-1}^{q_{d-1}} ~t^{q_d} ~,~
\eea
where $d=d_{\mathcal{F}^\flat}$ is the dimension of the master space ${}^{\text{Irr}}\mathcal{F}^\flat$.
The grading $(q_1, \dots, q_{d})$ is composed of charges $q_a$ coming from the $a$-th $U(1)$ factor of the global symmetry of the BFT master space. 
We note that the first $F_I-1$ of these $U(1)$ symmetries for $B=0$ and the first $F_I$ for $B>0$ BFTs are the $U(1)$ gauge symmetries that act as the baryonic part of the global symmetry of the master space ${}^{\text{Irr}}\mathcal{F}^\flat$.
Additionally, we choose the fugacity $y_d = t$ to correspond to the $U(1)_R$ symmetry.
\\

\paragraph{Hilbert Series for the BFT Mesonic Moduli Space.}
As the gauge-invariant subspace of the master space ${}^{\text{Irr}}\mathcal{F}^\flat$, the mesonic moduli space $\mathcal{M}^{mes}$ is invariant under the $F_I-1$ $U(1)$ gauge symmetries for $B=0$ BFTs and the $F_I$ $U(1)$ gauge symmetries for $B>0$ BFTs.
Accordingly, the Hilbert series of the mesonic moduli space $\mathcal{M}^{mes}$ is the generating function of gauge-invariant operators of the abelian BFT.
It can be obtained directly from the Hilbert series of the master space ${}^{\text{Irr}}\mathcal{F}^\flat$ in \eref{es10a25} by integrating out the fugacities corresponding to the $U(1)_b$ gauge symmetries that play the role of baryonic directions in the global symmetry of the master space ${}^{\text{Irr}}\mathcal{F}^\flat$.
We use the Molien integral formula \cite{Benvenuti:2006qr,Butti:2007jv} to obtain the Hilbert series of the mesonic moduli space $\mathcal{M}^{mes}$ as follows, 
\beal{es10a30}
g(y_{m+1}, \dots, y_{d-1}, t; \mathcal{M}^{mes})
= 
\prod_{i=1}^{G}
\oint_{|y_i|=1}
\frac{\ud y_i}{2\pi i y_i}
~
g(y_1, \dots, y_{d-1} , t; {}^{\text{Irr}}\mathcal{F}^\flat)
~,~
\eea
where $G$ is the number of $U(1)_b$ gauge symmetries in the abelian BFT.
Accordingly, $d-G$ is the dimension of the mesonic moduli space $\mathcal{M}^{mes}$.
The remaining fugacities $y_{k+1}, \dots, y_{d}$ correspond each to the $U(1)$ factors of the remaining global symmetry of the mesonic moduli space, including the $U(1)_x$ mesonic symmetry, $U(1)_f$ flavor symmetry, $U(1)_h$ boundary symmetry and the $U(1)_R$ symmetry as summarized in \eref{es05a25}.
\\

\paragraph{Hilbert Series and GLSM fields.}
The Hilbert series for the mesonic moduli space $\mathcal{M}^{mes}$ can be computed using the symplectic quotient description in \eref{es06a26b}.
The mesonic Hilbert series under the Molien integral formula \cite{Benvenuti:2006qr,Butti:2007jv} takes the following form,
\beal{es10a40}
g(y_a; \mathcal{M}^{mes})
= 
\prod_{i=1}^{c-d+G} \oint_{|z_i| = 1} \frac{\ud z_i}{2\pi i z_i}
\prod_{a=1}^{c}
\frac{1}{
(1-y_a \prod_{j=1}^{c-d+G} z_j^{(Q_t)_{ja}}) ~,~
}
\eea
where $c$ is the number of GLSM fields labelled by $a=1,\dots, c$,
$d=d_{\mathcal{F}^\flat}$ is the dimension of the master space ${}^{\text{Irr}}\mathcal{F}^\flat$, 
and $G$ is the number of $U(1)_b$ gauge symmetries in the abelian BFT.
Here, the fugacities $y_a$ correspond to GLSM fields $p_a$ and count their degrees.
The fugacities $y_a$ can be mapped to a new set of fugacities that count global symmetry charges carried by the GLSM fields.

By replacing the total charge matrix $Q_t$ in \eref{es10a40} with the $F$-term charge matrix $Q_F$, we can also calculate the Hilbert series of the master space ${}^{\text{Irr}}\mathcal{F}^\flat$ of the BFT. 
In the following work, we will focus on the computation of the Hilbert series of the mesonic moduli space $\mathcal{M}^{mes}$ for BFTs and leave the computation of the master space Hilbert series for future studies.

\paragraph{Plethystics.}
The Hilbert series contains information about the algebraic structure of the BFT moduli space.
We can make use of plethystics \cite{Benvenuti:2006qr,Feng:2007ur,Butti:2006au,Butti:2007jv,hanany2007counting} in order to obtain information about the generators of the moduli space as well as the defining relations that are formed amongst them. 
The plethystic logarithm of a Hilbert series $g(y_a; \mathcal{M})$ is defined as follows, 
\beal{es10a50}
PL[g(y_a; \mathcal{M})]
= 
\sum_{k=1}^{\infty}
\frac{\mu(k)}{k}
\log\left[
g(y_a^k; \mathcal{M})
\right]
~,~
\eea
where $\mu(k)$ is the M\"obius function.
We can identify from the plethystic logarithm of the Hilbert series whether the associated moduli space is a \textit{complete intersection} or not.
When the expansion of the plethystic logarithm is finite, then the moduli space is generated by a finite number of generators that are subject to a finite number of defining relations, which indicates that the associated moduli space is a complete intersection. 
The first positive terms of the expansion of the plethystic logarithm correspond to the generators of the moduli space, whereas any higher order terms correspond to relations among generators and relations among relations that are known as \textit{syzygies} \cite{Gray:2006jb,M2}. 
\\

In this work, we will focus on the computation of the Hilbert series for the mesonic moduli space $\mathcal{M}^{mes}$ of BFTs. 
For this computation, we will make use of both the description of the mesonic moduli space in terms of the chiral fields and the ideal formed by the binomial $F$-terms, as summarized in \eref{es05a20}, as well as the description in terms of GLSM fields and the symplectic quotient in terms of charge matrices $Q_F$ and $Q_D$ as summarized in \eref{es06a26b}.
We illustrate that for BFTs both descriptions of the mesonic moduli space $\mathcal{M}^{mes}$ are identical and that their Hilbert series are the same up to a change of fugacities. 
In certain BFT examples, we observe that a rearrangement and mapping of fugacities into characters of representations of hidden enhanced global symmetries in the Hilbert series allows us to identify such global symmetry enhancements in the mesonic moduli space $\mathcal{M}^{mes}$.
We note that these symmetry enhancements are not necessarily visible in the BFT Lagrangian.
In addition to detecting symmetry enhancements, we also summarize through the computation of the Hilbert series the generators and defining relations amongst generators for mesonic moduli spaces $\mathcal{M}^{mes}$ of a large class of BFTs.
We express the generators both in terms of chiral fields as well as the GLSM fields of the BFT and show how the global symmetry of the mesonic moduli space $\mathcal{M}^{mes}$ acts on them. 
\\

\section{BFTs on a disk \label{sec_e1}}

In this section, we study BFTs defined on a disk.
These BFTs have a tendency to have large mesonic moduli spaces $\mathcal{M}^{mes}$.
We observe in the first example that the global symmetry is enhanced.
We calculate the Hilbert series, generators and relations of the mesonic moduli spaces $\mathcal{M}^{mes}$, and express them in terms of representations of the full global symmetry.
\\

\subsection{$\text{BFT}[Q^{1,1,1}]$ Model \label{sec_e01}}

\begin{figure} [h]
\centering
\includegraphics{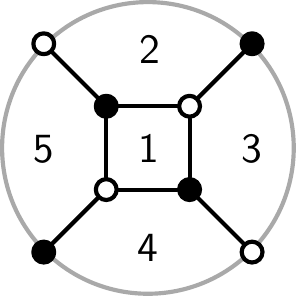}
\caption{
The bipartite graph for the $\text{BFT}[Q^{1,1,1}]$ Model.
\label{fig_10}
}
\end{figure}

\begin{figure} [ht!]
\centering
\includegraphics{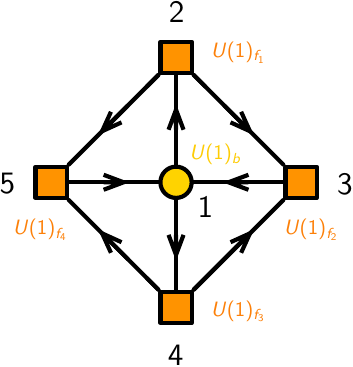}
\caption{
The quiver for the $\text{BFT}[Q^{1,1,1}]$ Model.
\label{fig_11}
}
\end{figure}

The bipartite graph and the quiver for the 
$\text{BFT}[Q^{1,1,1}]$ Model 
 are shown in \fref{fig_10} and \fref{fig_11}, respectively.
 The corresponding superpotential is given by,
\beal{es10a50}
W = X_{12} X_{23} X_{31} + X_{14} X_{45} X_{51} - X_{12} X_{25} X_{51} - X_{14} X_{43} X_{31} ~,~
\eea
where gauge and global symmetry charges on the chiral fields $X_{ij}$ are summarized in \tref{tab_10}.
We note here that one of the $U(1)_f$ flavor symmetries is not independent and decouples from the global symmetry of the BFT.
The $U(1)_R$ charges on chiral fields are obtained using $a$-maximization as summarized in section \sref{smod}.

\begin{table}[ht!]
\centering
\begin{tabular}{|c|c|c|c|c|c|c|c|}
\hline
\; & $U(1)_{b}$ & $U(1)_{f_1}$ & $U(1)_{f_2}$  & $U(1)_{f_3}$  & $U(1)_{f_4}$ & $U(1)_R$ & fugacity\\
\hline\hline
$X_{12}$  & $-1$ & + 1 & 0 & 0 & 0 & 1/2 & $b^{-1} f_1 \bar{t}$\\
$X_{14}$  &$-1$ & 0 & 0 & +1 & 0 & 1/2 & $b^{-1 } f_3 \bar{t}$\\
$X_{31}$  & $+1$ & 0 & $-1$ & 0 &0  &1/2 & $b f_2^{-1}\bar{t}$\\
$X_{51}$  & $+1$ & 0 & 0 & 0 & $-1$ &1/2 & $b f_4^{-1}\bar{t}$\\
\hline
$X_{23}$  & 0 & $-1$ & +1 & 0 & 0 & 1 & $f_1^{-1}f_2\bar{t}^2$\\
$X_{43}$  & 0 & 0 & +1 & $-1$ & 0 & 1& $f_2 f_3^{-1}\bar{t}^2$\\
$X_{25}$  & 0 & $-1$ & 0 & 0 & +1 & 1 &$f_1^{-1}f_4 \bar{t}^2$\\
$X_{45}$  & 0 & 0 & 0 & $-1$ & +1 & 1& $f_3^{-1} f_4 \bar{t}^2$\\
\hline
\end{tabular}
\caption{
Gauge and global symmetry charges on the chiral fields of the $\text{BFT}[Q^{1,1,1}]$ Model.
The $U(1)_R$-charges have been computed using $a$-maximization.
Note here that one of the $U(1)_f$ flavor symmetries is not independent and decouples from the global symmetry of the BFT.
}
\label{tab_10}
\end{table}

\paragraph{Forward Algorithm.}
The F-terms under the chiral fields corresponding to the internal edges of the BFT take the following form,
\beal{es10a51}
&
\partial_{X_{12}}W =
X_{23} X_{31}  - X_{24} X_{41} 
= 0 ~,~
\partial_{X_{31}}W =
X_{12} X_{23}  - X_{14} X_{43} 
= 0 ~,~
&
\nn\\
&
\partial_{X_{14}}W =
X_{45} X_{51}  - X_{43} X_{31} 
= 0 ~,~
\partial_{X_{51}}W =
X_{14} X_{45}  - X_{12} X_{25} 
= 0 ~.~
&
\eea
Under the above F-terms, we can see that the chiral fields are not all independent to each other.
We identify 5 independent chiral fields, 
\beal{es10a52}
v_1 = X_{14} ~,~
v_2 = X_{23} ~,~
v_3 = X_{25} ~,~
v_4 = X_{31} ~,~
v_5 = X_{45} ~,~
\eea
with which we can express the remaining chiral fields as follows,
\beal{es10a53}
X_{12} = \frac{v_1 v_5}{v_3} ~,~
X_{43} = \frac{v_2 v_5}{v_3} ~,~
X_{51} = \frac{v_2 v_4}{v_3} ~.~
\eea
We can encode the relations in \eref{es10a52} and \eref{es10a53} in the following $K$-matrix,
\beal{es10a60}
K=\left(
\begin{array}{c|ccccc}
& v_1 & v_2 & v_3 & v_4 & v_5\\
\hline
X_{12} & 1 & 0 & -1 & 0 & 1 \\
X_{14} & 1 & 0 & 0 & 0 & 0 \\
X_{31} & 0 & 0 & 0 & 1 & 0 \\
X_{51} & 0 & 1 & -1 & 1 & 0 \\
X_{23} & 0 & 1 & 0 & 0 & 0 \\
X_{43} & 0 & 1 & -1 & 0 & 1 \\
X_{25} & 0 & 0 & 1 & 0 & 0 \\
X_{45} & 0 & 0 & 0 & 0 & 1 \\
 \end{array}\right)~.~
 \eea
 
Using the forward algorithm for BFTs, we are able to obtain from the above $K$-matrix the perfect matching matrix as follows,
\beal{es10a61}
P=\left(
\begin{array}{c|cc|cc|c|cc}
\; & p_1 & p_2 & q & r & s & u_1 & u_2 \\
\hline
X_{12} & 0 & 1 & 0 & 0 & 0 & 1 & 0 \\
X_{14} & 1 & 0 & 0 & 0 & 0 & 1 & 0 \\
X_{31} & 0 & 0 & 1 & 0 & 0 & 0 & 1 \\
X_{51} & 0 & 0 & 0 & 1 & 0 & 0 & 1 \\
X_{23} & 1 & 0 & 0 & 1 & 1 & 0 & 0 \\
X_{43} & 0 & 1 & 0 & 1 & 1 & 0 & 0 \\
X_{25} & 1 & 0 & 1 & 0 & 1 & 0 & 0 \\
X_{45} & 0 & 1 & 1 & 0 & 1 & 0 & 0 \\
 \end{array}
\right)~.~
\eea
The perfect matchings in the BFT correspond to GLSM fields that can be used to parameterize the mesonic moduli space $\mathcal{M}^{mes}_{\text{BFT}[Q^{1,1,1}]}$.
In terms of the GLSM fields summarized in the $P$-matrix above, we can express the chiral fields as follows, 
\beal{es10a61b}
&
X_{12} = p_2 u_1 ~,~
X_{14} = p_1 u_1 ~,~
X_{31} = q u_2 ~,~
X_{51} = r u_2 ~,~
&
\nn\\
&
X_{23} = p_1 r s ~,~
X_{43} = p_2 r s ~,~
X_{25} = p_1 q s ~,~
X_{45} = p_2 q s 
~.~
&
\eea
Additionally, we can write down the $F$-term charge matrix,
\beal{es10a62}
Q_{F} =\left(
\begin{array}{cc|cc|c|cc}
p_1 & p_2 & q & r & s & u_1 & u_2 \\
\hline
 1 & 1 & 0 & 0 & -1 & -1 & 0 \\
 0 & 0 & 1 & 1 & -1 & 0 & -1 \\
\end{array}
\right)~.~
\eea
Similarly, we can obtain the D-term charges on the GLSM fields as follows,
\beal{es10a63}
Q_{D} =\left(
\begin{array}{cc|cc|c|cc}
p_1 & p_2 & q & r & s & u_1 & u_2 \\
\hline
 1 & 1 & -1 & -1 & 0 & 0 & 0 \\
\end{array}
\right)~.~
\eea

\begin{figure} [h]
\centering
\includegraphics[width=.35\textwidth]{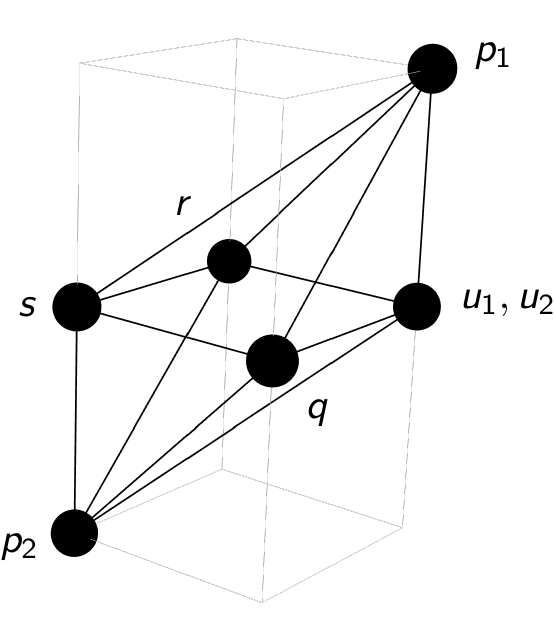}
\caption{
Toric diagram for the $\text{BFT}[Q^{1,1,1}]$ Model. 
}
\label{fig_12}
\end{figure}

The toric diagram for the $\text{BFT}[Q^{1,1,1}]$ Model is then given by, 
\beal{es10a65}
G_{t} =\left(
\begin{array}{cc|cc|c|cc}
p_1 & p_2 & q & r & s & u_1 & u_2 \\
\hline
 1 & 0 & 1 & 0 & 1 & 0 & 0 \\
 1 & 0 & 0 & 1 & 1 & 0 & 0 \\
 1 & -1 & 0 & 0 & 0 & 0 & 0 \\
 1 & 1 & 1 & 1 & 1 & 1 & 1 \\
\end{array}
\right)~,~
\eea
where \fref{fig_12} illustrates the toric diagram with perfect matching labels. 
\\

\paragraph{Hilbert Series.}
Using the $F$- and $D$-term charges on the GLSM fields $p_1, p_2, q, r, u_1, u_2$ summarized in \eref{es10a62} and \eref{es10a63}, we can compute the Hilbert series of the mesonic moduli space $\mathcal{M}^{mes}$ of the $\text{BFT}[Q^{1,1,1}]$ Model as follows, 
\beal{es10a70}
&&
g(t_1,t_2,y_q,y_r,y_s,y_u;\mathcal{M}^{mes}_{\text{BFT}[Q^{1,1,1}]})
=
(1-y_q y_r y_s^2 t_1 t_2 - y_q^2 y_s y_u t_1 t_2 
\nn\\
&&
\hspace{1cm}
-y_q y_r y_s y_u t_1^2-3  y_q y_r y_s y_u t_1 t_2- y_q y_r y_s y_u t_2^2-y_r^2 y_s y_u t_1 t_2 +2 y_q^2 y_r y_s^2 y_u t_1^2 t_2  
\nn\\
&&
\hspace{1cm}
+2 y_q^2 y_r y_s^2 y_u  t_1 t_2^2+2 y_q y_r^2 y_s^2 y_u   t_1^2 t_2+2 y_q y_r^2 y_s^2 y_u t_1 t_2^2 -y_q^2 y_r^2 y_s^3 y_u t_1^2 t_2^2 
\nn\\
&&
\hspace{1cm}
-y_q y_r y_u^2  t_1 t_2 +2  y_q^2 y_r y_s y_u^2  t_1^2 t_2 +2 y_q^2 y_r y_s y_u^2  t_1 t_2^2 +2 y_q y_r^2 y_s y_u^2  t_1^2 t_2  
\nn\\
&&
\hspace{1cm}
+2 y_q y_r^2 y_s y_u^2  t_1 t_2^2  -y_q^3 y_r y_s^2 y_u^2  t_1^2 t_2^2 -y_q^2 y_r^2 y_s^2 y_u^2 t_1^3 t_2 -3 y_q^2 y_r^2 y_s^2 y_u^2  t_1^2 t_2^2
\nn\\
&&
\hspace{1cm}
 -y_q^2 y_r^2 y_s^2 y_u^2  t_1 t_2^3 -y_q y_r^3 y_s^2 y_u^2  t_1^2 t_2^2 -y_q^2 y_r^2 y_s y_u^3 t_1^2 t_2^2 +y_q^3 y_r^3 y_s^3 y_u^3 t_1^3 t_2^3 )
 \nn\\
 &&
 \hspace{1cm} 
 \times \frac{1}{(1-y_q y_s t_1 ) (1- y_q y_s t_2 )  (1-y_r y_s t_1 )
 (1-y_r y_s t_2) (1-y_q y_u  t_1) }
 \nn\\
 &&\hspace{1cm}
 \times \frac{1}{(1-y_q y_u  t_2) (1-y_r y_u  t_1) (1-y_r y_u  t_2)}
 ~,~
 \eea
where the fugacities $t_1, t_2$ correspond to the GLSM fields $p_1,p_2$, and 
$y_q, y_r, y_s$ correspond to GLSM fields $q, r, s$. 
The fugacity $y_u = y_{u_1} y_{u_2}$ corresponds to the product of GLSM fields $u = u_1 u_2$.
The plethystic logarithm of the Hilbert series above is given by,
\beal{es10a71}
&&
PL[
g(t_1,t_2,y_q,y_r,y_s,y_u;\mathcal{M}^{mes}_{\text{BFT}[Q^{1,1,1}]})
]
= y_q y_s t_1+y_q y_s t_2 +y_r y_s t_1 + y_r y_s t_2 
\nn\\
&&
\hspace{1cm}
+ y_q y_u  t_1+y_q y_u  t_2 +y_r y_u  t_1 +y_r y_u t_2 -(y_q y_r y_s^2 t_1 t_2 +y_q y_r y_s y_u  t_1^2 
\nn\\
&&
\hspace{1cm} 
+3 y_q y_r y_s y_u  t_1 t_2 + y_q y_r y_s y_u t_2^2+y_q^2 y_s y_u t_1 t_2 +y_r^2 y_s y_u  t_1 t_2  
+ y_q y_r y_u^2  t_1 t_2 ) + \dots 
~,~
\nn\\
\eea
where the infinite expansion of the plethystic logarithm indicates that $\mathcal{M}^{mes}_{\text{BFT}[Q^{1,1,1}]}$ is a non-complete intersection.

\begin{table}[ht!]
\centering
\begin{tabular}{|l|c|c|c|c|c|}
\hline
\; & $SU(2)_{x}$ & $SU(2)_{y}$ & $SU(2)_{z}$ & $U(1)_{R}$ & fugacity\\
\hline\hline
$p_{1}$  & $+1$ & $0$& $0$ & $1/3$ &  $t_1 = x t^{1/3}$  \\
$p_{2}$  & $-1$ & $0$ & $0$&  $1/3$ &  $t_2 = x^{-1} t^{1/3}$\\
\hline
$q$  & $0$& $+1$  &  $0$  &  $1/3$& $y_q = y t^{1/3}$\\
$r$  & $0$& $-1$ & $0$ & $1/3$ &  $y_r = y^{-1} t^{1/3}$\\
\hline
$s$  & $0$ & $0$ & $+1$ & $1/3$&  $y_s = z t^{1/3}$\\
$u_1 $  & $0$ & $0$ & $-1/2$& $1/6$ & $y_{u_1} = z^{-1/2} t^{1/6}$ \\
$u_2 $  & $0$ & $0$& $-1/2$ & $1/6$ & $y_{u_2} = z^{-1/2} t^{1/6}$\\
\hline
\end{tabular}
\caption{Global symmetry charges under $SU(2)_x \times SU(2)_y \times SU(2)_z \times U(1)_R$ on perfect matchings and their corresponding fugacities for the $\text{BFT}[Q^{1,1,1}]$ Model.}
\label{tab01_2}
\end{table}

Using the following fugacity map, 
\beal{es10a75}
x=\frac{t_1^{1/2}}{t_2^{1/2}}~,~
y=\frac{y_q^{1/2}}{y_r^{1/2}}~,~
z=\frac{y_s^{1/2}}{y_{u_1}^{1/2} y_{u_2}^{1/2}}~,~
t=t_1^{1/2} t_2^{1/2} y_q^{1/2} y_r^{1/2} y_u^{1/2} y_{v_1}^{1/2} y_{v_2}^{1/2} ~,~
\eea
we can rewrite the Hilbert series in \eref{es10a70} in terms of characters of irreducible representations of the enhanced global symmetry $SU(2)_x \times SU(2)_y \times SU(2)_z \times U(1)_R$ as follows,
\beal{es10a76}
g(t,x,y,z;\mathcal{M}^{mes}_{\text{BFT}[Q^{1,1,1}]})=\sum_{n=0}^{\infty}[n;m;k]t^{3n}~,~
\eea
where $[n;m;k]=[n]_{SU(2)_x}[m]_{SU(2)_y}[k]_{SU(2)_z}$.

Here, we note that the fugacity map in \eref{es10a75} is based on the global symmetry charge assignment on GLSM fields summarized in \tref{tab01_2}.
The $U(1)_R$ charges on the GLSM fields come from the $U(1)_R$ charges on chiral fields summarized in \tref{tab_10}.
These have been computed via $a$-maximization.
\\

The plethystic logarithm of the refined Hilbert series in \eref{es10a76} takes the following form, 
\beal{es10a80}
PL[g(t,x,y,z;\mathcal{M}^{mes}_{\text{BFT}[Q^{1,1,1}]})]=[1;1;1]t^3-([2;0;0]+[0;2;0]+[0;0;2])t^6+\dots
~,~
\nn\\
\eea
where the first positive terms in the expansion correspond to generators of the mesonic moduli space, and the following negative terms correspond to the first order relations between the generators. 
We note from the infinite expansion of the plethystic logarithm that the mesonic moduli space of the BFT is not a complete intersection.
The generators of the form $A_{ijk}$ for the mesonic moduli space $\mathcal{M}^{mes}_{\text{BFT}[Q^{1,1,1}]}$ are summarized in \tref{tab01_3}.

\begin{table}[ht!]
\centering
\resizebox{0.95\textwidth}{!}{
\begin{tabular}{|c|c|c|c|c|c|c|c|}
\hline
generators  & GLSM fields & $SU(2)_{x}$ & $SU(2)_{y}$ & $SU(2)_{z}$  & $U(1)_R$ & Fugacity\\
\hline\hline
$A_{112}= X_{31} X_{12}$ & $p_1 q v_1 v_2 $ & $+1$ &$+1$&$-1$ & $3$ & $x y z^{-1} t^3$\\
$A_{212}= X_{41} X_{12}$ & $p_2 q v_1 v_2 $ & $-1$ &$+1$&$-1$ & $3$ & $x^{-1} y z^{-1} t^3$\\
$A_{122}=X_{31} X_{15}$ & $p_1 r v_1 v_2 $ & $+1$ &$-1$&$-1$ & $3$ & $x y^{-1} z^{-1} t^3$\\
$A_{222}=X_{41} X_{15}$ & $p_2 r v_1 v_2 $ & $-1$ &$-1$&$-1$ & $3$ & $x^{-1} y^{-1} z^{-1} t^3$\\
\hline
$A_{111}=X_{54}$ & $p_1 q u $ & $+1$ &$+1$&$+1$ & $3$ & $x y z t^3$\\
$A_{211}=X_{53}$ & $p_2 q u $ & $-1$ &$+1$&$+1$ & $3$ & $x^{-1} y z t^3$\\
$A_{121}=X_{24}$ & $p_1 r  u $ & $+1$ &$-1$&$+1$ & $3$ & $x y^{-1} z t^3$\\
$A_{221}=X_{23}$ & $p_2 r  u $ & $-1$ &$-1$&$+1$ & $3$ & $x^{-1} y^{-1} z t^ 3$\\
\hline
\end{tabular}}
\caption{
Generators of the mesonic moduli space of the $\text{BFT}[Q^{1,1,1}]$ Model. 
\label{tab01_3}
}
\end{table}

The generators in \tref{tab01_3} form $9$ quadratic relations that define the mesonic moduli space $\mathcal{M}^{mes}_{\text{BFT}[Q^{1,1,1}]}$. 
We can express the mesonic moduli space in terms of the generators and relations as follows, 
\beal{es10a82}
&&
\mathcal{M}^{mes}_{\text{BFT}[Q^{1,1,1}]}
= 
\text{Spec}~
\mathbb{C}[A_{ijk}] / \langle
\nn\\
&&
\hspace{1cm}
A_{122} A_{212} - A_{112} A_{222}~,~ 
A_{122} A_{221} - A_{121} A_{222}~,~ 
A_{122} A_{211} - A_{111} A_{222}~,~ 
\nn\\
&&
\hspace{1cm}
A_{212} A_{221} - A_{211} A_{222}~,~ 
A_{121} A_{211} - A_{111} A_{221}~,~ 
A_{121} A_{212} - A_{111} A_{222}~,~ 
\nn\\
&&
\hspace{1cm}
A_{112} A_{221} - A_{111} A_{222}~,~ 
A_{112} A_{121} - A_{111} A_{122}~,~ 
A_{112} A_{211} - A_{111} A_{212}~
\rangle
~.~
\nn\\
\eea
With the toric diagram in \fref{fig_12}, we identify the mesonic moduli space of the $\text{BFT}[Q^{1,1,1}]$ Model as the non-compact Calabi-Yau cone over $Q^{1,1,1}$ \cite{DAuria:1983sda,Nilsson:1984bj,Sorokin:1984ca,Franco:2015tna,Franco:2015tya}. 
\\

Let us also compute the Hilbert series using directly the binomial ideal formed by the $F$-terms of the BFT. 
Using the global symmetry charges on chiral fields summarized in \tref{tab_10}, 
we can compute the Hilbert series of the mesonic moduli space $\mathcal{M}^{mes}$ of the $\text{BFT}[Q^{1,1,1}]$ Model directly from the following description of the mesonic moduli space in terms of chiral fields $X_{ij}$, 
\beal{es10a85}
\mathcal{M}^{mes}_{\text{BFT}[Q^{1,1,1}]} = 
\text{Spec}~
\left(
\mathbb{C}[X_{ij}] / \mathcal{I}^{Irr}
\right)
// U(1)_b
~,~
\eea
where the irreducible ideal formed by the $F$-terms in \eref{es10a51} takes the form,
\beal{es10a86}
\mathcal{I}^{\text{Irr}} &= &\langle
X_{53} X_{24} - X_{54} X_{23} ~,~ X_{24} X_{41} - X_{23} X_{31} ~,~ X_{15} X_{53} - X_{12} X_{23} ~,~X_{15} X_{54} - X_{12} X_{24}\nn\\&&
X_{54} X_{41}-X_{53} X_{31}
\rangle ~.~ 
\eea
The resulting Hilbert series of the mesonic moduli space takes the following form,
\beal{es10a87}
&&
g(\bar{t},f_i;\mathcal{M}^{mes}_{\text{BFT}[Q^{1,1,1}]})
=
\nn\\
&&
\hspace{0.5cm}
\frac{
P(t,f_i)
}
{
(1-\frac{f_1}{f_2} t^3) (1-\frac{f_2}{f_1}t^3 ) (1-\frac{
   f_1}{f_3} t^3 ) (1-\frac{ f_3}{f_1} t^3 ) (1-\frac{f_2}{f_4} t^3 )(1-\frac{f_4}{f_2} t^3 )    (1-\frac{f_3}{f_4} t^3 ) (1-\frac{f_4}{f_3} t^3)}
\nn\\
\eea
where the numerator is 
\beal{es10a88}
&&
P(\bar{t},f_{i})
=
1-(3+\frac{f_2}{f_3}+\frac{f_3}{f_2}+\frac{f_1}{f_4}+\frac{f_2 f_3}{f_1 f_4}+\frac{f_4}{f_1}+\frac{f_1 f_4}{f_2 f_3}) \bar{t}^6 +(\frac{2 f_1}{f_2}+\frac{2 f_2}{f_1} 
\nn\\
&&
\hspace{1cm}
+\frac{2 f_1}{f_3} +\frac{2 f_3}{f_1}+\frac{2 f_2}{f_4}+\frac{2 f_3}{f_4}  +\frac{2 f_4}{f_2}+\frac{2 f_4}{f_3}) \bar{t}^9 -(3+\frac{f_2}{f_3}+\frac{f_3}{f_2}+\frac{f_1}{f_4}+\frac{f_2 f_3}{f_1 f_4}
\nn\\
&&
\hspace{1cm}
+\frac{f_4}{f_1}+\frac{f_1 f_4}{f_2 f_3}) \bar{t}^{12}+\bar{t}^{18} 
~.~
\eea
Here, the fugacity $\bar{t}$ keeps track of the $U(1)_R$ charge on the chiral fields and $f_i$ are the fugacities of the flavor $U(1)_{f_i}$ symmetries in the BFT's quiver, as summarized in \tref{tab_10}. 
The corresponding plethystic logarithm takes the form,
\beal{es10a89}
&&
PL[g(t,f_{i};\mathcal{M}^{mes}_{\text{BFT}[Q^{1,1,1}]})]
=
(
\frac{f_1}{f_2}+\frac{f_2}{f_1}+\frac{f_1}{f_3}+\frac{f_3}{f_1}+\frac{f_2}{f_4}+\frac{f_4}{f_2}+\frac{f_3}{f_4}+\frac{f_4}{f_3})\bar{t}^3
\nn\\
&&
\hspace{1cm}
-(3+\frac{f_2}{f_3}+\frac{f_3}{f_2}+\frac{f_1}{f_4}
+\frac{f_4}{f_1}+\frac{f_2 f_3}{f_1 f_4}+\frac{f_1 f_4}{f_2 f_3}) \bar{t}^6
+\dots~.~
\eea

Using the following fugacity map, 
\beal{es10a90}
x=\frac{f_3^{1/2}}{f_2^{1/2}}~,~
y=\frac{f_1^{1/2}}{f_4^{1/2}}~,~
z=\frac{f_2^{1/2} f_3^{1/2}}{f_1^{1/2} f_4^{1/2}}~,~
t= \bar{t}^{2}~,~
\eea
we can rewrite the Hilbert series in \eref{es10a87} in terms of characters of irreducible representations of the enhanced global symmetry $SU(2)_x \times SU(2)_y \times SU(2)_z \times U(1)_R$.
The resulting Hilbert series is identical to the one found in \eref{es10a76}. 

The unrefined Hilbert series with $f_i=1$ takes the following form,
\beal{es10a91}
g(\bar{t},f_{i}=1;\mathcal{M}^{mes}_{\text{BFT}[Q^{1,1,1}]})
=
\frac{1-9 \bar{t}^6+16 \bar{t}^9-9 \bar{t}^{12}+\bar{t}^{18}}{\left(1-\bar{t}^3\right)^8} 
~.~
\eea
One obtains the same unrefined Hilbert series when one sets the fugacities for GLSM fields in \eref{es10a70} as $t_1=t_2=y_q=y_r=y_s=t$ and $y_{u_i}=t^{1/2}$.
The numerator of the unrefined Hilbert series in \eref{es10a91} is palindromic, indicating that the mesonic moduli space is as expected Calabi-Yau. 
\\

\subsection{$\text{BFT}[\text{DM}_6]$ Model \label{sec_e40}}

\begin{figure} [ht!]
\centering
\includegraphics{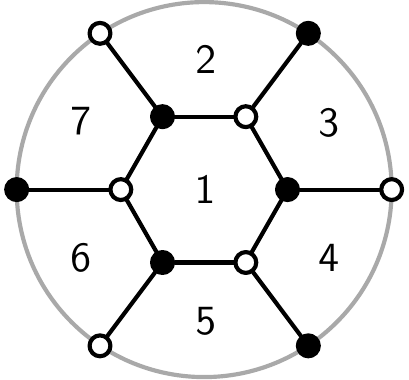}
\caption{
The BFT for $F_{I}=1$, $F_{E}=6$ on $\Sigma_{g=0,B=1}$.
\label{fig_40}
}
\end{figure}

\begin{figure} [h]
\centering
\includegraphics{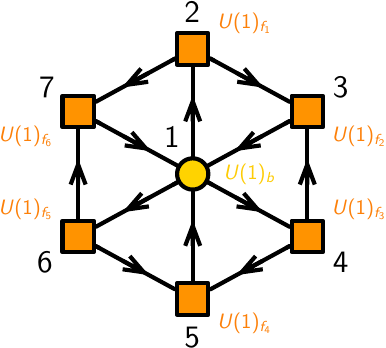}
\caption{
The quiver for the $F_{I}=6$, $F_{E}=6$ on $\Sigma_{g=0,B=1}$ Model.
\label{fig_41}
}
\end{figure}

\fref{fig_40} shows the bipartite graph on the disk corresponding to the $\text{BFT}[\text{DM}_6]$ Model.
The associated quiver diagram is shown in \fref{fig_41}.
The superpotential of the BFT is given by, 
\beal{es40a10}
W = 
&
X_{12} X_{23} X_{31}
+X_{14} X_{45} X_{51}
+X_{16} X_{67} X_{71}
&
\nn\\
&
-X_{12} X_{27} X_{71}
-X_{14} X_{43} X_{31}
-X_{16} X_{65} X_{51} 
~,~&
\eea
where \tref{tab_40} shows the gauge and global symmetry charges on the chiral fields $X_{ij}$.
We note here that one of the $U(1)_f$ flavor symmetries is not independent and decouples from the global symmetry of the BFT.
The $U(1)_R$ charges on the chiral fields are computed using $a$-maximization as summarized in section \sref{smod}.

\begin{table}[ht!]
\centering
\begin{tabular}{|c|c|c|c|c|c|c|c|c|c|}
\hline
\; & $U(1)_{b}$ & $U(1)_{f_1}$ & $U(1)_{f_2}$  & $U(1)_{f_3}$  & $U(1)_{f_4}$ & $U(1)_{f_5}$ & $U(1)_{f_6}$ & $U(1)_R$ & fugacity\\
\hline\hline
$X_{12}$  & $-1$ & + 1 & 0 & 0 & 0 & 0 & 0 & 2/3 & $b^{-1} f_1 \bar{t}$\\
$X_{14}$  & $-1$ & 0 & 0 & +1 & 0 & 0 & 0 &2/3& $b^{-1} f_3\bar{t}$\\
$X_{16}$  & $-1$ & 0 & 0 & 0 & 0 & +1 & 0 &2/3& $b^{-1} f_5\bar{t}$\\
$X_{31}$  & +1 & 0 & $-1$ & 0 & 0 & 0 & 0 & 2/3& $b f_2^{-1}\bar{t}$\\
$X_{51}$  & +1 & 0 & 0 & 0 & $-1$ & 0 & 0 & 2/3& $b f_4^{-1}\bar{t}$\\
$X_{71}$  & +1 & 0 & 0 & 0 & 0 & 0 & $-1$ & 2/3& $b f_6^{-1}\bar{t}$\\
\hline
$X_{23}$  & 0 & $-1$ & +1 & 0 & 0 & 0 & 0 & 2/3& $f_1^{-1} f_2 \bar{t}$\\
$X_{45}$  & 0 & 0 & 0 & $-1$ & +1 & 0 & 0 & 2/3& $f_3^{-1} f_4 \bar{t}$\\
$X_{67}$  & 0 & 0 & 0 & 0 & 0 & $-1$ & +1 & 2/3& $f_5^{-1} f_6 \bar{t}$\\
$X_{27}$  & 0 & $-1$ & 0 & 0 & 0 & 0 & +1 & 2/3& $f_1^{-1} f_6\bar{t}$\\
$X_{43}$  & 0 & 0 & +1 & $-1$ & 0 & 0 & 0 & 2/3& $f_2 f_3^{-1}\bar{t}$\\
$X_{65}$  & 0 & 0 & 0 & 0 & +1 & $-1$ & 0 & 2/3& $f_4 f_5^{-1}\bar{t}$\\
\hline
\end{tabular}
\caption{
Gauge and global symmetry charges on the chiral fields of the $\text{BFT}[\text{DM}_6]$ Model.
The $U(1)_R$-charges have been computed using $a$-maximization.
Note here that one of the $U(1)_f$ flavor symmetries is not independent and decouples from the global symmetry of the BFT.
}
\label{tab_40}
\end{table}

\paragraph{Forward Algorithm.}
We first summarize the $F$-terms of the $\text{BFT}[\text{DM}_6]$ Model as follows,
\beal{es40a11}
&
\partial_{X_{12}} W =
X_{23} X_{31} - X_{27} X_{71} ~,~
\partial_{X_{14}} W = 
X_{45} X_{51} - X_{43} X_{31}~,~
&
\nn\\
&
\partial_{X_{16}} W = 
X_{67} X_{71} - X_{65} X_{51}~,~
\partial_{X_{31}} W = 
X_{12} X_{23} - X_{14} X_{45} ~,~
&
\nn\\
&
\partial_{X_{51}} W = 
X_{14} X_{45} - X_{16} X_{65}  ~,~
\partial_{X_{71}} W = 
X_{16} X_{67} - X_{12} X_{27} ~.~
&
\eea
Under these $F$-terms, we can identify $7$ independent chiral fields, 
\beal{es40a12}
v_1 = X_{14} ~,~
v_2 = X_{16} ~,~
v_3 = X_{31} ~,~
v_4 = X_{51} ~,~
v_5 = X_{23} ~,~
v_6 = X_{27} ~,~
v_7 = X_{67} ~,~
&
\nn\\
\eea
which we can use to express the remaining chiral fields as follows,
\beal{es40a13}
X_{12} = \frac{v_2 v_7}{v_6}~,~
X_{71} = \frac{v_3 v_5}{v_6} ~,~
X_{43} = \frac{v_2 v_5 v_7}{v_1 v_6} ~,~
X_{45} = \frac{v_2 v_3 v_5 v_7}{v_1 v_4 v_6} ~,~
X_{65} = \frac{v_3 v_5v_7}{v_4 v_6} ~.~
\nn\\
\eea
The $K$-matrix encodes the above relations between independent chiral fields and the rest of the chiral fields of the BFT.
The $K$-matrix here takes the following form,
\beal{es40a60}
K=\left(
\begin{array}{c|ccccccc}
\; & v_1 & v_2 & v_3 & v_4 & v_5 & v_6 & v_7\\
\hline
X_{12} & 0 & 1 & 0 & 0 & 0 & -1 & 1 \\
X_{14} & 1 & 0 & 0 & 0 & 0 & 0 & 0 \\
X_{16} & 0 & 1 & 0 & 0 & 0 & 0 & 0 \\
X_{31} & 0 & 0 & 1 & 0 & 0 & 0 & 0 \\
X_{51} & 0 & 0 & 0 & 1 & 0 & 0 & 0 \\
X_{71} & 0 & 0 & 1 & 0 & 1 & -1 & 0 \\
X_{23} & 0 & 0 & 0 & 0 & 1 & 0 & 0 \\
X_{45} & -1 & 1 & 1 & -1 & 1 & -1 & 1 \\
X_{67} & 0 & 0 & 0 & 0 & 0 & 0 & 1 \\
X_{27} & 0 & 0 & 0 & 0 & 0 & 1 & 0 \\
X_{43} & -1 & 1 & 0 & 0 & 1 & -1 & 1 \\
X_{65} & 0 & 0 & 1 & -1 & 1 & -1 & 1 \\
\end{array}
\right)
~,~
 \eea

Using the $K$-matrix, the forward algorithm allow us to construct the $P$-matrix, summarizing the relationship between chiral fields of the BFT and the GLSM fields represented as perfect matchings of the bipartite graph of the BFT, 
\beal{es40a61}
P=\left(
\begin{array}{c|cccccccccccccccc|cc}
\; & p_1 & p_2 & p_3 & p_4 & p_5 & p_6 & p_7 & p_8 & p_9 & p_{10} & p_{11} & p_{12} & p_{13} & p_{14} & p_{15} & p_{16} & q_{1} & q_{2} \\
\hline
X_{12} & 1 & 1 & 1 & 1 & 0 & 0 & 0 & 0 & 0 & 0 & 0 & 0 & 0 & 0 & 0 & 0 & 1 & 0 \\
X_{14} & 1 & 0 & 0 & 0 & 1 & 1 & 1 & 0 & 0 & 0 & 0 & 0 & 0 & 0 & 0 & 0 & 1 & 0 \\
X_{16} & 0 & 1 & 0 & 0 & 1 & 0 & 0 & 1 & 1 & 0 & 0 & 0 & 0 & 0 & 0 & 0 & 1 & 0 \\
X_{31} & 0 & 0 & 0 & 0 & 0 & 0 & 0 & 0 & 1 & 0 & 0 & 0 & 1 & 1 & 1 & 0 & 0 & 1 \\
X_{51} & 0 & 0 & 0 & 1 & 0 & 0 & 0 & 0 & 0 & 0 & 1 & 1 & 0 & 1 & 0 & 0 & 0 & 1 \\
X_{71} & 0 & 0 & 0 & 0 & 0 & 0 & 1 & 0 & 0 & 0 & 0 & 1 & 0 & 0 & 1 & 1 & 0 & 1 \\
X_{23} & 0 & 0 & 0 & 0 & 1 & 1 & 1 & 1 & 0 & 1 & 1 & 1 & 0 & 0 & 0 & 1 & 0 & 0 \\
X_{45} & 0 & 1 & 1 & 0 & 0 & 0 & 0 & 1 & 1 & 1 & 0 & 0 & 1 & 0 & 1 & 1 & 0 & 0 \\
X_{67} & 1 & 0 & 1 & 1 & 0 & 1 & 0 & 0 & 0 & 1 & 1 & 0 & 1 & 1 & 0 & 0 & 0 & 0 \\
X_{27} & 0 & 0 & 0 & 0 & 1 & 1 & 0 & 1 & 1 & 1 & 1 & 0 & 1 & 1 & 0 & 0 & 0 & 0 \\
X_{43} & 0 & 1 & 1 & 1 & 0 & 0 & 0 & 1 & 0 & 1 & 1 & 1 & 0 & 0 & 0 & 1 & 0 & 0 \\
X_{65} & 1 & 0 & 1 & 0 & 0 & 1 & 1 & 0 & 0 & 1 & 0 & 0 & 1 & 0 & 1 & 1 & 0 & 0 \\
 \end{array}
\right)~.~
\eea
Each chiral field of the BFT can be expressed as a product of GLSM fields as follows,
\beal{es40a61b}
&
X_{12} = p_1 p_2 p_3 p_4 p_{17} ~,~
X_{14} = p_1 p_5 p_6 p_7 p_{17}~,~
X_{16} = p_2 p_5 p_8 p_9 p_{17} ~,~
&
\nn\\
&
X_{23} = p_5 p_6 p_7 p_8 p_{10} p_{11} p_{12} p_{16}~,~
X_{27} = p_5 p_6 p_8 p_9 p_{10} p_{11} p_{13} p_{14}~,~
X_{31} = p_9 p_{13} p_{14} p_{15} p_{18} ~,~
&
\nn\\
&
X_{43} = p_2 p_3 p_4 p_8 p_{10} p_{11} p_{12} p_{16} ~,~
X_{45} = p_2 p_3 p_8 p_9 p_{10} p_{13} p_{15} p_{16}~,~
X_{51} = p_4 p_{11} p_{12} p_{14} p_{18}~,~
&
\nn\\
&
X_{65} = p_1 p_3 p_6 p_7 p_{10} p_{13} p_{15} p_{16} ~,~
X_{67} =p_1 p_3 p_4 p_6 p_{10} p_{11} p_{13} p_{14} ~,~
X_{71}=p_7 p_{12} p_{15} p_{16} p_{18}~.~\nn\\
\eea
Using the GLSM fields, we can express the $F$-terms as a collection of $U(1)$ charges on the GLSM fields.
These charges are summarized in the following $Q_F$ charge matrix,
\beal{es40a62}
Q_{F} =\left(
\begin{array}{cccccccccccccccc|cc}
p_1 & p_2 & p_3 & p_4 & p_5 & p_6 & p_7 & p_8 & p_9 & p_{10} & p_{11} & p_{12} & p_{13} & p_{14} & p_{15} & p_{16} & q_{1} & q_{2} \\
\hline
 1 & 0 & 0 & 0 & 0 & 0 & 0 & 0 & 1 & 0 & 0 & 0 & 0 & -1 & -1 & 0 & -1 & 1 \\
 0 & 1 & 0 & 0 & 0 & 0 & 1 & 0 & 0 & 0 & 0 & 0 & 0 & 0 & 0 & -1 & -1 & 0 \\
 0 & 0 & 1 & 0 & 0 & 0 & 1 & 0 & 1 & 0 & 0 & 0 & 0 & -1 & -1 & -1 & -1 & 1 \\
 0 & 0 & 0 & 1 & 0 & 0 & 1 & 0 & 1 & 0 & 0 & 0 & 0 & -1 & 0 & -1 & -1 & 0 \\
 0 & 0 & 0 & 0 & 1 & 0 & -1 & 0 & -1 & 0 & 0 & 0 & 0 & 0 & 1 & 0 & 0 & 0 \\
 0 & 0 & 0 & 0 & 0 & 1 & -1 & 0 & 0 & 0 & 0 & 0 & 0 & -1 & 0 & 0 & 0 & 1 \\
 0 & 0 & 0 & 0 & 0 & 0 & 0 & 1 & -1 & 0 & 0 & 0 & 0 & 0 & 1 & -1 & 0 & 0 \\
 0 & 0 & 0 & 0 & 0 & 0 & 0 & 0 & 0 & 1 & 0 & 0 & 0 & -1 & 0 & -1 & 0 & 1 \\
 0 & 0 & 0 & 0 & 0 & 0 & 0 & 0 & 0 & 0 & 1 & 0 & 0 & -1 & 1 & -1 & 0 & 0 \\
 0 & 0 & 0 & 0 & 0 & 0 & 0 & 0 & 0 & 0 & 0 & 1 & 0 & 0 & 1 & -1 & 0 & -1 \\
 0 & 0 & 0 & 0 & 0 & 0 & 0 & 0 & 0 & 0 & 0 & 0 & 1 & -1 & -1 & 0 & 0 & 1 \\
\end{array}
\right)~.~
\eea
We can also obtain the charge matrix coming from the $D$-terms of the BFT as follows, 
\beal{es40a63}
Q_{D} =\left(
\begin{array}{cccccccccccccccc|cc}
p_1 & p_2 & p_3 & p_4 & p_5 & p_6 & p_7 & p_8 & p_9 & p_{10} & p_{11} & p_{12} & p_{13} & p_{14} & p_{15} & p_{16} & q_{1} & q_{2} \\
\hline
 0 & 0 & 0 & -1 & 0 & 0 & -1 & 0 & -1 & 0 & 0 & 1 & 0 & 1 & 1 & 0 & 0 & 0 \\
\end{array}
\right)~.~
\eea

As a result of the forward algorithm, we can compute the toric diagram of the mesonic moduli space of the $\text{BFT}[\text{DM}_6]$ Model using the $Q_F$- and $Q_D$-matrices.
The toric diagram of the mesonic moduli space is summarized by the following toric diagram matrix, 
\beal{es40a65}
G_{t} =\left(
\begin{array}{cccccccccccccccc|cc}
p_1 & p_2 & p_3 & p_4 & p_5 & p_6 & p_7 & p_8 & p_9 & p_{10} & p_{11} & p_{12} & p_{13} & p_{14} & p_{15} & p_{16} & q_{1} & q_{2} \\
\hline
 1 & 0 & 0 & 0 & 0 & 0 & 0 & -1 & 1 & -1 & -1 & -1 & 1 & 1 & 1 & -1 & 1 & 1 \\
 1 & -1 & 0 & 0 & 0 & 1 & 0 & -1 & 0 & 0 & 0 & -1 & 1 & 1 & 0 & -1 & 0 & 0 \\
  0 & 0 & 0 & -1 & 0 & 0 & 0 & 0 & 1 & 0 & -1 & -1 & 1 & 0 & 1 & 0 & 0 & 0 \\
 0 & 0 & 0 & 0 & 1 & 1 & 0 & 1 & 1 & 1 & 1 & 0 & 1 & 1 & 0 & 0 & 0 & 0 \\
 0 & 0 & 0 & 0 & 0 & 0 & -1 & 0 & 1 & 0 & 0 & -1 & 1 & 1 & 0 & -1 & 0 & 0 \\
 1 & 1 & 1 & 1 & 1 & 1 & 1 & 1 & 1 & 1 & 1 & 1 & 1 & 1 & 1 & 1 & 1 & 1 \\
\end{array}
\right)~.~
\eea
We can see here that the mesonic moduli space of the $\text{BFT}[\text{DM}_6]$ Model is a toric Calabi-Yau 6-fold.
The toric diagram has vertices labelled by the GLSM fields $p_1, \dots, p_{16}$ and $q_{1}, q_{2}$.
\\

\paragraph{Hilbert Series.}
We can calculate the Hilbert series of the mesonic moduli space using the GLSM fields as a basis.
The mesonic moduli space of the $\text{BFT}[\text{DM}_6]$ Model can be expressed in terms of the $Q_F$ and $Q_D$ charge matrices as follows, 
\beal{es40a65}
\mathcal{M}^{mes}_{\text{BFT}[\text{DM}_6]}
= \text{Spec}
\left(\mathbb{C}[p_1,\dots, p_{16}, q_1, q_2] // Q_F \right) // Q_D
~.~
\eea
The resulting Hilbert series computed using the Molien integral formula summarized in section \sref{shs} takes the following form,
\beal{es40a70}
&&
g(t_i;\mathcal{M}^{mes}_{\text{BFT}[\text{DM}_6]})
=
\frac{P(t_i)}{
(1-t_1 t_3 t_4 t_6 t_{10} t_{11} t_{13} t_{14})(1-t_5 t_6 t_8 t_9 t_{10} t_{11} t_{13} t_{14})
}
\nn\\
&&
\hspace{1cm}
\times
\frac{1}{
(1-t_2 t_3 t_4 t_8 t_{10} t_{11} t_{12} t_{16})(1-t_5 t_6 t_7 t_8 t_{10} t_{11} t_{12} t_{16})(1-t_1 t_3 t_6 t_7 t_{10} t_{13} t_{15} t_{16})
}
\nn\\
&&
\hspace{1cm}
\times
\frac{1}{
(1-t_2 t_3 t_8 t_9 t_{10} t_{13} t_{15} t_{16})(1-t_1 t_2 t_3 t_4^2 t_{11} t_{12} t_{14} y_{q})
(1-t_1 t_4 t_5 t_6 t_7 t_{11} t_{12} t_{14} y_{q})
}
\nn\\
&&
\hspace{1cm}
\times
\frac{1}{
(1-t_2 t_4 t_5 t_8 t_9 t_{11} t_{12} t_{14} y_{q})
(1-t_1 t_2 t_3 t_4 t_9 t_{13} t_{14} t_{15} y_{q})(1-t_1 t_5 t_6 t_7 t_9 t_{13} t_{14} t_{15} y_{q})
}
\nn\\
&&
\hspace{1cm}
\times
\frac{1}{
(1-t_2 t_5 t_8 t_9^2 t_{13} t_{14} t_{15} y_{q})(1-t_1 t_2 t_3 t_4 t_7 t_{12} t_{15} t_{16} y_{q})
(1-t_1 t_5 t_6 t_7^2 t_{12} t_{15} t_{16} y_{q})
}
\nn\\
&&
\hspace{1cm}
\times
\frac{1}{(1-t_2 t_5 t_7 t_8 t_9 t_{12} t_{15} t_{16} y_{q})}
~,~
\eea
where the numerator of the Hilbert series $P(t_i)$ is not written in full, but can be identified to be palindromic. 
The fugacities $t_a$ correspond to the GLSM fields $p_a$, and the fugacity $y_q = y_{q_1} y_{q_2}$ counts the product of GLSM fields $q= q_1 q_2$.
The plethystic logarithm of the Hilbert series above takes the following form, 
\beal{es40a71}
&&
PL[g(t_i;\mathcal{M}^{mes}_{\text{BFT}[\text{DM}_6]})]
=
t_1 t_3 t_4 t_6 t_{10} t_{11} t_{13} t_{14} + t_5 t_6 t_8 t_9 t_{10} t_{11} t_{13} t_{14}
\nn\\
&&
\hspace{1cm}
+t_2 t_3 t_4 t_8 t_{10} t_{11} t_{12} t_{16}
+t_5 t_6 t_7 t_8 t_{10} t_{11} t_{12} t_{16}+t_1 t_3 t_6 t_7 t_{10} t_{13} t_{15} t_{16}
\nn\\
&&
\hspace{1cm}
+t_2 t_3 t_8 t_9 t_{10} t_{13} t_{15} t_{16} 
+t_1 t_2 t_3 t_4^2 t_{11} t_{12} t_{14} y_{q}+t_1 t_4 t_5 t_6 t_7 t_{11} t_{12} t_{14} y_{q}
\nn\\
&&
\hspace{1cm}
+t_2 t_4 t_5 t_8 t_9 t_{11} t_{12} t_{14} y_{q}
+t_1 t_2 t_3 t_4 t_9 t_{13} t_{14} t_{15} y_{q}+t_1 t_5 t_6 t_7 t_9 t_{13} t_{14} t_{15} y_{q}
\nn\\
&&
\hspace{1cm}
+t_2 t_5 t_8 t_9^2 t_{13} t_{14} t_{15} y_{q}
t_1 t_2 t_3 t_4 t_7 t_{12} t_{15} t_{16} y_{q}+t_1 t_5 t_6 t_7^2 t_{12} t_{15} t_{16} y_{q}
\nn\\
&&
\hspace{1cm}
+t_2 t_5 t_7 t_8 t_9 t_{12} t_{15} t_{16} y_{q}
-(t_1 t_2 t_3^2 t_4 t_5 t_6^2 t_7 t_8^2 t_9 t_{10}^3 t_{11}^2 t_{12} t_{13}^2 t_{14} t_{15} t_{16}^2
\nn\\
&&
\hspace{1cm}
+t_1 t_2 t_3 t_4^2 t_5 t_6 t_8 t_9 t_{10} t_{11}^2 t_{12} t_{13} t_{14}^2 y_{q}
+t_1 t_2 t_3 t_4 t_5 t_6 t_8 t_9^2 t_{10} t_{11} t_{13}^2 t_{14}^2 t_{15} y_{q}
\nn\\
&&
\hspace{1cm}
+t_1 t_2 t_3 t_4^2 t_5 t_6 t_7 t_8 t_{10} t_{11}^2 t_{12}^2 t_{14} t_{16} y_{q}
+t_1^2 t_2 t_3^2 t_4^2 t_6 t_7 t_{10} t_{11} t_{12} t_{13} t_{14} t_{15} t_{16} y_{q}
\nn\\
&&
\hspace{1cm}
+t_1^2 t_3 t_4 t_5 t_6^2 t_7^2 t_{10} t_{11} t_{12} t_{13} t_{14} t_{15} t_{16} y_{q}
+t_1 t_2^2 t_3^2 t_4^2 t_8 t_9 t_{10} t_{11} t_{12} t_{13} t_{14} t_{15} t_{16} y_{q} 
\nn\\
&&
\hspace{1cm}
+t_1 t_2 t_5^2 t_6 t_7^2 t_8 t_9^2 t_{12} t_{13} t_{14} t_{15}^2 t_{16} y_{q}^2
+t_1 t_5^2 t_6^2 t_7^2 t_8 t_9 t_{10} t_{11} t_{12} t_{13} t_{14} t_{15} t_{16} y_{q}
\nn\\
&&
\hspace{1cm}
+t_2^2 t_3 t_4 t_5 t_8^2 t_9^2 t_{10} t_{11} t_{12} t_{13} t_{14} t_{15} t_{16} y_{q}
+t_2 t_5^2 t_6 t_7 t_8^2 t_9^2 t_{10} t_{11} t_{12} t_{13} t_{14} t_{15} t_{16} y_{q}
\nn\\
&&
\hspace{1cm}
+t_1 t_2 t_3 t_5 t_6 t_7 t_8 t_9^2 t_{10} t_{13}^2 t_{14} t_{15}^2 t_{16} y_{q}
+t_1 t_2 t_3 t_4 t_5 t_6 t_7^2 t_8 t_{10} t_{11} t_{12}^2 t_{15} t_{16}^2 y_{q}
\nn\\
&&
\hspace{1cm}
+t_1 t_2 t_3 t_5 t_6 t_7^2 t_8 t_9 t_{10} t_{12} t_{13} t_{15}^2 t_{16}^2 y_{q}
+t_1^2 t_2 t_3 t_4^2 t_5 t_6 t_7 t_9 t_{11} t_{12} t_{13} t_{14}^2 t_{15} y_{q}^2
\nn\\
&&
\hspace{1cm}
+t_1 t_2^2 t_3 t_4^2 t_5 t_8 t_9^2 t_{11} t_{12} t_{13} t_{14}^2 t_{15} y_{q}^2
+t_1 t_2 t_4 t_5^2 t_6 t_7 t_8 t_9^2 t_{11} t_{12} t_{13} t_{14}^2 t_{15} y_{q}^2
\nn\\
&&
\hspace{1cm}
+t_1^2 t_2 t_3 t_4^2 t_5 t_6 t_7^2 t_{11} t_{12}^2 t_{14} t_{15} t_{16} y_{q}^2
+t_1 t_2^2 t_3 t_4^2 t_5 t_7 t_8 t_9 t_{11} t_{12}^2 t_{14} t_{15} t_{16} y_{q}^2
\nn\\
&&
\hspace{1cm}
+t_1 t_2 t_4 t_5^2 t_6 t_7^2 t_8 t_9 t_{11} t_{12}^2 t_{14} t_{15} t_{16} y_{q}^2
+t_1^2 t_2 t_3 t_4 t_5 t_6 t_7^2 t_9 t_{12} t_{13} t_{14} t_{15}^2 t_{16} y_{q}^2
\nn\\
&&
\hspace{1cm}
+t_1 t_2^2 t_3 t_4 t_5 t_7 t_8 t_9^2 t_{12} t_{13} t_{14} t_{15}^2 t_{16} y_{q}^2
+5 t_1 t_2 t_3 t_4 t_5 t_6 t_7 t_8 t_9 t_{10} t_{11} t_{12} t_{13} t_{14} t_{15} t_{16} y_{q})
\nn\\
&&
\hspace{1cm}
+\dots~,~
\eea
where we can see that the infinite expansion of the plethystic logarithm indicates that the mesonic moduli space of the $\text{BFT}[\text{DM}_6]$ Model is a non-complete intersection.

Using the following fugacity map, 
\beal{es40a75}
&
\bar{f}_1
= \frac{t_1^{1/4} t_2^{1/4}}{t_3^{1/4} t_4^{1/4}} ~,~
\bar{f}_2
= \frac{t_5^{1/4} t_6^{1/4}}{t_7^{1/4} t_8^{1/4}} ~,~
\bar{f}_3
= \frac{t_9^{1/4} t_{10}^{1/4}}{t_{11}^{1/4} t_{12}^{1/4}} ~,~
\bar{f}_4
= \frac{t_{13}^{5/12} t_{14}^{5/12}}{t_{15}^{1/12} t_{16}^{1/12} y_{q}^{1/12}} ~,~
\bar{f}_5
= \frac{t_{13}^{1/3} t_{14}^{1/3}}{t_{15}^{1/3} t_{16}^{1/3} y_{q}^{1/6}} ~,~
&
\nn\\
&
t =
t_1^{1/24} t_2^{1/24} t_3^{1/24}
t_4^{1/24} t_5^{1/24} t_6^{1/24}
t_7^{1/24} t_8^{1/24} t_9^{1/24}
t_{10}^{1/24} t_{11}^{1/24} t_{12}^{1/24}
t_{13}^{1/24} t_{14}^{1/24} t_{15}^{1/24}
t_{16}^{1/24}
y_{q}^{1/24}
~,~
&
\nn\\
\eea
we can express the Hilbert series of the mesonic moduli space of the $\text{BFT}[\text{DM}_6]$ Model in terms of fugacities associated to the global symmetry $U(1)_{\bar{f}_1} \times U(1)_{\bar{f}_2}  \times  U(1)_{\bar{f}_3}  \times  U(1)_{\bar{f}_4}  \times U(1)_{\bar{f}_5}  \times U(1)_{R}$.
By keeping just the fugacity $t$ for the $U(1)_R$ symmetry, unrefined Hilbert series becomes
\beal{es40a76}
&&
g(t,\bar{f}_i=1;\mathcal{M}^{mes}_{\text{BFT}[\text{DM}_6]})
=
\nn\\
&&
\hspace{1cm}
\frac{
1 + 5 t^8 + 19 t^{16} + 37 t^{24} + 50 t^{32} + 37 t^{40} + 19 t^{48} + 5 t^{56} + t^{64}
}{
(1 - t^8) (1 - t^{16})^5
}~,~
\eea
where we see clearly the palindromic nature of the numerator of the Hilbert series.
This indicates that the mesonic moduli space of the $\text{BFT}[\text{DM}_6]$ Model is Calabi-Yau. 

The plethystic logarithm of the refined Hilbert series in \eref{es40a65} takes the form, 
\beal{es40a77}
&&
PL[g(t,\bar{f}_i;\mathcal{M}^{mes}_{\text{BFT}[\text{DM}_6]})]
=
\Big(
\frac{\bar{f}_{5}}{\bar{f}_{1} \bar{f}_{2} \bar{f}_{3} \bar{f}_{4}}+\frac{\bar{f}_{2} \bar{f}_{4}^2}{\bar{f}_{1}}+\frac{\bar{f}_{3}^2 \bar{f}_{5}^2}{\bar{f}_{2} \bar{f}_{4}}+\bar{f}_{2} \bar{f}_{3} \bar{f}_{4}^2+\frac{\bar{f}_{3} \bar{f}_{5}^2}{\bar{f}_{4}}+\frac{\bar{f}_{5}}{\bar{f}_{3} \bar{f}_{4}}
\Big) t^{8}
\nn\\
&&
\hspace{1cm}
+
\Big(
\frac{\bar{f}_{1} \bar{f}_{2} \bar{f}_{3} \bar{f}_{4}}{\bar{f}_{5}}+\frac{\bar{f}_{1}}{\bar{f}_{2} \bar{f}_{4}^2}+\frac{\bar{f}_{4}}{\bar{f}_{1} \bar{f}_{3}^2 \bar{f}_{5}^2}+\frac{\bar{f}_{1} \bar{f}_{3}^2 \bar{f}_{4}}{\bar{f}_{5}}+\frac{\bar{f}_{1}}{\bar{f}_{3} \bar{f}_{4}^2}+\frac{\bar{f}_{2} \bar{f}_{4}}{\bar{f}_{3}^2 \bar{f}_{5}^2}+\frac{1}{\bar{f}_{2} \bar{f}_{3} \bar{f}_{4}^2}+\frac{\bar{f}_{4}}{\bar{f}_{3} \bar{f}_{5}^2}
\nn\\
&&
\hspace{1cm}
+\frac{\bar{f}_{3} \bar{f}_{4}}{\bar{f}_{5}}
\Big) t^{16}
-
\Big(
\frac{\bar{f}_{1} \bar{f}_{3} \bar{f}_{5}^2}{\bar{f}_{2} \bar{f}_{4}^3}
+\frac{\bar{f}_{2} \bar{f}_{4}^3}{\bar{f}_{1} \bar{f}_{3} \bar{f}_{5}^2}
+\bar{f}_{1} \bar{f}_{2}
+\frac{1}{\bar{f}_{1} \bar{f}_{2}}
+\bar{f}_{1} \bar{f}_{3}^3 \bar{f}_{5}
+\frac{1}{\bar{f}_{1} \bar{f}_{3}^3 \bar{f}_{5}}
+\frac{\bar{f}_{3} \bar{f}_{5}^3}{\bar{f}_{1}}
\nn\\
&&
\hspace{1cm}
+\bar{f}_{1} \bar{f}_{3}+\frac{1}{\bar{f}_{1} \bar{f}_{3}}+\frac{\bar{f}_{2} \bar{f}_{3}^2 \bar{f}_{4}^3}{\bar{f}_{5}}+\frac{\bar{f}_{5}}{\bar{f}_{2} \bar{f}_{3}^2 \bar{f}_{4}^3}+\frac{\bar{f}_{3}}{\bar{f}_{2}}+\frac{\bar{f}_{2}}{\bar{f}_{3}}+5
\Big) t^{24}
- 
\Big(
\frac{\bar{f}_{1}^2 \bar{f}_{3}}{\bar{f}_{4} \bar{f}_{5}}+\frac{\bar{f}_{1} \bar{f}_{3}}{\bar{f}_{2} \bar{f}_{4} \bar{f}_{5}}
\nn\\
&&
\hspace{1cm}
+\frac{\bar{f}_{1} \bar{f}_{2} \bar{f}_{4}^2}{\bar{f}_{5}^3}+\frac{\bar{f}_{1}}{\bar{f}_{3}^2 \bar{f}_{4} \bar{f}_{5}^2}+\frac{\bar{f}_{1}}{\bar{f}_{4}
   \bar{f}_{5}}+\frac{1}{\bar{f}_{2} \bar{f}_{3}^2 \bar{f}_{4} \bar{f}_{5}^2}+\frac{\bar{f}_{2} \bar{f}_{4}^2}{\bar{f}_{3}
   \bar{f}_{5}^3}+\frac{1}{\bar{f}_{3}^3 \bar{f}_{4} \bar{f}_{5}^2}+\frac{\bar{f}_{4}^2}{\bar{f}_{5}^3}
\Big) t^{32}+ \dots
~,~
\nn\\
\eea
which further indicates that the mesonic moduli space is not a complete intersection.
From here, we can identify the generators of the mesonic moduli space, which are summarized in \tref{tab40_3}.

\begin{table}[ht!]
\centering
\begin{tabular}{|l|c|c|c|c|c|c|c|}
\hline
\; & $U(1)_{\bar{f}_1}$ & $U(1)_{\bar{f}_2}$ & $U(1)_{\bar{f}_3}$ & $U(1)_{\bar{f}_4}$ & $U(1)_{\bar{f}_5}$ & $U(1)_{R}$ & fugacity\\
\hline\hline
$p_1$ &  +1 & 0 & 0 & 0 & 0 &  $1/12$ & $t_1 = \bar{f}_1 t$\\
$p_2$ &  +1 & 0 & 0 & 0 & 0 &  $1/12$ & $t_2 = \bar{f}_1 t$\\
$p_3$ &  $-1$ & 0 & 0 & 0 & 0 &  $1/12$ & $t_3 = \bar{f}_1^{-1} t$\\
$p_4$ &  $-1$ & 0 & 0 & 0 & 0 &  $1/12$ & $t_4 = \bar{f}_1^{-1} t$\\
$p_5$ &  0 & +1 & 0 & 0 & 0 &  $1/12$ & $t_5 =\bar{f}_2 t$\\
$p_6$ &  0 & +1 & 0 & 0 & 0 &  $1/12$ & $t_6 = \bar{f}_2 t$\\
$p_7$ &  0 & $-1$ & 0 & 0 & 0 &  $1/12$ & $t_7 = \bar{f}_2^{-1} t$\\
$p_8$ &  0 & $-1$ & 0 & 0 & 0 &  $1/12$ & $t_8 = \bar{f}_2^{-1} t$\\
$p_9$ &  0 & 0 & +1 & 0 & 0 &  $1/12$ & $t_9 =\bar{f}_3 t$\\
$p_{10}$ & 0 & 0 & +1 & 0 & 0 &  $1/12$ & $t_{10} =\bar{f}_3 t$\\
$p_{11}$ &  0 & 0 & $-1$ & 0 & 0 &  $1/12$ & $t_{11} =\bar{f}_3^{-1} t$\\
$p_{12}$ &  0 & 0 & $-1$ & 0 & 0 &  $1/12$ & $t_{12} =\bar{f}_3^{-1} t$\\
$p_{13}$ &  0 & 0 & 0 & +1 & 0 &  $1/12$ & $t_{13} =\bar{f}_4  t$\\
$p_{14}$ &  0 & 0 & 0 & +1 & 0 &  $1/12$ & $t_{14} =\bar{f}_4 t$\\
$p_{15}$ &  0 & 0 & 0 & $-1$ & +1 &  $1/12$ & $t_{15} =\bar{f}_4^{-1} \bar{f}_5 t$\\
$p_{16}$ &  0 & 0 & 0 & $-1$ & +1 &  $1/12$ & $t_{16} = \bar{f}_4^{-1} \bar{f}_5 t$\\
\hline
$q_1$ &  0 & 0 & 0 & 0 & $-1$ &  $1/3$ & $y_{q_1} = \bar{f}_5^{-1} t^4$\\
$q_2$ &  0 & 0 & 0 & 0 & $-1$ &  $1/3$ & $y_{q_2} = \bar{f}_5^{-1} t^4$\\
\hline
\end{tabular}
\caption{Global symmetry charges according to $U(1)_{\bar{f}_1}\times U(1)_{\bar{f}_2} \times U(1)_R$ on perfect matchings of the $\text{BFT}[\text{DM}_6]$ Model.}
\label{tab40_2}
\end{table}

\begin{table}[ht!]
\centering
\resizebox{1\textwidth}{!}{
\begin{tabular}{|c|c|c|c|c|c|c|c|c|c|}
\hline
generators & GLSM fields & $U(1)_{\bar{f}_1}$ & $U(1)_{\bar{f}_2}$ & $U(1)_{\bar{f}_3}$ & $U(1)_{\bar{f}_4}$ & $U(1)_{\bar{f}_5}$ & $U(1)_{R}$ & fugacity\\
\hline \hline
$A_{11} = X_{31} X_{12}$ & $p_1 p_2 p_3 p_4 p_9 p_{13} p_{14} p_{15} q_1 q_2$
& 0 & 0 & +1 & +1 & $-1$ & 16 & 
$\bar{f}_{3} \bar{f}_{4} \bar{f}_{5}^{-1} t^{16}$
\\
$A_{12} = X_{31} X_{14}$ & $p_1 p_5 p_6 p_7 p_9 p_{13} p_{14} p_{15} q_1 q_2$
& +1 & +1 & +1 & +1 & $-1$ & 16 &
$\bar{f}_{1} \bar{f}_{2} \bar{f}_{3} \bar{f}_{4} \bar{f}_{5}^{-1} t^{16}$
\\
$A_{13} = X_{31} X_{16}$ & $p_2 p_5 p_8 p_9^2 p_{13} p_{14} p_{15} q_1 q_2$
& +1 & 0 & +2 & +1 & $-1$ & 16 &
$\bar{f}_{1} \bar{f}_{3}^2 \bar{f}_{4} \bar{f}_{5}^{-1} t^{16}$
\\
$A_{21} = X_{51} X_{12}$ & $p_1 p_2 p_3 p_4^2 p_{11} p_{12} p_{14} q_1 q_2$
& $-1$ & 0 & -2 & +1 & -2 & 16 &
$\bar{f}_{4} \bar{f}_{1}^{-1} \bar{f}_{3}^{-2} \bar{f}_{5}^{-2} t^{16}$
\\
$A_{22} = X_{51} X_{14}$ & $p_1 p_4 p_5 p_6 p_7 p_{11} p_{12} p_{14} q_1 q_2$
& 0 & +1 & -2 & +1 & -2 & 16 &
$\bar{f}_{2} \bar{f}_{4} \bar{f}_{3}^{-2} \bar{f}_{5}^{-2} t^{16}$
\\
$A_{23} = X_{51} X_{16}$ & $p_2 p_4 p_5 p_8 p_9 p_{11} p_{12} p_{14} q_1 q_2$
& 0 & 0 & $-1$ & +1 & -2 & 16 &
$\bar{f}_{4} \bar{f}_{3}^{-1} \bar{f}_{5}^{-2} t^{16}$
\\
$A_{31} = X_{71} X_{12}$ & $p_1 p_2 p_3 p_4 p_7 p_{12} p_{15} p_{16} q_1 q_2$
& 0 & $-1$ & $-1$ & $-1$ & 0 & 16 &
$\bar{f}_{2}^{-1} \bar{f}_{3}^{-1} \bar{f}_{4}^{-2} t^{16}$
\\
$A_{32} = X_{71} X_{14}$ & $p_1 p_5 p_6 p_7^2 p_{12} p_{15} p_{16} q_1 q_2$
& +1 & 0 & $-1$ & -2 & 0 & 16 &
$\bar{f}_{1} \bar{f}_{3}^{-1} \bar{f}_{4}^{-2} t^{16}$
\\
$A_{33} = X_{71} X_{16}$ & $p_2 p_5 p_7 p_8 p_9 p_{12} p_{15} p_{16} q_1 q_2$
& +1 & $-1$ & 0 & -2 & 0 & 16 &
$\bar{f}_{1} \bar{f}_{2}^{-1} \bar{f}_{4}^{-2} t^{16}$
\\
\hline
$B_{1} = X_{23}$ & $p_5 p_6 p_7 p_8 p_{10} p_{11} p_{12} p_{16}$
& 0 & 0 & $-1$ & $-1$ & +1 & 8 &
$\bar{f}_{5} \bar{f}_{3}^{-1} \bar{f}_{4}^{-1} t^{8}$
\\
$B_{2} = X_{45}$ & $p_2 p_3 p_8 p_9 p_{10} p_{13} p_{15} p_{16}$
& 0 & $-1$ & +2 & $-1$ & +2 & 8 &
$\bar{f}_{3}^2 \bar{f}_{5}^2 \bar{f}_{2}^{-1} \bar{f}_{4}^{-1} t^{8}$
\\
$B_{3} = X_{67}$ & $p_1 p_3 p_4 p_6 p_{10} p_{11} p_{13} p_{14}$
& $-1$ & +1 & 0 & +2 & 0 & 8 &
$\bar{f}_{2} \bar{f}_{4}^2 \bar{f}_{1}^{-1} t^{8}$
\\
$B_{4} = X_{27}$ & $p_5 p_6 p_8 p_9 p_{10} p_{11} p_{13} p_{14}$
& 0 & +1 & +1 & +2 & 0 & 8 &
$\bar{f}_{2} \bar{f}_{3} \bar{f}_{4}^2 t^{8}$
\\
$B_{5} = X_{43}$ & $p_2 p_3 p_4 p_8 p_{10} p_{11} p_{12} p_{16}$
& $-1$ & $-1$ & $-1$ & $-1$ & +1 & 8 &
$\bar{f}_{5} \bar{f}_{1}^{-1} \bar{f}_{2}^{-1} \bar{f}_{3}^{-1} \bar{f}_{4}^{-1} t^{8}$
\\
$B_{6} = X_{65}$ & $p_1 p_3 p_6 p_7 p_{10} p_{13} p_{15} p_{16}$
& 0 & 0 & +1 & $-1$ & +2 & 8 &
$\bar{f}_{3} \bar{f}_{5}^2 \bar{f}_{4}^{-1} t^{8}$
\\
\hline
\end{tabular}}
\caption{
Generators of the mesonic moduli space of the $\text{BFT}[\text{DM}_6]$  Model. 
\label{tab40_3}
}
\end{table}

In terms of the generators in \tref{tab40_3}, we can express the mesonic moduli space of the $\text{BFT}[\text{DM}_6]$ Model as follows, 
\beal{es40a81}
&&
\mathcal{M}^{mes}_{\text{BFT}[\text{DM}_6]}
=
\text{Spec}
~
\mathbb{C}[
A_{ij}, B_{k}
]
/
\langle
A_{12} B_{5}-A_{23} B_{6}~,~
A_{31} B_{4}-A_{23} B_{6}~,~
A_{33} B_{3}-A_{23} B_{6}~,~
\nn\\
&&
\hspace{1cm}
A_{31} B_{3}-A_{21} B_{6}~,~
A_{23} B_{3}-A_{21} B_{4}~,~
A_{13} B_{3}-A_{11} B_{4}~,~
A_{32} B_{2}-A_{33} B_{6}~,~
\nn\\
&&
\hspace{1cm}
A_{23} B_{2}-A_{13} B_{5}~,~
A_{22} B_{2}-A_{23} B_{6}~,~
A_{21} B_{2}-A_{11} B_{5}~,~
A_{12} B_{2}-A_{13} B_{6}~,~
\nn\\
&&
\hspace{1cm}
A_{31} B_{1}-A_{32} B_{5}~,~
A_{21} B_{1}-A_{22} B_{5}~,~
A_{13} B_{1}-A_{33} B_{4}~,~
A_{12} B_{1}-A_{32} B_{4}~,~
\nn\\
&&
\hspace{1cm}
A_{11} B_{1}-A_{23} B_{6}~,~
B_{1} B_{2} B_{3} -B_{4} B_{5} B_{6}~,~
A_{23} A_{32} -A_{22} A_{33}~,~
A_{13} A_{32} -A_{12} A_{33}~,~
\nn\\
&&
\hspace{1cm}
A_{23} A_{31} -A_{21} A_{33}~,~
A_{22} A_{31} -A_{21} A_{32}~,~
A_{13} A_{31} -A_{11} A_{33}~,~
A_{12} A_{31} -A_{11} A_{32}~,~
\nn\\
&&
\hspace{1cm}
A_{13} A_{22} -A_{12} A_{23}~,~
A_{13} A_{21} -A_{11} A_{23}~,~
A_{12} A_{21} -A_{11} A_{22}~
\rangle
~.~
\eea
Combined with the information about the toric diagram in \eref{es40a65}, we can see from here that the mesonic moduli space of the $\text{BFT}[\text{DM}_6]$ Model is a toric Calabi-Yau 6-fold with 15 generators and 26 defining first order relations between the generators.
We call this non-compact toric Calabi-Yau 6-fold based on a BFT defined on a disk as $\text{DM}_6$.
\\

We can also calculate the Hilbert series using the binomial ideal formed by the $F$-terms of the BFT in \eref{es40a11}. 
Using the global symmetry charges on chiral fields summarized in \tref{tab_40}, we can identify the appropriate grading for the following quotient description of the mesonic moduli space in order to calculate the corresponding Hilbert series, 
\beal{es40a85}
\mathcal{M}^{mes}_{\text{BFT}[\text{DM}_6]}
= \text{Spec}
\left(\mathbb{C}[X_{ij}] / \mathcal{I}^{\text{Irr}}_{\partial W} \right) // U(1)_b
\eea
where the coherent component of the ideal obtained using primary decomposition takes the form,
\beal{es40a86}
\mathcal{I}^{\text{Irr}}_{\partial W}=
&
\langle
~
X_{23} X_{31} - X_{27} X_{71} ~,~
X_{45} X_{51} - X_{43} X_{31}~,~
X_{67} X_{71} - X_{65} X_{51}~,~
&
\nn\\
&
X_{12} X_{23} - X_{14} X_{45} ~,~
X_{14} X_{45} - X_{16} X_{65}  ~,~
X_{16} X_{67} - X_{12} X_{27}  ~
\rangle 
~.~
&
\eea
The resulting Hilbert series of the mesonic moduli space takes the form, 
\beal{es40a87}
g(\bar{t}, f_i = 1; \mathcal{M}^{mes}_{\text{BFT}[\text{DM}_6]})&=&
\frac{
1 + 5 \bar{t} + 19 \bar{t}^2 + 37 \bar{t}^3 + 50 \bar{t}^4 + 37 \bar{t}^5 + 19 \bar{t}^6 + 5 \bar{t}^7 + \bar{t}^8
}{
(1 - \bar{t}) (1- \bar{t}^2)^5
}
~,~
\nn\\
\eea
where for conciseness we have unrefined the Hilbert series by setting the flavor fugacities $f_i=1$.
We see that the numerator is palindromic and that the unrefined Hilbert series is identical to the one obtained in \eref{es40a76} up to $\bar{t}=t^8$.
\\

\section{BFTs on a cylinder \label{sec_e2}}

In this section, we consider BFTs that are defined on a cylinder with $B=2$ boundaries. 
We identify two infinite families of BFTs on the cylinder whose mesonic moduli spaces are toric Calabi-Yau 3-folds and complete intersections. 
We also consider an example where the mesonic moduli space is a toric Calabi-Yau 5-fold.
In all examples, we calculate the Hilbert series of the mesonic moduli spaces and summarize the generators and defining relations formed amongst them. 
\\

\subsection{Family of $\text{BFT}[\mathbb{C}^2/\mathbb{Z}_n \times \mathbb{C}]$ Models \label{sec_f01}}

We introduce in this section an infinite family of BFTs defined on a cylinder whose mesonic moduli spaces are the orbifolds of the form $\mathbb{C}^2/\mathbb{Z}_n \times \mathbb{C}$ \cite{Douglas:1996sw,Douglas:1997de,Johnson:1996py}. 
We note that this family of BFTs very much resembles the worldvolume theories of D-branes probing $\mathbb{C}^2/\mathbb{Z}_n \times \mathbb{C}$ \cite{Hanany:1998it,Davey:2010px,Hanany:2010ne}.

\subsubsection{$\text{BFT}[\mathbb{C}^2/\mathbb{Z}_2 \times \mathbb{C}]$ Model \label{sec_e09}}

\begin{figure} [h]
\centering
\includegraphics{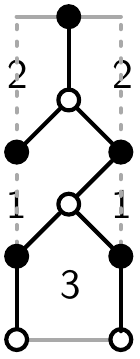}
\caption{
The bipartite graph for the $\text{BFT}[\mathbb{C}^2/\mathbb{Z}_2 \times \mathbb{C}]$ Model.
\label{fig_90}
}
\end{figure}

\begin{figure} [h]
\centering
\includegraphics{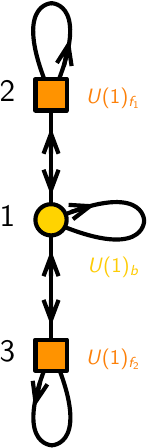}
\caption{
The quiver for the $\text{BFT}[\mathbb{C}^2/\mathbb{Z}_2 \times \mathbb{C}]$ Model.
\label{fig_91}
}
\end{figure}

The bipartite graph and the quiver diagram for the $\text{BFT}[\mathbb{C}^2/\mathbb{Z}_2 \times \mathbb{C}]$ Model is shown in \fref{fig_90} and \fref{fig_91}, respectively.
The superpotential of the BFT takes the following form,
\beal{es90a10}
W = 
X_{12} X_{22} X_{21}
+ X_{31} X_{11} X_{13}
- X_{21} X_{11} X_{12}
- X_{13} X_{33} X_{31}
~,~
\eea
where the gauge and flavor symmetry charges on the chiral fields $X_{ij}$ are summarized in \tref{tab_90}. 
Even though we note that the quiver diagram in \fref{fig_91} indicates a global symmetry of $U(1)_{f_1} \times U(1)_{f_2} \times U(1)_R$, we note that given $B=2$ that there is an additional $U(1)_h$ global symmetry as discussed in section \sref{smod}. 
As a result, we can rewrite the global symmetry as $U(1)_{f} \times U(1)_{h} \times U(1)_R$, where the independent product $U(1)_{f}\times U(1)_h$ can be chosen instead of $U(1)_{f_1}
U(1)_{f_2}$.
For now, we can keep the global symmetry as $U(1)_{f_1} \times U(1)_{f_2} \times U(1)_R$ as summarized in \tref{tab_90}. 
The $U(1)_R$ charges on the chiral fields are obtained via $a$-maximization as discussed in section \sref{smod}.
\\

\begin{table}[ht!]
\centering
\begin{tabular}{|c|c|c|c|c|c|c|}
\hline
\; & $U(1)_{b}$ & $U(1)_{f_1}$ & $U(1)_{f_2}$ & $U(1)_R$ & fugacity \\
\hline\hline
$X_{11}$  & 0 & 0 & 0 & 2/3 & $\bar{t}$\\
\hline
$X_{12}$  & $-1$ & $+1$ & 0 & 2/3 & $b^{-1} f_1 \bar{t}$\\
$X_{21}$  & $+1$ & $-1$ & 0 & 2/3 & $b f_1^{-1}\bar{t}$\\
$X_{13}$  & $-1$ & 0 & $+1$ & 2/3 & $b^{-1} f_2 \bar{t}$\\
$X_{31}$  & $+1$ & 0 & $-1$ & 2/3 & $b f_2^{-1}\bar{t}$\\
\hline
$X_{22}$  & 0 & 0 & 0 & 2/3& $\bar{t}$\\
$X_{33}$  & 0 & 0 & 0 & 2/3& $\bar{t}$\\
\hline
\end{tabular}
\caption{
Gauge and global symmetry charges on the chiral fields of the $\text{BFT}[\mathbb{C}^2/\mathbb{Z}_2 \times \mathbb{C}]$ Model.
}
\label{tab_90}
\end{table}

\paragraph{Forward Algorithm.}
The $F$-terms from the superpotential in \eref{es90a10} take the following form,
\beal{es90a11}
&
\partial_{X_{11}} W =
X_{13} X_{31} - X_{12} X_{21} ~,~
\partial_{X_{12}} = 
X_{22} X_{21} - X_{21} X_{11}~,~
&
\nn\\
&
\partial_{X_{21}} W = 
X_{12} X_{22} - X_{11} X_{12}~,~
\partial_{X_{13}} W = 
X_{31} X_{11} - X_{33} X_{31} ~,~
&
\nn\\
&
\partial_{X_{31}} W = 
X_{11} X_{13} - X_{13} X_{33} ~.~
&
\eea
where we identify $4$ independent chiral fields subject to the above $F$-terms,
\beal{es90a12}
v_1 = X_{11} ~,~
v_2 = X_{12} ~,~
v_3 = X_{13} ~,~
v_4 = X_{21} ~.~
\eea
Using the above independent fields, we can express the remaining chiral fields as follows,
\beal{es90a13}
X_{22} = X_{33} = v_1~,~
X_{31} = \frac{v_2 v_4}{v_3} ~.~
\eea
These relations can be summarized in the following $K$-matrix,
\beal{es90a60}
K=\left(
\begin{array}{c|cccc}
\; & v_1 & v_2 & v_3 & v_4 \\
\hline
X_{11} & 1 & 0 & 0 & 0 \\
X_{12} & 0 & 1 & 0 & 0 \\
X_{21} & 0 & 0 & 0 & 1 \\
X_{13} & 0 & 0 & 1 & 0 \\
X_{31} & 0 & 1 & -1 & 1 \\
X_{22} & 1 & 0 & 0 & 0 \\
X_{33} & 1 & 0 & 0 & 0 \\
\end{array}
\right)
~.~
 \eea
 
One of the main results of the forward algorithm is the computation of the perfect matching matrix from the $K$-matrix.
The perfect matching matrix takes the following form,
 \beal{es90a61}
P=\left(
\begin{array}{c|ccc|cc}
\; & p_1 & p_2 & p_3 & s_1 & s_2 \\
\hline
X_{11} & 0 & 0 & 1 & 0 & 0 \\
X_{12} & 1 & 0 & 0 & 0 & 1 \\
X_{21} & 0 & 1 & 0 & 1 & 0 \\
X_{13} & 0 & 1 & 0 & 0 & 1 \\
X_{31} & 1 & 0 & 0 & 1 & 0 \\
X_{22} & 0 & 0 & 1 & 0 & 0 \\
X_{33} & 0 & 0 & 1 & 0 & 0 \\
 \end{array}
\right)~.~
\eea
The perfect matchings correspond to GLSM fields that can be used to parameterize the mesonic moduli space of the $\text{BFT}[\mathbb{C}^2/\mathbb{Z}_2 \times \mathbb{C}]$ Model.
Each of the chiral fields can be expressed as a product of the GLSM fields as shown below, 
\beal{es90a61b}
&
X_{12} = p_1 s_2 ~,~
X_{21} = p_2 s_1 ~,~
X_{13} = p_2 s_2 ~,~
X_{31} = p_1 s_1 ~,~
X_{11} = X_{22} = X_{33} = p_3
~.~
&
\nn\\
\eea
In terms of the GLSM fields, the $F$-terms can be encoded in terms of a collection of $U(1)$ charges on the GLSM fields as summarized in the $Q_F$ charge matrix,
\beal{es90a62}
Q_{F} =\left(
\begin{array}{ccc|cc}
 p_1 & p_2 & p_3 & s_1 & s_2 \\
\hline
 1 & 1 & 0 & -1 & -1 \\
\end{array}
\right)~.~
\eea
The $D$-terms can also be summarized as charges on the GLSM fields as shown below, 
\beal{es90a63}
Q_{D} =\left(
\begin{array}{ccc|cc}
 p_1 & p_2 & p_3 & s_1 & s_2 \\
\hline
 1 & 1 & 0 & -2 & 0 \\
\end{array}
\right)~.~
\eea

\begin{figure} [h]
\centering
\includegraphics[width=.3\textwidth]{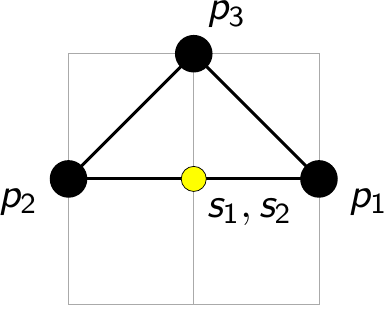}
\caption{
Toric diagram for the $\text{BFT}[\mathbb{C}^2/\mathbb{Z}_2 \times \mathbb{C}]$ Model.
}
\label{fig_92}
\end{figure}

The toric diagram of the $\text{BFT}[\mathbb{C}^2/\mathbb{Z}_2 \times \mathbb{C}]$ Model can be obtained from the $Q_F$ and $Q_D$ charge matrices.
It is given by the toric diagram matrix as follows,
\beal{es90a65}
G_{t} =\left(
\begin{array}{ccc|cc}
 p_1 & p_2 & p_3 & s_1 & s_2 \\
\hline
 1 & -1 & 0 & 0 & 0 \\
 0 & 0 & 1 & 0 & 0 \\
 1 & 1 & 1 & 1 & 1 \\
\end{array}
\right)~,~
\eea
where each of the perfect matchings correspond to vertices in the $2$-dimensional toric diagram as shown in \fref{fig_92}.
We can see from the toric diagram that the mesonic moduli space of the $\text{BFT}[\mathbb{C}^2/\mathbb{Z}_2 \times \mathbb{C}]$ Model is $\mathbb{C}^2/\mathbb{Z}_2 \times \mathbb{C}$.
\\

\paragraph{Hilbert Series.}
In terms of the GLSM fields, the mesonic moduli space of the $\text{BFT}[\mathbb{C}^2/\mathbb{Z}_2 \times \mathbb{C}]$ Model can be expressed as the following symplectic quotient, 
\beal{es90a66}
\mathcal{M}^{mes}_{\text{BFT}[\mathbb{C}^2/\mathbb{Z}_2 \times \mathbb{C}]}
= \text{Spec}
\left(\mathbb{C}[p_1,p_2,p_3, s_1, s_2] // Q_F \right) // Q_D
\eea
Using this description of the mesonic moduli space, we can calculate the corresponding Hilbert series as follows,
\beal{es90a70}
&&
g(t_1,t_2, t_3, y_s;\mathcal{M}^{mes}_{\text{BFT}[\mathbb{C}^2/\mathbb{Z}_2 \times \mathbb{C}]})
=
\frac{1- y_s t_1 t_2}{
(1-t_3)
(1-y_s t_1^2)
(1- y_s t_2^2)
}
~,~
\eea
where the fugacity $t_a$ corresponds to GLSM field $p_a$, and the fugacity $y_s$ corresponds to the product of GLSM fields $s=s_1 s_2$.
The plethystic logarithm of the Hilbert series is given by,
\beal{es90a71}
&&
PL[g(t_1,t_2, t_3, y_s;\mathcal{M}^{mes}_{\text{BFT}[\mathbb{C}^2/\mathbb{Z}_2 \times \mathbb{C}]})]
= 
t_3
+ y_s t_1^2 
+ y_s t_1 t_2
+ y_s t_2^2
- y_s^2 t_1^2 t_2^2
~,~
\eea
where we note that the expansion is finite, indicating that the mesonic moduli space $\mathbb{C}^2/\mathbb{Z}_2 \times \mathbb{C}$ is as expected a complete intersection. 

\begin{table}[ht!]
\centering
\begin{tabular}{|l|c|c|c|c|}
\hline
\; & $SU(2)_{x}$ & $U(1)_{\bar{f}}$ & $U(1)_{R}$ & fugacity\\
\hline\hline
$p_1$ & $+1$ & $+1$ & $2/3$ & $t_1 = x \bar{f} t$\\
$p_2$ & $-1$ & $+1$& $2/3$ & $t_2 = x^{-1} \bar{f} t$\\
$p_3$ &  0 & $-2$ & $2/3$ & $t_3 = \bar{f}^{-2} t$\\
\hline
\end{tabular}
\caption{Global symmetry charges according to $SU(2)_x\times U(1)_f \times U(1)_R$ on perfect matchings of the $\text{BFT}[\mathbb{C}^2/\mathbb{Z}_2 \times \mathbb{C}]$ Model.}
\label{tab90_2}
\end{table}

Using the following fugacity map,
\beal{es90a75}
&
x = 
\frac{t_1^{1/2}}{t_2^{1/2}}
~,~
\bar{f} =
\frac{t_1^{1/4} t_2^{1/4}}{t_3^{1/4}}
~,~
t = t_1^{1/3} t_2^{1/3} t_3^{1/3}
~,~
&
\eea
we can rewrite the Hilbert series in terms of an enhanced global symmetry of the form $SU(2)_{x} \times U(1)_{\bar{f}}\times U(1)_{R}$. 
We see that the new fugacity $x$ corresponding to the non-abelian factor $SU(2)_{x}$ in the global symmetry forms characters of irreducible representations of $SU(2)_{x}$ as follows,
\beal{es90a76}
g(t,x,\bar{f};\mathcal{M}^{mes}_{\text{BFT}[\mathbb{C}^2/\mathbb{Z}_2 \times \mathbb{C}]})
=\sum_{n,m=0} [2n]_x \bar{f}^{2n-2m} t^{2n+m} ~,~
\eea
where $[2n]_x = [2n]_{SU(2)_x}$.
As a result, we can confirm from the Hilbert series computation that as expected for $\mathbb{C}^2/\mathbb{Z}_2 \times \mathbb{C}$, there is a symmetry enhancement in the form of $SU(2)_{x} \times U(1)_{\bar{f}}\times U(1)_{R}$. 
The corresponding plethystic logarithm of the form,
\beal{es90a77}
PL[g(t,x,\bar{f};\mathcal{M}^{mes}_{\text{BFT}[\mathbb{C}^2/\mathbb{Z}_2 \times \mathbb{C}]})]
= 
\bar{f}^{-2} t 
+ [2]_x t^2
- \bar{f}^4 t^4 
~,~
\eea
shows that generators and relations of the mesonic moduli space also transform in representations of $SU(2)_{x} \times U(1)_{\bar{f}}\times U(1)_{R}$.

\begin{table}[ht!]
\centering
\begin{tabular}{|l|l|c|c|c|c|}
\hline
generator & GLSM fields & $SU(2)_x$ & $U(1)_{\bar{f}}$ & $U(1)_R$ & fugacities \\
\hline\hline
$A_{+} = X_{31} X_{12}$ & $p_1^2 s_1 s_2$ & $+2$ & $+2$ & 4/3  & $x^2 \bar{f}^2 t^2$ \\
$A_0=X_{21} X_{12} = X_{31} X_{13}$ & $p_1 p_2 s_1 s_2$ & 0 & $+2$ & 4/3 & $\bar{f}^2 t^2$ \\
$A_{-}=X_{21} X_{13}$ & $p_2^2 s_1 s_2$ & $-2$ & $+2$ & 4/3 & $x^{-2} \bar{f}^2 t^2$\\
$B=X_{11} = X_{22} = X_{33}$ & $p_3$ & 0 & $-2$ & 2/3 & $\bar{f}^{-2} t$ \\
\hline
\end{tabular}
\caption{Generators of the mesonic moduli space of the $\text{BFT}[\mathbb{C}^2/\mathbb{Z}_2 \times \mathbb{C}]$ Model.}
\label{tab90_3}
\end{table}

From the plethystic logarithm, we identify $4$ generators of the form $A_+, A_0, A_-, B$ as summarized in \tref{tab90_3}. 
The generators $A_+, A_0, A_-$ in the adjoint representation of $SU(2)_{x}$ form the defining equation for $\mathbb{C}^2/\mathbb{Z}_2$, whereas the generator $B$ corresponds to the $\mathbb{C}$ factor of the moduli space. Accordingly, we can write the mesonic moduli space of the $\text{BFT}[\mathbb{C}^2/\mathbb{Z}_2 \times \mathbb{C}]$ Model as the following complete intersection,
\beal{es90a80}
\mathcal{M}^{mes}_{\text{BFT}[\mathbb{C}^2/\mathbb{Z}_2 \times \mathbb{C}]} 
&=&
\text{Spec}~  \mathbb{C}[A_1, A_2,A_3,B] /  \langle A_{+} A_{-} - A_0^2 \rangle 
\nn\\
&=& \mathbb{C}^2 /\mathbb{Z}_2 \times \mathbb{C}
~.~
\eea
\\

We can also calculate the Hilbert series using the ideal formed by the $F$-terms of the BFT.
In terms of the $F$-term ideal, the mesonic moduli space of the $\text{BFT}[\mathbb{C}^2/\mathbb{Z}_2 \times \mathbb{C}]$ Model can be written as, 
\beal{es90a85}
\mathcal{M}^{mes}_{\text{BFT}[\mathbb{C}^2/\mathbb{Z}_2 \times \mathbb{C}]} 
= \text{Spec}~
(\mathbb{C}[X_{ij}] / \mathcal{I}^{\text{Irr}}_{\partial W} ) // U(1)_b
~,~
\eea
where the coherent component of the ideal formed by the $F$-terms takes the following form,
\beal{es90a86}
\mathcal{I}^{\text{Irr}}_{\partial W}=
&
\langle
~
X_{13} X_{31} - X_{12} X_{21} ~,~
X_{22} X_{21} - X_{21} X_{11}~,~
X_{12} X_{22} - X_{11} X_{12}~,~
&
\nn\\
&
X_{31} X_{11} - X_{33} X_{31} ~,~
X_{11} X_{13} - X_{13} X_{33} 
~
\rangle 
~.~
&
\eea
Using the global symmetry charges on chiral fields summarized in \tref{tab_90}, we can obtain the following refined Hilbert series,
\beal{es90a87}
g(\bar{t}, f_1,f_2; \mathcal{M}^{mes}_{\text{BFT}[\mathbb{C}^2/\mathbb{Z}_2 \times \mathbb{C}]})&=&
\frac{
1 - \bar{t}^4
}{
(1 - \bar{t}) (1 - f_1 f_2^{-1} \bar{t}^2 ) (1 - \bar{t}^2) (1 - f_1^{-1} f_2 \bar{t}^2 )
}
~,~
\eea
where now $\bar{t}$ corresponds to the $U(1)_R$ symmetry on the chiral fields, and $f_1,f_2$ correspond to the $U(1)_{f_1} \times U(1)_{f_2}$ flavor symmetry on the chiral fields.
The corresponding plethystic logarithm takes the form,
\beal{es90a88}
PL[g(\bar{t}, f_1,f_2; \mathcal{M}^{mes}_{\text{BFT}[\mathbb{C}^2/\mathbb{Z}_2 \times \mathbb{C}]})]
&=&
\bar{t} + ( f_1 f_2^{-1} + 1 + f_1^{-1} f_2)  \bar{t}^2 - \bar{t}^4 ~.~
\eea
We see that the Hilbert series above is identical to the one in \eref{es90a70} up to a fugacity map, hence describing the same complete intersection mesonic moduli space of the $\text{BFT}[\mathbb{C}^2/\mathbb{Z}_2 \times \mathbb{C}]$ Model.
\\

\subsubsection{$\text{BFT}[\mathbb{C}^2/\mathbb{Z}_3 \times \mathbb{C}]$ Model \label{sec_e13}}

\begin{figure} [h]
\centering
\includegraphics{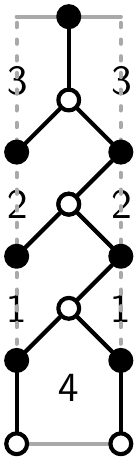}
\caption{
The bipartite graph for the $\text{BFT}[\mathbb{C}^2/\mathbb{Z}_3 \times \mathbb{C}]$ Model.
\label{fig_130}
}
\end{figure}

\begin{figure} [h]
\centering
\includegraphics{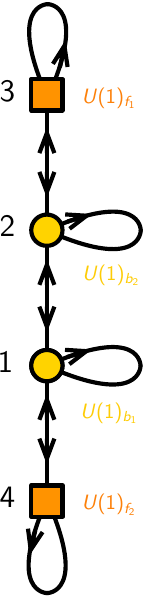}
\caption{
The quiver for the $\text{BFT}[\mathbb{C}^2/\mathbb{Z}_3 \times \mathbb{C}]$ Model.
\label{fig_131}
}
\end{figure}

\fref{fig_130} shows the bipartite graph on a cylinder that describes the $\text{BFT}[\mathbb{C}^2/\mathbb{Z}_3 \times \mathbb{C}]$ Model.
The corresponding quiver diagram is shown in \fref{fig_131} and the associated superpotential takes the following form,
\beal{es130a10}
W = 
&
X_{12} X_{22} X_{21}
+ X_{23} X_{33} X_{32}
+ X_{41} X_{11} X_{14} 
&
\nn\\
&
- X_{21} X_{11} X_{12}
- X_{32} X_{22} X_{23} 
- X_{14} X_{44} X_{41}
&
~,~
\eea
where the charges on the chiral fields $X_{ij}$ under the gauge and global symmetries of the BFT are summarized in \tref{tab_130}.
Similarly to the $\text{BFT}[\mathbb{C}^2/\mathbb{Z}_2 \times \mathbb{C}]$ Model, we note here that the flavor symmetry factors $U(1)_{f_1} \times U(1)_{f_2}$ can be replaced by $U(1)_{f} \times U(1)_{h}$, where $U(1)_{h}$ originates from the fact that the BFT has $B=2$ as discussed in section \sref{smod}.
For the moment, we keep the global symmetry to be $U(1)_{f_1}\times U(1)_{f_2} \times U(1)_R$.

\begin{table}[ht!]
\centering
\begin{tabular}{|c|c|c|c|c|c|c|}
\hline
\; & $U(1)_{b_1}$ & $U(1)_{b_2}$ & $U(1)_{f_1}$ & $U(1)_{f_2}$ & $U(1)_R$ & fugacity\\
\hline\hline
$X_{11}$  & 0 & 0 & 0 & 0& 2/3 & $\bar{t}$\\
$X_{12}$  & $-1$ & $+1$ & 0&  0 & 2/3& $b_1^{-1} b_2\bar{t}$\\
$X_{21}$  & $+1$ & $-1$ & 0& 0 & 2/3& $b_1 b_2^{-1} \bar{t}$\\
$X_{22}$  & 0 & 0 & 0 & 0 & 2/3 & $\bar{t}$\\
\hline
$X_{23}$  & 0& $-1$ & $$+1$$ & 0& 2/3 & $ b_2^{-1} f_1 \bar{t}$\\
$X_{32}$  & 0& $+1$ & $$-1$$ & 0& 2/3 & $b_2 f_1^{-1}\bar{t}$\\
$X_{14}$  & $-1$ & 0 & 0& $+1$& 2/3 & $b_1^{-1} f_2 \bar{t}$\\
$X_{41}$  & $+1$ & 0 & 0& $-1$& 2/3& $b_1 f_2^{-1}\bar{t}$\\
\hline
$X_{33}$  & 0 & 0 & 0& 0 & 2/3& $\bar{t}$\\
$X_{44}$  & 0 & 0 & 0& 0  & 2/3& $\bar{t}$\\
\hline
\end{tabular}
\caption{Gauge and global symmetry charges on the chiral fields of the $\text{BFT}[\mathbb{C}^2/\mathbb{Z}_3 \times \mathbb{C}]$ Model.}
\label{tab_130}
\end{table}

\paragraph{Forward Algorithm.}
The $F$-terms of the $\text{BFT}[\mathbb{C}^2/\mathbb{Z}_3 \times \mathbb{C}]$ Model take the following form,
\beal{es130a11}
&
\partial_{X_{11}} W =
X_{14} X_{41} - X_{12} X_{21}
~,~
\partial_{X_{12}} W =
X_{22} X_{21} - X_{21} X_{11}
~,~
&
\nn\\
&
\partial_{X_{21}} W =
X_{12} X_{22} - X_{11} X_{12}
~,~
\partial_{X_{22}} W =
X_{21} X_{12} - X_{23} X_{32}
~,~
&
\nn\\
&
\partial_{X_{23}} W =
X_{33} X_{32} - X_{32} X_{22}
~,~
\partial_{X_{32}} W =
X_{23} X_{33} - X_{22} X_{23}
~,~
&
\nn\\
&
\partial_{X_{14}} W =
X_{41} X_{11} - X_{44} X_{41}
~,~
\partial_{X_{41}} W =
X_{11} X_{14} - X_{14} X_{44}
~.~
&
\eea
Under the above $F$-terms, we identify the following independent chiral fields of the BFT,
\beal{es130a12}
v_1 = X_{11} ~,~
v_2 = X_{12} ~,~
v_3 = X_{14} ~,~
v_4 = X_{21} ~,~
v_5 = X_{23} ~,~
\eea
which we can use in order to express the remaining chiral fields as follows,
\beal{es130a13}
X_{22} = X_{33} = X_{44} =v_1~,~
X_{32} = \frac{v_2 v_4}{v_5} ~,~
X_{41} = \frac{v_2 v_4}{v_3} ~.~
\eea
The above relations can be summarized in the following $K$-matrix,
\beal{es130a60}
K=\left(
\begin{array}{c|ccccc}
\; & v_1 & v_2 & v_3 & v_4 & v_5 \\
\hline
X_{11} & 1 & 0 & 0 & 0 & 0 \\
X_{12} & 0 & 1 & 0 & 0 & 0 \\
X_{21} & 0 & 0 & 0 & 1 & 0 \\
X_{22} & 1 & 0 & 0 & 0 & 0 \\
X_{23} & 0 & 0 & 0 & 0 & 1 \\
X_{32} & 0 & 1 & 0 & 1 & -1 \\
X_{14} & 0 & 0 & 1 & 0 & 0 \\
X_{41} & 0 & 1 & -1 & 1 & 0 \\
X_{33} & 1 & 0 & 0 & 0 & 0 \\
X_{44} & 1 & 0 & 0 & 0 & 0 \\
\end{array}
\right)
~.~
 \eea
 
Using the $K$-matrix, we can identify the perfect matchings of the BFT which correspond to GLSM fields. 
They are summarized in the following $P$-matrix,
 \beal{es130a61}
P=\left(
\begin{array}{c|ccc|ccc|ccc}
\; & p_1 & p_2 & p_3 & s_1 & s_2 & s_3 & u_1 & u_2 & u_3 \\
\hline
X_{11} & 0 & 0 & 1 & 0 & 0 & 0 & 0 & 0 & 0 \\
X_{12} & 0 & 1 & 0 & 0 & 1 & 1 & 0 & 0 & 1 \\
X_{21} & 1 & 0 & 0 & 1 & 0 & 0 & 1 & 1 & 0 \\
X_{22} & 0 & 0 & 1 & 0 & 0 & 0 & 0 & 0 & 0 \\
X_{23} & 0 & 1 & 0 & 1 & 0 & 1 & 0 & 1 & 0 \\
X_{32} & 1 & 0 & 0 & 0 & 1 & 0 & 1 & 0 & 1 \\
X_{14} & 1 & 0 & 0 & 0 & 0 & 1 & 0 & 1 & 1 \\
X_{41} & 0 & 1 & 0 & 1 & 1 & 0 & 1 & 0 & 0 \\
X_{33} & 0 & 0 & 1 & 0 & 0 & 0 & 0 & 0 & 0 \\
X_{44} & 0 & 0 & 1 & 0 & 0 & 0 & 0 & 0 & 0 \\
 \end{array}
\right)~.~
\eea
The chiral fields of the BFT can be expressed as products of GLSM fields as shown below,
\beal{es130a61b}
&
X_{12} = p_2 s_2 s_3  u_3 ~,~
X_{21} = p_1 s_1 u_1 u_2 ~,~
X_{23} = p_2 s_1 s_3 u_2 ~,~
X_{32} = p_1 s_2 u_1 u_3 ~,~
&
\nn\\
&
X_{14} = p_1 s_3 u_2 u_3 ~,~
X_{41} = p_2 s_1 s_2 u_1 ~,~
X_{11}  = X_{22} =  X_{33} = X_{44} = p_3 ~.~
&
\eea
Using the GLSM fields, we can express the $F$-terms of the BFT as a collection of $U(1)$ charges that are given by the following charge matrix,
\beal{es130a62}
Q_{F} =\left(
\begin{array}{ccc|ccc|ccc}
 p_1 & p_2 & p_3 & s_1 & s_2 & s_3 & u_1 & u_2 & u_3 \\
\hline
 1 & 0 & 0 & 0 & 0 & 1 & 0 & -1 & -1 \\
 0 & 1 & 0 & 0 & 0 & -2 & -1 & 1 & 1 \\
 0 & 0 & 0 & 1 & 0 & -1 & -1 & 0 & 1 \\
 0 & 0 & 0 & 0 & 1 & -1 & -1 & 1 & 0 \\
\end{array}
\right)~.~
\eea
We also have a charge matrix for the $D$-terms, which takes the following form,
\beal{es130a63}
Q_{D} =\left(
\begin{array}{ccc|ccc|ccc}
 p_1 & p_2 & p_3 & s_1 & s_2 & s_3 & u_1 & u_2 & u_3 \\
\hline
 1 & 0 & 0 & 0 & 1 & 0 & -2 & 0 & 0 \\
 0 & 0 & 0 & 1 & -1 & 0 & 0 & 0 & 0 \\
\end{array}
\right)~.~
\eea

\begin{figure} [h]
\centering
\includegraphics[width=.45\textwidth]{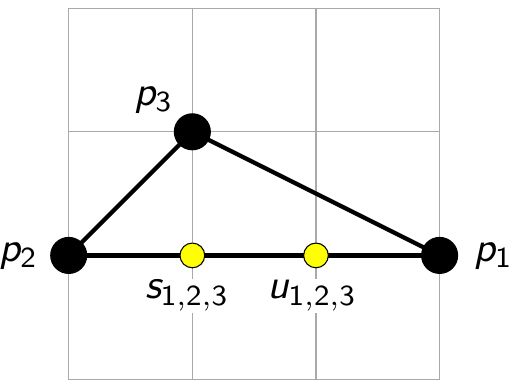}
\caption{
Toric diagram for the $\text{BFT}[\mathbb{C}^2/\mathbb{Z}_3 \times \mathbb{C}]$ Model.
}
\label{fig_132}
\end{figure}

Using the $F$- and $D$-term charge matrices, we can calculate the coordinates of the vertices in the toric diagram of the mesonic moduli space of the BFT. 
The toric diagram is summarized in the following coordinate matrix,
\beal{es130a65}
G_{t} =\left(
\begin{array}{ccc|ccc|ccc}
 p_1 & p_2 & p_3 & s_1 & s_2 & s_3 & u_1 & u_2 & u_3 \\
\hline
 2 & -1 & 0 & 0 & 0 & 0 & 1 & 1 & 1 \\
 0 & 0 & 1 & 0 & 0 & 0 & 0 & 0 & 0 \\
 1 & 1 & 1 & 1 & 1 & 1 & 1 & 1 & 1 \\
\end{array}
\right)~,~
\eea
where the corresponding toric diagram is shown in \fref{fig_132}.
We see from the toric diagram that the mesonic moduli space of the $\text{BFT}[\mathbb{C}^2/\mathbb{Z}_3 \times \mathbb{C}]$ Model is a toric Calabi-Yau 3-fold. We identified as the abelian orbifold of the form $\mathbb{C}^2/\mathbb{Z}_3 \times \mathbb{C}$.
\\

\paragraph{Hilbert Series.}
Using the basis given in terms of the GLSM fields, we can calculate the Hilbert series of the mesonic moduli space of the BFT as follows, 
\beal{es130a70}
&&
g(t_1,t_2, t_3, y_s,y_u;\mathcal{M}^{mes}_{\text{BFT}[\mathbb{C}^2/\mathbb{Z}_3 \times \mathbb{C}]})
=
\frac{1- y_s^3 y_u^3 t_1^3 t_2^3}{
(1-t_3)
(1- y_s y_u t_1 t_2)
(1-y_s^2 y_u t_2^3)
(1- y_s y_u^2 t_1^3)
}
~,~
\nn\\
\eea
where the fugacities $t_a$ correspond to the GLSM fields $p_a$, and the fugacities $y_s$ and $y_u$ correspond to the products of GLSM fields $s=s_1 s_2$ and $u=u_1 u_2$, respectively. 
The corresponding plethystic logarithm takes the following form,
\beal{es130a71}
PL[g(t_1,t_2, t_3, y_s,y_u;\mathcal{M}^{mes}_{\text{BFT}[\mathbb{C}^2/\mathbb{Z}_3 \times \mathbb{C}]})]
= 
&&
t_3
+ y_s y_u t_1 t_2
+ y_s^2 y_u t_2^3 
+ y_s y_u^2 t_1^3
- y_s^3 y_u^3 t_1^3 t_2^3
~,~
\nn\\
\eea
where we can see that the expansion is finite. 
This indicates that the mesonic moduli space of the $\text{BFT}[\mathbb{C}^2/\mathbb{Z}_3 \times \mathbb{C}]$ Model is a complete intersection, as we expect for the abelian orbifold of the form $\mathbb{C}^2/\mathbb{Z}_3 \times \mathbb{C}$.

\begin{table}[ht!]
\centering
\begin{tabular}{|l|c|c|c|l|}
\hline
\; & $U(1)_{\bar{f}_1}$ & $U(1)_{\bar{f}_2}$ & $U(1)_{R}$ & fugacity\\
\hline\hline
$p_1$ & +1 & +1 & $2/3$ & $t_1 = \bar{f}_1 \bar{f}_2 t$\\
$p_2$ & -1 & +1& $2/3$ & $t_2 = \bar{f}_1^{-1} \bar{f}_2 t$\\
$p_3$ &  0 & -2 & $2/3$ & $t_3 = \bar{f}_2^{-2} t$\\
\hline
\end{tabular}
\caption{Global symmetry charges according to $U(1)_{\bar{f}_1}\times U(1)_{\bar{f}_2} \times U(1)_R$ on perfect matchings of the $\text{BFT}[\mathbb{C}^2/\mathbb{Z}_3 \times \mathbb{C}]$ Model.}
\label{tab130_2}
\end{table}

Using the global symmetry charge assignment on GLSM fields as summarized in \tref{tab130_2}, we can obtain the following fugacity map
\beal{es130a75}
&
\bar{f}_1 = 
\frac{t_1^{1/2}}{t_2^{1/2}}
~,~
\bar{f}_2 = 
\frac{t_1^{1/4} t_2^{1/4}}{t_3^{1/4}}
~,~
t = t_1^{1/3} t_2^{1/3} t_3^{1/3}
~,~
&
\eea
that allows us to rewrite the Hilbert series in \eref{es130a70} in terms of fugacities corresponding to the global symmetry $U(1)_{\bar{f}_1} \times U(1)_{\bar{f}_2} \times U(1)_{R}$ of the BFT. 
The resulting refined Hilbert series in terms of global symmetry fugacities takes the following form, 
\beal{es130a76}
g(t, \bar{f}_1, \bar{f}_2 ;\mathcal{M}^{mes}_{\text{BFT}[\mathbb{C}^2/\mathbb{Z}_3 \times \mathbb{C}]})=
\frac{1- \bar{f}_2^5 t^6}{
(1-\bar{f}_2^{-2} t)
(1-\bar{f}_2^{2} t^2)
(1-\bar{f}_1^{-3} \bar{f}_2^3 t^3)
(1-\bar{f}_1^{3} \bar{f}_2^3 t^3)
}
 ~,~
 \nn\\
\eea
with the corresponding plethystic logarithm being,
\beal{es130a76}
PL[
g(t, \bar{f}_1, \bar{f}_2 ;\mathcal{M}^{mes}_{\text{BFT}[\mathbb{C}^2/\mathbb{Z}_3 \times \mathbb{C}]})
]
= 
\bar{f}_2^{-2} t
+\bar{f}_2^2 t^2 
+ (\bar{f}_1^3 \bar{f}_2^{3} + \bar{f}_1^{-3} \bar{f}_2^{3}) t^3
- \bar{f}_2^6 t^6
~.~
\eea
We see that there is no additional global symmetry enhancement that can be obtained from an appropriate re-definition of global symmetry fugacities.

\begin{table}[ht!]
\centering
\begin{tabular}{|l|l|c|c|c|c|}
\hline
generator & GLSM fields & $U(1)_{\bar{f}_1}$ & $U(1)_{\bar{f}_2}$ & $U(1)_R$ & fugacities \\
\hline\hline
$A_{+} = X_{41} X_{12} X_{23}$ & $p_2^3 s_1^2 s_2^2 s_3^2 u_1 u_2 u_3$ 
& -3&  +3 &  2 & $\bar{f}_1^{-3} \bar{f}_2^3 t^3$  \\
$A_0=X_{12} X_{21} $ & $p_1 p_2 s_1 s_2 s_3 u_1 u_2 u_3$ 
&0 &+2 & 4/3 & $\bar{f}_2^2 t^2$\\
$= X_{41}  X_{14} = X_{32} X_{23}$ & & & & & \\
$A_{-}=X_{32} X_{21} X_{14}$ & $p_1^3 s_1 s_2 s_3 u_1^2 u_2^2 u_3^2$ 
& +3 &+3   & 2  & $\bar{f}_1^3 \bar{f}_2^3 t^3$\\
$B=X_{11} = X_{22} = X_{33} = X_{44}$ & $p_3$ 
&  0 & -2 & 2/3  &$\bar{f}_{2}^{-2} t$\\
\hline
\end{tabular}
\caption{Generators of the mesonic moduli space of the $\text{BFT}[\mathbb{C}^2/\mathbb{Z}_3 \times \mathbb{C}]$ Model.}
\label{tab130_3}
\end{table}

From the plethystic logarithm of the Hilbert series, we can identify 4 generators of the mesonic moduli space that satisfy a defining quadratic equation. 
The generators and their global symmetry charges are summarized in \tref{tab130_3}. 
In terms of the generators and the defining relation amongst them, we can express the mesonic moduli space as, 
\beal{es130a80}
\mathcal{M}^{mes}_{\text{BFT}[\mathbb{C}^2/\mathbb{Z}_3 \times \mathbb{C}]} 
&=&
\text{Spec}~  \mathbb{C}[A_+, A_-,A_0,B] /  \langle A_{+} A_{-} - A_0^3 \rangle 
\nn\\
&=& \mathbb{C}^2 /\mathbb{Z}_3 \times \mathbb{C}
~.~
\eea
\\

We can also calculate the Hilbert series using the binomial ideal formed by the $F$-terms of the BFT. 
The mesonic moduli space in terms of the binomial ideal can be expressed as follows, 
\beal{es130a85}
\mathcal{M}^{mes}_{\text{BFT}[\mathbb{C}^2/\mathbb{Z}_3 \times \mathbb{C}]} 
= \text{Spec}~
(\mathbb{C}[X_{ij}] / \mathcal{I}^{\text{Irr}}_{\partial W} ) // (U(1)_{b_1} \times U(1)_{b_2})
\eea
where the coherent component of the ideal under primary decomposition takes the form,
\beal{es130a86}
\mathcal{I}^{\text{Irr}}_{\partial W}=
&
\langle
~
X_{33}-X_{44}~,~
X_{22}-X_{44}~,~
X_{11}-X_{44}~,~
&
\nn\\
&
X_{23} X_{32}-X_{14} X_{41}~,~
X_{12} X_{21}-X_{14} X_{41}
~
\rangle 
~.~
&
\eea
Using this description of the mesonic moduli space expressed in terms of the relations formed by the chiral fields of the BFT, we can calculate the Hilbert series as follows,
\beal{es130a87}
g(\bar{t}, f_1, f_2 ;\mathcal{M}^{mes}_{\text{BFT}[\mathbb{C}^2/\mathbb{Z}_3 \times \mathbb{C}]})=
\frac{1 - \bar{t}^6
}{
(1 - \bar{t})(1 - \bar{t}^2) (1 - f_1 f_2^{-1} \bar{t}^3) (1 - f_1^{-1} f_2 \bar{t}^3)
}
~,~
\eea
where the fugacities $f_1,f_2, \bar{t}$ correspond to the global symmetry charge assignment on the chiral fields of the BFT, as summarized in \tref{tab_130}.
The corresponding plethystic logarithm takes the following form,
\beal{es130a88}
PL[g(\bar{t}, f_1, f_2 ;\mathcal{M}^{mes}_{\text{BFT}[\mathbb{C}^2/\mathbb{Z}_3 \times \mathbb{C}]})]
=
\bar{t} 
+ \bar{t}^2
+ 
(f_1 f_2^{-1} 
+ f_1^{-1} f_2 ) \bar{t}^3
- \bar{t}^6
 ~.~
\eea
We see that the above Hilbert series is identical up to a fugacity map to the one in \eref{es130a70} for the $\text{BFT}[\mathbb{C}^2/\mathbb{Z}_3 \times \mathbb{C}]$ Model. 
Hence, they correspond to the same complete intersection mesonic moduli space for the $\text{BFT}[\mathbb{C}^2/\mathbb{Z}_3 \times \mathbb{C}]$ Model.
\\

\subsubsection{$\text{BFT}[\mathbb{C}^2/\mathbb{Z}_{n} \times \mathbb{C}]$ Model \label{sec_e1n}}

\begin{figure} [h]
\centering
\includegraphics{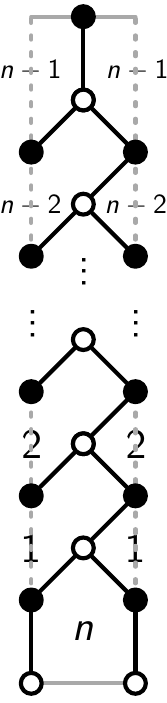}
\caption{
The bipartite graph for the $\text{BFT}[\mathbb{C}^2/\mathbb{Z}_n \times \mathbb{C}]$ Model.
\label{fig_130n}
}
\end{figure}

The BFT models $\text{BFT}[\mathbb{C}^2/\mathbb{Z}_2 \times \mathbb{C}]$ and $\text{BFT}[\mathbb{C}^2/\mathbb{Z}_3 \times \mathbb{C}]$ are two BFT models out of an infinite family of BFT models that we denote as $\text{BFT}[\mathbb{C}^2/\mathbb{Z}_n \times \mathbb{C}]$.
This family is parameterized by the integer $n\geq 2$ and is given by the bipartite graph on a cylinder shown in \fref{fig_130n}. 
The abelian BFT has $U(1)_{b_1}, \dots, U(1)_{b_{n-2}}$ gauge groups, and $U(1)_{f_1}$ and $U(1)_{f_2}$ flavor. 
The corresponding superpotential takes the form,
\beal{es130na10}
W = 
&
X_{12} X_{22} X_{21}
+ X_{23} X_{33} X_{32}
+ \dots
+ X_{n-2.n-1} X_{n-1.n-1} X_{n-1.n-2}
+ X_{n.1} X_{11} X_{1.n} 
&
\nn\\
&
- X_{21} X_{11} X_{12}
- X_{32} X_{22} X_{23}
- \dots 
- X_{n-1.n-2} X_{n-2.n-2} X_{n-2.n-1}
- X_{1.n} X_{n.n} X_{n.1}
&
~.~
\nn\\
\eea

\begin{figure} [h]
\centering
\includegraphics[width=.55\textwidth]{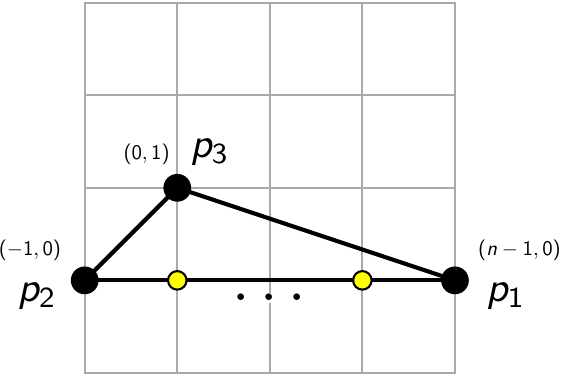}
\caption{
Toric diagram for the $\text{BFT}[\mathbb{C}^2/\mathbb{Z}_n \times \mathbb{C}]$ Model.
}
\label{fig130n}
\end{figure}

The unrefined Hilbert series of the mesonic moduli space of the $\text{BFT}[\mathbb{C}^2/\mathbb{Z}_n \times \mathbb{C}]$ Model takes the following form,
\beal{es130na70}
&&
g(t_1,t_2, t_3, y_q=1;\mathcal{M}^{mes}_{\text{BFT}[\mathbb{C}^2/\mathbb{Z}_n \times \mathbb{C}]})
=
\frac{1- t_1^n t_2^n}{
(1-t_3)
(1- t_1 t_2)
(1- t_1^n)
(1- t_2^n)
}
~,~
\nn\\
\eea
where the fugacities $t_1,t_2,t_3$ correspond the GLSM fields $p_1,p_2,p_3$.
These GLSM fields are associated to the extremal vertices of the toric diagram of the mesonic moduli space as illustrated in \fref{fig130n}.
 The plethystic logarithm of the Hilbert series, 
\beal{es130na71}
PL[g(t_1,t_2, t_3, y_q=1;\mathcal{M}^{mes}_{\text{BFT}[\mathbb{C}^2/\mathbb{Z}_n \times \mathbb{C}]})]
= 
&&
t_3
+ t_1 t_2
+ t_1^n
+ t_2^n 
- t_1^n t_2^n
~,~
\nn\\
\eea
shows that the mesonic moduli space is a complete intersection. 

\begin{table}[ht!]
\centering
\begin{tabular}{|l|l|c|c|c|c|c|}
\hline
generator & GLSM fields & fugacities \\
\hline\hline
$A_{+}=X_{n.1}X_{12}X_{23} \cdots X_{n-2.n-1}$ & $p_2^n$ & $t_2^n$ \\
$A_{0}= X_{1.n}X_{n.1}=X_{21} X_{12} = X_{32} X_{23} $ & $p_1 p_2$ & $t_1 t_2$   \\
$ = \dots = X_{n-1.n-2}X_{n-2.n-1}$ &  & \\
$A_{-}=X_{n-1.n-2} \cdots X_{32} X_{21} X_{1.n}$ & $p_1^n$ & $t_1^n$\\
$B=X_{11}=X_{22}=\dots=X_{n.n}$ & $p_3$ & $t_3$\\
\hline
\end{tabular}
\caption{Generators of the mesonic moduli space of the $\text{BFT}[\mathbb{C}^2/\mathbb{Z}_n \times \mathbb{C}]$ Model.}
\label{tab130n_3}
\end{table}

The mesonic moduli space has always 4 generators as summarized in \tref{tab130n_3}.
They satisfy a single defining relation, which allows us to write the mesonic moduli space of the $\text{BFT}[\mathbb{C}^2/\mathbb{Z}_n \times \mathbb{C}]$ Model as follows,
\beal{es130na80}
\mathcal{M}^{mes}_{\text{BFT}[\mathbb{C}^2/\mathbb{Z}_n \times \mathbb{C}]} 
&=&
\text{Spec}~  \mathbb{C}[A_+, A_-,A_0,B] /  \langle A_{+} A_{-} - A_0^n \rangle 
\nn\\
&=& \mathbb{C}^2 /\mathbb{Z}_n \times \mathbb{C}
~,~
\eea
where we note that the mesonic moduli space of the BFT is an abelian orbifold of the form $\mathbb{C}^2 /\mathbb{Z}_n \times \mathbb{C}$. 
\\

\subsection{Family of $\text{BFT}[L_{n-1,n,n-1}]$ Models \label{sec_f02}}

We introduce in this section another infinite family of BFTs defined on a cylinder whose mesonic moduli spaces are toric Calabi-Yau 3-folds and are known as $L_{n-1,n,n-1}$ \cite{Benvenuti:2005ja,Butti:2005sw,Franco:2005sm}. 
We note that for the same family of toric Calabi-Yau 3-folds, there are also corresponding brane tilings defined on a 2-torus \cite{Franco:2005sm}.
\\

\subsubsection{$\text{BFT}[\text{SPP}]$ Model \label{sec_e08}}

\begin{figure} [h]
\centering
\includegraphics{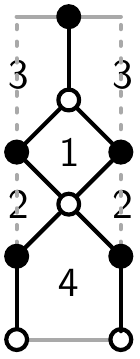}
\caption{
The bipartite graph for the $\text{BFT}[\text{SPP}]$ Model.
\label{fig_80}
}
\end{figure}

\begin{figure} [h]
\centering
\includegraphics{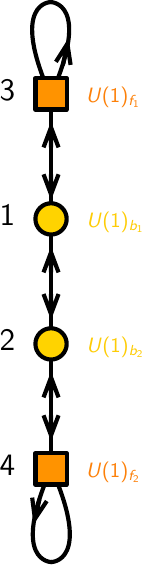}
\caption{
The quiver for the $\text{BFT}[\text{SPP}]$ Model.
\label{fig_81}
}
\end{figure}

The bipartite graph on the cylinder and quiver diagram for the $\text{BFT}[\text{SPP}]$ Model are shown in \fref{fig_80} and \fref{fig_81}, respectively. 
The corresponding superpotential takes the following form,
\beal{es80a10}
W = 
X_{13} X_{33} X_{31} 
+ X_{12} X_{24} X_{42} X_{21}
- X_{12} X_{21} X_{13} X_{31} 
- X_{24} X_{44} X_{42}
~,~
\eea
where the gauge and flavor symmetry charges on the chiral fields $X_{ij}$ are summarized in \tref{tab_80}.
We note that the global symmetry of the mesonic moduli space based on the quiver diagram in \fref{fig_81} is $U(1)_{f_1} \times U(1)_{f_2} \times U(1)_R$.
However, since $B=2$ for this BFT, we have an additional $U(1)_h$ global symmetry that allows us to rewrite the global symmetry as $U(1)_{f} \times U(1)_{h} \times U(1)_R$ as discussed in section \sref{smod}.
For now, we keep the global symmetry $U(1)_{f_1} \times U(1)_{f_2} \times U(1)_R$ based on the quiver diagram as summarized in \tref{tab_80}.
The $U(1)_R$ charges on the chiral fields shown in \tref{tab_80} are obtained using $a$-maximization as discussed in section \sref{smod}.

\begin{table}[ht!]
\centering
\begin{tabular}{|c|c|c|c|c|c|c|}
\hline
\; & $U(1)_{b_1}$ & $U(1)_{b_2}$ & $U(1)_{f_1}$ & $U(1)_{f_2}$ & $U(1)_R$ & fugacity \\
\hline\hline
$X_{12}$  & $-1$ & $+ 1$ & 0 & 0 & $\frac{15-\sqrt{43}}{21}$ & $b_1^{-1} b_2 \bar{t}_1$\\
$X_{21}$  & $+1$ & $-1$& 0 & 0 &  $\frac{15-\sqrt{43}}{21}$ & $b_1 b_2^{-1} \bar{t}_1$\\
\hline
$X_{13}$  & $-1$ & 0 & $+1$ & 0 & $\frac{6+\sqrt{43}}{21}$ & $b_1^{-1} b_3 \bar{t}_2$\\ 
$X_{31}$  & $+1$ & 0 & $-1$& 0 & $\frac{6+\sqrt{43}}{21}$ & $b_1 b_3^{-1} \bar{t}_2$\\ 
$X_{24}$  & 0 & $-1$ & 0 & $+1$& $\frac{6+\sqrt{43}}{21}$ & $b_2^{-1} b_4 \bar{t}_2$\\ 
$X_{42}$  & 0 & $+1$ & 0 & $-1$ & $\frac{6+\sqrt{43}}{21}$ & $b_1 b_3^{-1} \bar{t}_2$\\ 
\hline
$X_{33}$  & 0 & 0 & 0 & 0 & $\frac{2(15-\sqrt{43})}{21}$ & $\bar{t}_1^2$ \\
$X_{44}$  & 0 & 0 & 0 & 0 & $\frac{2(15-\sqrt{43})}{21}$ & $\bar{t}_1^2$ \\
\hline
\end{tabular}
\caption{Gauge and global symmetry charges on the chiral fields of the $\text{BFT}[\text{SPP}]$ Model.}
\label{tab_80}
\end{table}

\paragraph{Forward Algorithm.}
Based on the superpotential of the $\text{BFT}[\text{SPP}]$ Model in \eref{es80a10}, 
the $F$-terms take the following form,
\beal{es80a11}
&
\partial_{X_{12}} W =
X_{24} X_{42} X_{21}  - X_{21} X_{13} X_{31} ~,~
\partial_{X_{13}} = 
X_{33} X_{31} - X_{31} X_{12} X_{21} ~,~
&
\nn\\
&
\partial_{X_{21}} W = 
X_{12} X_{24} X_{42} - X_{13} X_{31} X_{12}~,~
\partial_{X_{31}} W = 
X_{13} X_{33} - X_{12} X_{21} X_{13} ~,~
&
\nn\\
&
\partial_{X_{42}} W = 
X_{21} X_{12} X_{24} - X_{24} X_{44} ~.~
&
\eea
Under these $F$-terms, we can identify the following independent chiral fields,
\beal{es80a12}
v_1 = X_{12} ~,~
v_2 = X_{13} ~,~
v_3 = X_{21} ~,~
v_4 = X_{24} ~,~
v_5 = X_{31} ~,~
\eea
which we can use to express the remaining chiral fields as follows,
\beal{es80a13}
X_{33} = X_{44} = v_1 v_3~,~
X_{42} = \frac{v_2 v_5}{v_4} ~.~
\eea
These relations between chiral fields can be summarized in the following $K$-matrix,
\beal{es80a60}
K=\left(
\begin{array}{c|ccccc}
\; & v_1 & v_2 & v_3 & v_4 & v_5 \\
\hline
X_{12} & 1 & 0 & 0 & 0 & 0 \\
X_{21} & 0 & 0 & 1 & 0 & 0 \\
X_{13} & 0 & 1 & 0 & 0 & 0 \\
X_{31} & 0 & 0 & 0 & 0 & 1 \\
X_{24} & 0 & 0 & 0 & 1 & 0 \\
X_{42} & 0 & 1 & 0 & -1 & 1 \\
X_{33} & 1 & 0 & 1 & 0 & 0 \\
X_{44} & 1 & 0 & 1 & 0 & 0 \\
\end{array}
\right)
~.~
 \eea

The perfect matchings of the $\text{BFT}[\text{SPP}]$ Model can be obtained from the $K$-matrix.
They correspond to GLSM fields of the BFT and are summarized in the $P$-matrix as follows,
\beal{es80a61}
P=\left(
\begin{array}{c|cccc|cc}
\; & p_1 & p_2 & p_3 & p_4 & s_1 & s_2 \\
\hline
X_{12} & 0 & 0 & 1 & 0 & 0 & 0 \\
X_{21} & 0 & 1 & 0 & 0 & 0 & 0 \\
X_{13} & 1 & 0 & 0 & 0 & 0 & 1 \\
X_{31} & 0 & 0 & 0 & 1 & 1 & 0 \\
X_{24} & 0 & 0 & 0 & 1 & 0 & 1 \\
X_{42} & 1 & 0 & 0 & 0 & 1 & 0 \\
X_{33} & 0 & 1 & 1 & 0 & 0 & 0 \\
X_{44} & 0 & 1 & 1 & 0 & 0 & 0 \\
 \end{array}
\right)~.~
\eea
We note that the chiral fields can be expressed as products of GLSM fields as shown below, 
\beal{es80a61b}
&
X_{12} = p_3 ~,~
X_{21} = p_2 ~,~
X_{13} = p_1 s_2 ~,~
X_{31} = p_4 s_1 ~,~
&
\nn\\
&
X_{24} = p_4 s_2 ~,~
X_{42} = p_1 s_1 ~,~
X_{33} = X_{44} = p_2 p_3
~.~
&
\eea
Using the basis of GLSM fields, the $F$-terms can be encoded as a collection of $U(1)$ charges, which are summarized in the $F$-term charge matrix as follows,
\beal{es80a62}
Q_{F} =\left(
\begin{array}{cccc|cc}
 p_1 & p_2 & p_3 & p_4 & s_1 & s_2 \\
\hline
 1 & 0 & 0 & 1 & -1 & -1 \\
\end{array}
\right)~.~
\eea
The $D$-term charge matrix takes the following form,
\beal{es80a63}
Q_{D} =\left(
\begin{array}{cccc|cc}
p_1 & p_2 & p_3 & p_4 & s_1 & s_2 \\
\hline
 1 & 0 & 0 & 1 & -2 & 0 \\
 0 & 1 & -1 & 1 & -1 & 0 \\
\end{array}
\right)~.~
\eea

The toric diagram of the $\text{BFT}[\text{SPP}]$ Model can be obtained from the $F$- and $D$-term charge matrices. 
It is summarized in the toric diagram matrix, which is shown below,
\beal{es80a65}
G_{t} =\left(
\begin{array}{cccc|cc}
p_1 & p_2 & p_3 & p_4 & s_1 & s_2 \\
\hline
 1 & 1 & 0 & -1 & 0 & 0 \\
 0 & 1 & 1 & 0 & 0 & 0 \\
 1 & 1 & 1 & 1 & 1 & 1 \\
\end{array}
\right)~.~
\eea
The toric diagram of the mesonic moduli space of the $\text{BFT}[\text{SPP}]$ Model is shown in \fref{fig_82}.
Based on the toric diagram, the mesonic moduli space of the BFT is the suspended pinch point (SPP) \cite{Morrison:1998cs,Franco:2005rj}, which is equivalent to $L_{121}$ \cite{Benvenuti:2005ja,Butti:2005sw,Franco:2005sm}.
This toric Calabi-Yau 3-fold is known to be a complete intersection with 4 generators satisfying a defining relation. We will derive these properties of the mesonic moduli space by calculating the corresponding Hilbert series. 

\begin{figure} [h]
\centering
\includegraphics[width=.3\textwidth]{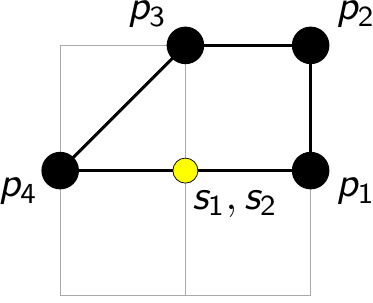}
\caption{
Toric diagram for the $\text{BFT}[\text{SPP}]$ Model. 
}
\label{fig_82}
\end{figure}

\paragraph{Hilbert Series.}
Using the $F$- and $D$-term charge matrices in \eref{es80a62} and \eref{es80a63}, respectively, we can calculate the Hilbert series of the mesonic moduli space of the $\text{BFT}[\text{SPP}]$ Model as follows,
\beal{es80a70}
&&
g(t_1,\dots, t_4, y_s;\mathcal{M}^{mes}_{\text{BFT}[\text{SPP}]})
=
\frac{1- y_s^2 t_1^2 t_2 t_3 t_4^2}{
(1-t_2 t_3)
(1-y_s t_1^2 t_2)
(1- y_s t_1 t_4)
(1-y_s t_3 t_4^2)
}
~,~
\nn\\
\eea
where the fugacities $t_a$ correspond to the GLSM fields $p_a$, 
and the fugacity $y_s$ corresponds to the product of GLSM fields $s=s_1 s_2$.
The plethystic logarithm of the Hilbert series takes the following form,
\beal{es80a71}
&&
PL[g(t_1,\dots, t_4, y_s;\mathcal{M}^{mes}_{\text{BFT}[\text{SPP}]})]
= 
t_2 t_3 + y_s t_1 t_4
+ y_s t_1^2 t_2 + y_s t_3 t_4^2
- y_s^2 t_1^2 t_2 t_3 t_4^2 
~,~
\nn\\
\eea
where the finite expansion of the plethystic logarithm indicates that the mesonic moduli space is a complete intersection. 

\begin{table}[ht!]
\centering
\begin{tabular}{|l|c|c|c|c|l|}
\hline
\; & $U(1)_{\bar{f}_1}$ & $U(1)_{\bar{f}_2}$ & $U(1)_{R}$ & fugacity\\
\hline\hline
$p_1$ & +1&  0 & $\frac{6+\sqrt{43}}{21}$   & $t_1=\bar{f}_1 t^a$\\
$p_2$ & - 1& +1& $\frac{15-\sqrt{43}}{21}$  & $t_2=\bar{f}_1^{-1}  \bar{f}_2 t^{1-a}$\\
$p_3$ &  0 & +1& $\frac{15-\sqrt{43}}{21}$  & $t_3=\bar{f}_2 t^{1-a}$\\
$p_4$ &  0 & -2 & $\frac{6+\sqrt{43}}{21}$   & $t_4=\bar{f}_2^{-2} t^{a}$\\
\hline
\end{tabular}
\caption{Global symmetry charges according to $U(1)_{\bar{f}_1}\times U(1)_{\bar{f}_2} \times U(1)_R$ on perfect matchings and their corresponding fugacities in the \text{BFT}[\text{SPP}].}
\label{tab80_2}
\end{table}

Based on the global symmetry charge assignment on the GLSM fields, as summarized in \tref{tab80_2}, we can identify the following fugacity map
\beal{es80a75}
\bar{f}_1 = \frac{t_3}{t_2}
~,~
\bar{f}_2 = \frac{t_1^{1/2} t_2^{1/2}}{t_3^{1/2} t_4^{1/2}}
~,~
t = 
t_1^{1/2} t_2^{1/2} t_3^{1/2} t_4^{1/2}
~,~
\eea
that allows us to rewrite the Hilbert series of the mesonic moduli space in terms of global symmetry fugacities as shown below, 
\beal{es80a76}
g(t, \bar{f}_1, \bar{f}_2 ;\mathcal{M}^{mes}_{\text{BFT}[\text{SPP}]})=
\frac{1- \bar{f}_1\bar{f}_2^{-2} t^{2+2a}}
{
(1-\bar{f}_1^{-1} \bar{f}_2^2  t^{2-2a})
(1-\bar{f}_1 \bar{f}_2^{-2} t^{2a})
(1-\bar{f}_1 \bar{f}_2 t^{1+a})
(1-\bar{f}_2^{-3} t^{1+a})
}
 ~.~
 \nn\\
\eea
Here $t$ corresponds to the $U(1)_R$ symmetry and $a=\frac{6+\sqrt{43}}{21}$ is taken as a unit of $U(1)_R$ charge on the GLSM fields of the BFT.
The corresponding plethystic logarithm takes the following form,
\beal{es80a76}
PL[
g(t, \bar{f}_1, \bar{f}_2 ;\mathcal{M}^{mes}_{\text{BFT}[\text{SPP}]})
]
= 
\bar{f}_1^{-1} \bar{f}_2^2 t^{2-2a}
+\bar{f}_1 \bar{f}_2^{-2} t^{2a} 
+ (\bar{f}_1 \bar{f}_2+\bar{f}_2 ^{-3}) t^{1+a}
- \bar{f}_1 \bar{f}_2^{-2} t^{2+2a}
~,~\nn\\
\eea
where we can identify the generators of the mesonic moduli space and their corresponding global symmetry charges as summarized in \tref{tab80_3}. 

\begin{table}[ht!]
\centering
\begin{tabular}{|l|l|c|c|c|c|}
\hline
generator & GLSM fields & $U(1)_{\bar{f}_1}$ & $U(1)_{\bar{f}_2}$ & $U(1)_R$ & fugacities \\
\hline\hline
$A_{+} = X_{12} X_{21} = X_{33} = X_{44}$ & $p_2 p_3$ 
& $-1$ &  $+2$ & $2-2a$ & $\bar{f}_1^{-1} \bar{f}_2^2 t^{2-2a} $\\
$A_{-}=X_{13} X_{31}=X_{24}X_{42} $ & $p_1 p_4 s_1 s_2$ 
& $+1$ & $-2$ & $2a$ & $\bar{f}_1 \bar{f}_2^{-2} t^{2a}$  \\
$B_{+}=X_{42} X_{21} X_{13}$ & $p_1^{2} p_2 s_1 s_2$ 
& $+1$ & $+1$  & $1+a$  & $\bar{f}_1 \bar{f}_2 t^{1+a}$ \\
$B_{-}=X_{31}X_{12}X_{24} $ & $p_3 p_4^2 s_1 s_2$ 
& $0$ & $-3$  & $1+a$  & $\bar{f}_2 ^{-3} t^{1+a}$ \\
\hline
\end{tabular}
\caption{Generators of the mesonic moduli space of the $\text{BFT}[\text{SPP}]$ Model.}
\label{tab80_3}
\end{table}

In terms of the 4 generators and the defining relation formed amongst them, we can write the mesonic moduli space of the $\text{BFT}[\text{SPP}]$ Model as follows,
\beal{es80a80}
\mathcal{M}^{mes}_{\text{BFT}[\text{SPP}]}
&=&
\text{Spec}~  \mathbb{C}[A_+, A_-,B_+,B_-] /  \langle A_{+} A_{-}^2 - B_+ B_- \rangle 
\nn\\
&=& \text{SPP} ~.~
\eea
As identified above through the Hilbert series, the mesonic moduli space of the BFT is the suspended pinch point (SPP).
\\

We can also express the mesonic moduli space of the BFT in terms of the binomial ideal that is formed by the $F$-terms in \eref{es80a11}. 
The mesonic moduli space of the $\text{BFT}[\text{SPP}]$ Model can be written as,
\beal{es80a85}
\mathcal{M}^{mes}_{\text{BFT}[\text{SPP}]}
= \text{Spec}~
(\mathbb{C}[X_{ij}] / \mathcal{I}^{\text{Irr}}_{\partial W} ) // (U(1)_{b_1} \times U(1)_{b_2})
~,~
\eea
where the coherent component of the ideal obtained via primary decomposition takes the following form,
\beal{es80a86}
\mathcal{I}^{\text{Irr}}_{\partial W}=
&
\langle
~
X_{33}-X_{44}~,~
X_{13} X_{31}-X_{24} X_{42}~,~
X_{12} X_{21}-X_{44}
~
\rangle 
~.~
&
\eea
Using fugacities corresponding to the global symmetry charges on the chiral fields summarized in \tref{tab_80}, we can calculate the refined Hilbert series as follows,
\beal{es80a87}
g(\bar{t}_1,\bar{t}_2,\bar{t}_3, f_1,f_2; \mathcal{M}^{mes}_{\text{BFT}[\text{SPP}]})
&=&
\frac{
1 - \bar{t}_1^2 \bar{t}_2^4
}{
(1 - \bar{t}_1^2) 
(1 - \bar{t}_2^2) 
(1 - f_1 f_2^{-1} \bar{t}_1 \bar{t}_2^2) 
(1 - f_1^{-1} f_2 \bar{t}_1 \bar{t}_2^2)
}
~,~
\nn\\
\eea
where the corresponding plethystic logarithm takes the form,
\beal{es80a88}
PL[g(\bar{t}_1,\bar{t}_2,\bar{t}_3, f_1,f_2; \mathcal{M}^{mes}_{\text{BFT}[\text{SPP}]})]
&=&
\bar{t}_1^2
+ \bar{t}_2^2
+ f_1 f_2^{-1} \bar{t}_1 \bar{t}_2^2
+ f_1^{-1} f_2 \bar{t}_1 \bar{t}_2^2
- \bar{t}_1^2 \bar{t}_2^4
~.~
\eea
Here, the fugacities can be mapped so that the above Hilbert series is identical to the one in \eref{es80a70}.
This indicates that both Hilbert series refer to the same mesonic moduli space, which is for the  $\text{BFT}[\text{SPP}]$ Model the suspended pinch point (SPP).
\\

\subsubsection{$\text{BFT}[L_{232}]$ Model \label{sec_e14}}

\begin{figure} [h]
\centering
\includegraphics{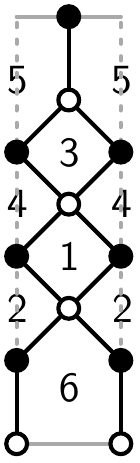}
\caption{
The bipartite graph for the $\text{BFT}[L_{232}]$ Model.
\label{fig_140}
}
\end{figure}

\begin{figure} [h]
\centering
\includegraphics{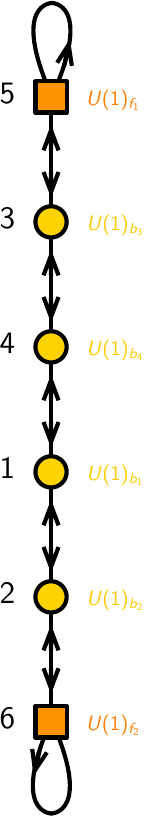}
\caption{
The quiver for the $\text{BFT}[L_{232}]$ Model.
\label{fig_141}
}
\end{figure}

In \fref{fig_140}, we have the bipartite graph on the cylinder which corresponds to the $\text{BFT}[L_{232}]$ Model.
The corresponding quiver diagram is shown in \fref{fig_141} and the superpotential is given by,
\beal{es140a10}
W = 
&
X_{12} X_{26} X_{62} X_{21} 
+  X_{34} X_{41} X_{14} X_{43} 
+  X_{53} X_{35} X_{55} 
&
\nn\\
&
-  X_{53} X_{34} X_{43} X_{35} 
-  X_{41} X_{12} X_{21} X_{14} 
-  X_{26} X_{66} X_{62}
&
~,~
\eea
where the gauge and flavor symmetry charges on the chiral fields $X_{ij}$ are summarized in \tref{tab_140}.
As we have seen above, because the BFT has $B=2$, the global symmetry has an additional $U(1)_h$ symmetry that allows us to rewrite the global symmetry based on the quiver diagram $U(1)_{f_1} \times U(1)_{f_2} \times U(1)_R$ with $U(1)_h$.
For now, we keep the global symmetry as $U(1)_{f_1} \times U(1)_{f_2} \times U(1)_R$.
The chiral fields carry $U(1)_R$ charges as shown in \tref{tab_140}. These $U(1)_R$ charges are obtained via $a$-maximization. 
\\

\begin{table}[ht!]
\centering
\begin{tabular}{|c|c|c|c|c|c|c|c|c|}
\hline
\; & $U(1)_{b_1}$ & $U(1)_{b_2}$ & $U(1)_{b_3}$ & $U(1)_{b_4}$ & $U(1)_{f_1}$ & $U(1)_{f_2}$ & $U(1)_R$ & fugacity\\
\hline\hline
$X_{12}$  & $-1$ & + 1 & 0 & 0 & 0 & 0 &$\frac{18-\sqrt{79}}{21}$ & $b_1^{-1} b_2\bar{t}_1$\\
$X_{21}$  & +1 & -1& 0 & 0 & 0 & 0 &$\frac{18-\sqrt{79}}{21}$ & $b_1 b_2^{-1} \bar{t}_1$\\
$X_{14}$  & $-1$ & 0 & 0 & +1 & 0 & 0 &$\frac{3+\sqrt{79}}{21}$ & $b_1^{-1} b_4\bar{t}_2$\\
$X_{41}$  & +1 & 0& 0 & $-1$ & 0 & 0 &$\frac{3+\sqrt{79}}{21}$ & $b_1 b_4^{-1}\bar{t}_2$\\
$X_{34}$  & 0 & 0 & $-1$ & +1 & 0 & 0 &$\frac{18-\sqrt{79}}{21}$ & $b_3^{-1} b_4 \bar{t}_1$\\
$X_{43}$  & 0 & 0& +1 & $-1$ & 0 & 0 &$\frac{18-\sqrt{79}}{21}$ & $b_3 b_4^{-1} \bar{t}_1$\\
\hline
$X_{35}$  & 0 & 0 & $-1$ & 0 & +1 & 0 &$\frac{3+\sqrt{79}}{21}$ & $b_3^{-1} f_1 \bar{t}_2$\\
$X_{53}$  & 0 & 0 & +1 & 0 & -1& 0 &$\frac{3+\sqrt{79}}{21}$ & $b_3 f_1^{-1} \bar{t}_2$\\
$X_{26}$  & 0 & $-1$ & 0 & 0 & 0 & +1 &$\frac{3+\sqrt{79}}{21}$ & $b_2^{-1} f_2 \bar{t}_2$\\
$X_{62}$  & 0 & +1 & 0 & 0 & 0 & $-1$ &$\frac{3+\sqrt{79}}{21}$& $b_2 f_2^{-1} \bar{t}_2$\\
\hline
$X_{55}$  & 0 & 0 & 0 & 0 & 0 & 0 &$\frac{2(18-\sqrt{79})}{21}$ & $\bar{t}_1^2$\\
$X_{66}$  & 0 & 0 & 0 & 0 & 0 & 0 &$\frac{2(18-\sqrt{79})}{21}$ & $\bar{t}_1^2$\\
\hline
\end{tabular}
\caption{Gauge and flavor symmetry charges on the chiral fields of the $\text{BFT}[L_{232}]$ Model.}
\label{tab_140}
\end{table}

\paragraph{Forward Algorithm.} As part of the forward algorithm, we obtain the $F$-terms from the superpotential in \eref{es140a10} as follows, 
\beal{es140a11}
&
\partial_{X_{12}}W =   X_{26} X_{62}    X_{21} -X_{21} X_{14} X_{41}~,~ 
\partial_{X_{21}}W =  X_{12} X_{26} X_{62}  -   X_{14} X_{41} X_{12} ~,~ 
&
\nn\\
&
\partial_{X_{14}}W =   X_{43} X_{34} X_{41} -X_{41} X_{12} X_{21} ~,~ 
\partial_{X_{41}}W =  X_{14} X_{43} X_{34} -X_{12} X_{21} X_{14} ~,~ 
&
\nn\\
&
\partial_{X_{34}}W =  X_{41} X_{14} X_{43}  - X_{43} X_{35} X_{53} ~,~ 
\partial_{X_{43}}W =  X_{34} X_{41} X_{14}  -   X_{35} X_{53} X_{34} ~,~
&
\nn\\
&
\partial_{X_{35}}W =  X_{55} X_{53}  -   X_{53} X_{34}  X_{43} ~,~ 
\partial_{X_{53}}W =  X_{35} X_{55}  -   X_{34} X_{43} X_{35} ~,~ 
&
\nn\\
&
\partial_{X_{26}}W =  X_{62} X_{21} X_{12} -X_{66} X_{62}  ~,~ 
\partial_{X_{62}}W = X_{21} X_{12} X_{26}  -X_{26} X_{66} ~.~
&
\eea
These $F$-terms show that one can identify 7 independent chiral fields, which are
\beal{es140a12}
v_1 = X_{12} ~,~
v_2 = X_{14} ~,~
v_3 = X_{21} ~,~
v_4 = X_{26} ~,~
v_5 = X_{34} ~,~
v_6 = X_{35} ~,~
v_7 = X_{41} ~.~
\nn\\
\eea
In terms of these independent chiral fields, we can express the remaining chiral fields as follows,
\beal{es140a13}
X_{43} = \frac{v_1 v_3}{v_5}~,~
X_{53} = \frac{v_2 v_7}{v_6}~,~
X_{62} = \frac{v_2 v_7}{v_4}~,~
X_{55} = X_{66} = v_1 v_3~.~
\eea
These relations between chiral fields can be summarized in the $K$-matrix, which takes the following form,
\beal{es140a60}
K=\left(
\begin{array}{c|ccccccc}
\; & v_1 & v_2 & v_3 & v_4 & v_5 & v_6 & v_7\\
\hline
X_{12} & 1 & 0 & 0 & 0 & 0 & 0 & 0 \\
X_{21} & 0 & 0 & 1 & 0 & 0 & 0 & 0 \\
X_{14} & 0 & 1 & 0 & 0 & 0 & 0 & 0 \\
X_{41} & 0 & 0 & 0 & 0 & 0 & 0 & 1 \\
X_{34} & 0 & 0 & 0 & 0 & 1 & 0 & 0 \\
X_{43} & 1 & 0 & 1 & 0 & -1 & 0 & 0 \\
X_{35} & 0 & 0 & 0 & 0 & 0 & 1 & 0 \\
X_{53} & 0 & 1 & 0 & 0 & 0 & -1 & 1 \\
X_{26} & 0 & 0 & 0 & 1 & 0 & 0 & 0 \\
X_{62} & 0 & 1 & 0 & -1 & 0 & 0 & 1 \\
X_{55} & 1 & 0 & 1 & 0 & 0 & 0 & 0 \\
X_{66} & 1 & 0 & 1 & 0 & 0 & 0 & 0 \\
\end{array}
\right)
~.~
 \eea

Using the $K$-matrix, we can obtain the perfect matching matrix of the $\text{BFT}[L_{232}]$ Model.
These perfect matchings correspond to GLSM fields of the BFT and are summarized in the following $P$-matrix,
\beal{es140a61}
P=\left(
\begin{array}{c|cccc|cc|ccc|ccc}
\; & p_1 & p_2 & p_3 & p_4 & s_1 & s_2 & q_1 & q_2 & q_3 & u_1 & u_2 & u_3\\
\hline
X_{12} & 0 & 1 & 0 & 0 & 0 & 1 & 0 & 0 & 0 & 0 & 0 & 0 \\
X_{21} & 0 & 0 & 1 & 0 & 1 & 0 & 0 & 0 & 0 & 0 & 0 & 0 \\
X_{14} & 0 & 0 & 0 & 1 & 0 & 0 & 0 & 0 & 1 & 0 & 1 & 1 \\
X_{41} & 1 & 0 & 0 & 0 & 0 & 0 & 1 & 1 & 0 & 1 & 0 & 0 \\
X_{34} & 0 & 1 & 0 & 0 & 1 & 0 & 0 & 0 & 0 & 0 & 0 & 0 \\
X_{43} & 0 & 0 & 1 & 0 & 0 & 1 & 0 & 0 & 0 & 0 & 0 & 0 \\
X_{35} & 0 & 0 & 0 & 1 & 0 & 0 & 0 & 1 & 0 & 1 & 0 & 1 \\
X_{53} & 1 & 0 & 0 & 0 & 0 & 0 & 1 & 0 & 1 & 0 & 1 & 0 \\
X_{26} & 1 & 0 & 0 & 0 & 0 & 0 & 0 & 1 & 1 & 0 & 0 & 1 \\
X_{62} & 0 & 0 & 0 & 1 & 0 & 0 & 1 & 0 & 0 & 1 & 1 & 0 \\
X_{55} & 0 & 1 & 1 & 0 & 1 & 1 & 0 & 0 & 0 & 0 & 0 & 0 \\
X_{66} & 0 & 1 & 1 & 0 & 1 & 1 & 0 & 0 & 0 & 0 & 0 & 0 \\
 \end{array}
\right)~.~
\eea
The chiral fields of the BFT can be written in terms of the GLSM fields as follows,
\beal{es140a61b}
&
X_{12} = p_2 s_2 ~,~
X_{21} = p_3  s_1~,~
X_{14} = p_4 q_3 u_2 u_3 ~,~
X_{41} = p_1 q_1 q_2 u_1 ~,~
&
\nn\\
&
X_{34} = p_2 s_1 ~,~
X_{43} = p_3 s_2 ~,~
X_{35} = p_4 q_2 u_1 u_3 ~,~
X_{53} = p_1 q_1 q_3 u_2 ~,~
&
\nn\\
&
X_{26} = p_1 q_2 q_3 u_3 ~,~
X_{62} = p_4 q_1 u_1 u_2 ~,~
X_{55} = X_{66} = p_2 p_3 s_1 s_2 ~.~
&
\eea
In terms of the GLSM fields, the $F$-terms of the BFT can be expressed as a collection of $U(1)$ charges summarized by the $Q_F$-matrix given as follows, 
\beal{es140a62}
Q_{F} =\left(
\begin{array}{cccc|cc|ccc|ccc}
p_1 & p_2 & p_3 & p_4 & s_1 & s_2 & q_1 & q_2 & q_3 & u_1 & u_2 & u_3\\
\hline
 1 & 0 & 0 & 0 & 0 & 0 & 0 & 0 & -2 & -1 & 1 & 1 \\
 0 & 1 & 1 & 0 & -1 & -1 & 0 & 0 & 0 & 0 & 0 & 0 \\
 0 & 0 & 0 & 1 & 0 & 0 & 0 & 0 & 1 & 0 & -1 & -1 \\
 0 & 0 & 0 & 0 & 0 & 0 & 1 & 0 & -1 & -1 & 0 & 1 \\
 0 & 0 & 0 & 0 & 0 & 0 & 0 & 1 & -1 & -1 & 1 & 0 \\
\end{array}
\right)~.~
\eea
We also have the $D$-term charge matrix, which takes the following form,
\beal{es140a63}
Q_{D} =\left(
\begin{array}{cccc|cc|ccc|ccc}
p_1 & p_2 & p_3 & p_4 & s_1 & s_2 & q_1 & q_2 & q_3 & u_1 & u_2 & u_3\\
\hline
 1 & 0 & 0 & 0 & 0 & 0 & -2 & 0 & 0 & 0 & 1 & 0 \\
 0 & 0 & 1 & 0 & 0 & -1 & 1 & 0 & 0 & 0 & -1 & 0 \\
 0 & 0 & 0 & 0 & 1 & -1 & 0 & 0 & 0 & 0 & 0 & 0 \\
 0 & 0 & 0 & 0 & 0 & 0 & 0 & 0 & 0 & 1 & -1 & 0 \\
\end{array}
\right)~.~
\eea

Using both the $F$- and $D$-term charge matrices, we can obtain the toric diagram of the mesonic moduli space of the $\text{BFT}[L_{232}]$ Model.
The toric diagram is summarized in terms of the following matrix,
\beal{es140a65}
G_{t} =\left(
\begin{array}{cccc|cc|ccc|ccc}
p_1 & p_2 & p_3 & p_4 & s_1 & s_2 & q_1 & q_2 & q_3 & u_1 & u_2 & u_3\\
\hline
 1 & 1 & -1 & -2 & 0 & 0 & 0 & 0 & 0 & -1 & -1 & -1 \\
 0 & 1 & 1 & 0 & 1 & 1 & 0 & 0 & 0 & 0 & 0 & 0 \\
 1 & 1 & 1 & 1 & 1 & 1 & 1 & 1 & 1 & 1 & 1 & 1 \\
\end{array}
\right)~,~
\eea
which shows that the mesonic moduli space of the BFT is a toric Calabi-Yau 3-fold.
The toric diagram is shown in \fref{fig_82}.
According to the toric diagram, the mesonic moduli space is toric Calabi-Yau 3-fold known as $L_{232}$ \cite{Benvenuti:2005ja,Butti:2005sw,Franco:2005sm}.
\\

\begin{figure} [h]
\centering
\includegraphics[width=.45\textwidth]{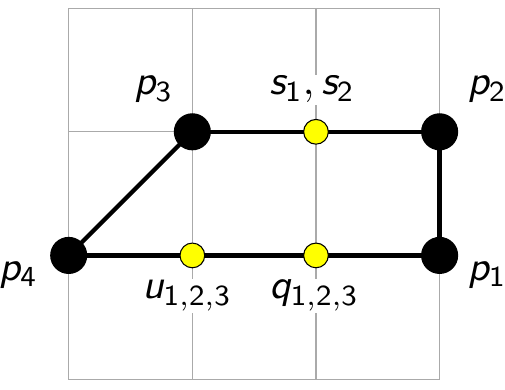}
\caption{
Toric diagram for the $\text{BFT}[L_{232}]$ Model.
}
\label{fig_82}
\end{figure}

\paragraph{Hilbert Series.}
We can verify the identity of the mesonic moduli space by calculating the corresponding Hilbert series refined in terms of fugacities corresponding to the BFT's GLSM fields,
\beal{es140a70}
&&
g(t_1,\dots, t_4, y_s,y_q,y_u;\mathcal{M}^{mes}_{\text{BFT}[L_{232}]})
=
\nn\\
&&
\hspace{2cm}
\frac{
1 - y_s^2 y_q^3 y_u^3 t_1^3 t_2^2 t_3^2 t_4^3 
}{
(1 - y_st_2 t_3) 
(1 - y_q y_u t_1 t_4) 
(1 - y_s y_q^2 y_u t_1^3 t_2^2) 
(1 - y_s  y_q y_u^2 t_3^2 t_4^3 )
}
~,~
\eea
where the fugacities $t_a$ correspond to the GLSM fields $p_a$, and the fugacities $y_s, y_q, y_u$ correspond to the products of GLSM fields $s=s_1 s_2$, $q=q_1 q_2 q_3$ and $u=u_1 u_2 u_3$, respectively. 
The plethystic logarithm of the Hilbert series takes the following form,
\beal{es140a71}
PL[g(t_1,\dots, t_4, y_s,y_q,y_u;\mathcal{M}^{mes}_{\text{BFT}[L_{232}]})]
= 
&&
y_st_2 t_3
+ y_q y_u t_1 t_4
+ y_s y_q^2 y_u t_1^3 t_2^2
+ y_s  y_q y_u^2 t_3^2 t_4^3
\nn\\
&&
-  y_s^2 y_q^3 y_u^3 t_1^3 t_2^2 t_3^2 t_4^3 
~,~
\eea
where the finite expansion of the Hilbert series indicates that the mesonic moduli space of the $\text{BFT}[L_{232}]$ Model is a complete intersection.

\begin{table}[ht!]
\centering
\begin{tabular}{|l|c|c|c|c|l|}
\hline
\; & $U(1)_{\bar{f}_1}$ & $U(1)_{\bar{f}_2}$ & $U(1)_{R}$ & fugacity\\
\hline\hline
$p_1$ &+1 & 0   &$\frac{3+\sqrt{79}}{21}$  & $t_1=\bar{f}_1 t^a$\\
$p_2$ &-1  & +1   &$\frac{18-\sqrt{79}}{21}$ & $t_2=\bar{f}_1^{-1} \bar{f}_2 t^{1-a}$\\
$p_3$ &0   & +1    &$\frac{18-\sqrt{79}}{21}$ & $t_3=\bar{f}_2 t^{1-a}$\\
$p_4$ &0   & -2     &$\frac{3+\sqrt{79}}{21}$  & $t_4=\bar{f}_2^{-2} t^a$\\
\hline
\end{tabular}
\caption{Global symmetry charges according to $U(1)_{\bar{f}_1}\times U(1)_{\bar{f}_2} \times U(1)_R$ on perfect matchings of the $\text{BFT}[L_{232}]$ Model.}
\label{tab140_2}
\end{table}

Using the following fugacity map based on the global symmetry charge assignment on the GLSM fields shown in \tref{tab140_2},
\beal{es140a75}
\bar{f}_1 = \frac{t_3}{t_2}
~,~
\bar{f}_2 = \frac{t_1^{1/2} t_2^{1/2}}{t_3^{1/2} t_4^{1/2}}
~,~
t=
t_1^{1/2} t_2^{1/2} t_3^{1/2} t_4^{1/2}
~,~
\eea
the Hilbert series in \eref{es140a70} can be expressed in terms of global symmetry fugacities as follows,
\beal{es140a76}
&&
g(t, \bar{f}_1, \bar{f}_2 ;\mathcal{M}^{mes}_{\text{BFT}[L_{232}]}))=
\frac{1- \bar{f}_1\bar{f}_2^{-2} t^{4+2a}}
{
(1-\bar{f}_1^{-1} \bar{f}^2  t^{2-2a})
(1-\bar{f}_1 \bar{f}_2^{-2} t^{2a})
}
\nn\\
&&
\hspace{1.5cm}
\times
\frac{1}{
(1-\bar{f}_1 \bar{f}_2^2 t^{2+a})
(1-\bar{f}_2^{-4} t^{2+a})
}
 ~.~
\eea
The corresponding plethystic logarithm takes the form,
\beal{es140a76}
PL[
g(t, \bar{f}_1, \bar{f}_2 ;\mathcal{M}^{mes}_{\text{BFT}[L_{232}]}))
]
= 
\bar{f}_1^{-1} \bar{f}^2  t^{2-2a}
+\bar{f}_1 \bar{f}_2^{-2} t^{2a}
+ (\bar{f}_1 \bar{f}_2^2+\bar{f}_2^{-4}) t^{2+a}
- \bar{f}_1\bar{f}_2^{-2} t^{4+2a}
~,~\nn\\
\eea
where we identify that the mesonic moduli space of the $\text{BFT}[L_{232}]$ Model has 4 generators and one defining relation charged under the global symmetry of the form $U(1)_{\bar{f}_1} \times U(1)_{\bar{f}_2} \times U(1)_{R}$.
The generators with their global symmetry charges are summarized in \tref{tab140_3}.

\begin{table}[ht!]
\centering
\begin{tabular}{|l|l|c|c|c|l|}
\hline
generator & GLSM fields & $U(1)_{\bar{f}_1}$ & $U(1)_{\bar{f}_2}$ & $U(1)_R$ & fugacities \\
\hline\hline
$A_{+} = X_{12} X_{21} =X_{34} X_{43} $ & $p_2 p_3 s_1 s_2$ 
& $-1$ &  $+2$ & $2-2a$ & $\bar{f}_1^{-1} \bar{f}^2  t^{2-2a}$\\
$= X_{55} = X_{66}$ &  
&  &  & & \\
$A_{-}=X_{14} X_{41} = X_{26} X_{62} $ & $p_1 p_4 q_1 q_2 q_3 u_1 u_2 u_3$ 
& +1 & $-2$ & $2a$ & $\bar{f}_1 \bar{f}_2^{-2} t^{2a}$  \\
$=X_{35}X_{53} $ &  
&  &  &  &  \\
$B_{+}=X_{53} X_{34} X_{41} X_{12} X_{26} $ & $p_1^{3} p_2^{2} s_1 s_2 q_1^{2} q_2^{2} q_3^{2} u_1 u_2 u_3$ 
&+1  & $+2$  & $2+a$  & $\bar{f}_1 \bar{f}_2^2 t^{2+a}$ \\
$B_{-}=X_{62} X_{21} X_{14} X_{43} X_{35} $ & $p_3^2 p_4^3 s_1 s_2 q_1 q_2 q_3 u_1^2 u_2^2 u_3^2$ 
& 0 & $-4$  & $2+a$  & $\bar{f}_2^{-4} t^{2+a}$ \\
\hline
\end{tabular}
\caption{Generators of the mesonic moduli space of the $\text{BFT}[L_{232}]$ Model.}
\label{tab140_3}
\end{table}

in terms of the generators summarized in \tref{tab140_3}, we can write the mesonic moduli space of the $\text{BFT}[L_{232}]$ Model as the following complete intersection,
\beal{es140a80}
\mathcal{M}^{mes}_{L_{232}}
&=&
\text{Spec}~  \mathbb{C}[A_+, A_-,B_+,B_-] /  \langle A_{+}^2 A_{-}^3 - B_+ B_- \rangle 
\nn\\
&=& L_{232} ~.~
\eea
We can see here that the defining relation amongst the 4 generators corresponds to the toric Calabi-Yau 3-fold known as $L_{232}$ whose toric diagram is shown in \fref{fig_82}.
\\

We can also calculate the Hilbert series of the mesonic moduli space by expressing it in terms of the binomial ideal formed by the $F$-terms as shown below,
\beal{es140a85}
\mathcal{M}^{mes}_{\text{BFT}[L_{232}]}
= \text{Spec}~
(\mathbb{C}[X_{ij}] / \mathcal{I}^{\text{Irr}}_{\partial W} ) // (U(1)_{b_1} \times U(1)_{b_2} \times U(1)_{b_3} \times U(1)_{b_4})
~,~
\eea
where the coherent component of the ideal obtained via primary decomposition takes the following form,
\beal{es140a86}
\mathcal{I}^{\text{Irr}}_{\partial W}=
&
\langle
~
X_{55}-X_{66}~,~
X_{35} X_{53}-X_{26} X_{62}~,~
X_{34} X_{43}-X_{66}~,~
&
\nn\\
&
X_{14} X_{41}-X_{26} X_{62}~,~
X_{12} X_{21}-X_{66}
~
\rangle 
~.~
&
\eea
Using the fugacities corresponding to the global symmetry charges on chiral fields shown in \tref{tab_140}, we can calculate the Hilbert series as follows,
\beal{es140a87}
g(\bar{t}, f_1, f_2 ;\mathcal{M}^{mes}_{\text{BFT}[L_{232}]})
=
\frac{
1 - \bar{t}_1^4 \bar{t}_2^6
}{
(1 - \bar{t}_1^2) 
(1 - \bar{t}_2^2) 
(1 - f_1 f_2^{-1} \bar{t}_1^2 \bar{t}_2^3 ) 
(1 - f_1^{-1} f_2 \bar{t}_1^2 \bar{t}_2^3 )
}
~,~
\eea
where the corresponding plethystic logarithm takes the form,
\beal{es140a88}
PL[g(\bar{t}, f_1, f_2 ;\mathcal{M}^{mes}_{\text{BFT}[L_{232}]})]
=
 \bar{t}_1^2
+ \bar{t}_2^2
+ f_1 f_2^{-1} \bar{t}_1^2 \bar{t}_2^3 
+ f_1^{-1} f_2 \bar{t}_1^2 \bar{t}_2^3 
- \bar{t}_1^4 \bar{t}_2^6
 ~.~
\eea
We see here that under an appropriate fugacity map, the Hilbert series in \eref{es140a87} is identical to the one in \eref{es140a70}.
This shows that the Hilbert series refer to the same mesonic moduli space of the $\text{BFT}[L_{232}]$ Model, which we identify as the toric Calabi-Yau 3-fold known as $L_{232}$ \cite{Benvenuti:2005ja,Butti:2005sw,Franco:2005sm}.
\\

\subsubsection{$\text{BFT}[L_{n-1.n.n-1}]$ Model \label{sec_e14n}}

\begin{figure} [h]
\centering
\includegraphics{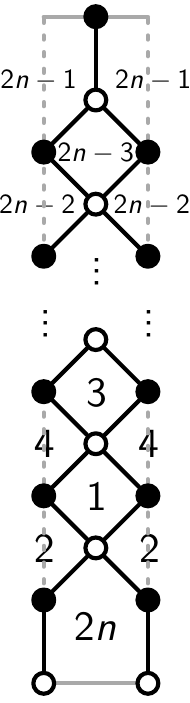}
\caption{
The bipartite graph for the $\text{BFT}[L_{n-1.n.n-1}]$ Model.
\label{fig_140n}
}
\end{figure}

We note that the BFT models $\text{BFT}[SPP]$ and $\text{BFT}[L_{232}]$ are the first two out of an infinite family of BFT models on a cylinder, which we call as $\text{BFT}[L_{n-1.n.n-1}]$.
This family of BFT models has a single integer parameter $n\geq 2$.
The corresponding bipartite graph on the cylinder is shown in \fref{fig_140n}.
We see from the bipartite graph that this family of BFTs has $U(1)_{b_1}, \dots, U(1)_{b_{2n-2}}$ gauge symmetries, and $U(1)_{f_1}$ and $U(1)_{f_2}$ flavor symmetry. 
The corresponding general form of the superpotential is, 
\beal{es140na10}
&&
W = 
X_{12} X_{2.2n} X_{2n.2} X_{21}
+ X_{34} X_{41} X_{14} X_{43}
\nn\\
&&
\hspace{0.3cm}
+ \dots
+ X_{2n-3.2n-2} X_{2n-2.2n-5} X_{2n-5.2n-2} X_{2n-2.2n-3}
+ X_{2n-3.2n-1} X_{2n-1.2n-1} X_{2n-1.2n-3}
\nn\\
&&
\hspace{0.3cm}
- X_{12} X_{21} X_{14} X_{41}
- X_{34} X_{43} X_{35} X_{53}
\nn\\
&&
\hspace{0.3cm}
- \dots
- X_{2n-3.2n-2} X_{2n-2.2n-3} X_{2n-3.2n-1} X_{2n-1.2n-3}
- X_{2.2n} X_{2n.2n} X_{2n.2}
~.~
\eea

\begin{figure} [h]
\centering
\includegraphics[width=.55\textwidth]{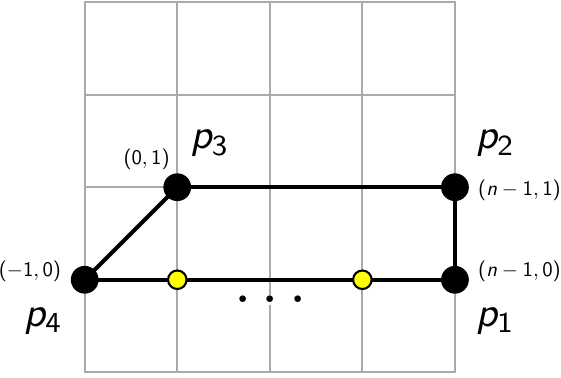}
\caption{
Toric diagram for the $\text{BFT}[L_{n-1.n.n-1}]$ Model.
}
\label{fig140n}
\end{figure}

The unrefined Hilbert series of the mesonic moduli space of the $\text{BFT}[L_{n-1.n.n-1}]$ Model takes the following form,
\beal{es140na70}
&&
g(t_1,\dots, t_4, y_q=1;\mathcal{M}^{mes}_{\text{BFT}[L_{n-1.n.n-1}]})
=
\frac{
1 - t_1^{n} t_2^{n-1} t_3^{n-1} t_4^{n}
}{
(1 - t_1 t_4)
(1 - t_2 t_3)  
(1 - t_1^{n} t_2^{n-1}) 
(1 - t_3^{n-1} t_4^{n})
}
~,~
\nn\\
\eea
 where the fugacities $t_1,\dots, t_4$ correspond to the GLSM fields $p_1,\dots,p_4$.
 These GLSM fields are associated to the extremal vertices of the toric diagram of the mesonic moduli space as illustrated in \fref{fig140n}.
The plethystic logarithm of the Hilbert series given by,
\beal{es140na71}
PL[g(t_1,\dots, t_4, y_q=1;\mathcal{M}^{mes}_{\text{BFT}[L_{n-1.n.n-1}]})]
= 
&&
t_1 t_4
+ t_2 t_3
+ t_1^{n} t_2^{n-1}
+ t_3^{n-1} t_4^{n}
\nn\\
&&
 - t_1^{n} t_2^{n-1} t_3^{n-1} t_4^{n}
~,~
\eea
shows that the mesonic moduli space is a complete intersection.

\begin{table}[ht!]
\centering
\begin{tabular}{|l|l|c|}
\hline
generator & GLSM fields & fugacities \\
\hline\hline
$A_{+}=X_{12}X_{21} =X_{34}X_{43} = \dots$ & $p_2 p_3$ & $t_2 t_3$ \\
$= X_{2n-3.2n-2}X_{2n-2.2n-3} =X_{2n-1.2n-1} = X_{2n.2n}$ & & \\
$A_{-}=X_{2.2n}X_{2n.2}=X_{14}X_{41} = \dots$ & $p_1 p_4$ & $t_1 t_4$\\
$=X_{2n-5.2n-2} X_{2n-2.2n-5}=X_{2n-3.2n-1} X_{2n-1.2n-3}$ &  & \\
$B_{+}=X_{2n-1.2n-3} X_{2n-3.2n-2} \cdots X_{34} X_{41} X_{12} X_{2.2n}$ & $p_1^{n} p_2^{n-1}$ & $t_1^{n} t_2^{n-1}$\\
$B_{-}=X_{2n.2}X_{21}X_{14} X_{43} \cdots X_{2n-2.2n-3} X_{2n-3.2n-1}$ & $p_3^{n-1} p_4^{n}$ & $t_3^{n-1} t_4^{n}$\\
\hline
\end{tabular}
\caption{Generators of the mesonic moduli space of the $\text{BFT}[L_{n-1.n.n-1}]$ Model.}
\label{tab140n_3}
\end{table}

The mesonic moduli space of the $\text{BFT}[L_{n-1.n.n-1}]$ Model has 4 generators as summarized in \tref{tab140n_3}.
They satisfy a single defining relation.
In terms of the generators and the defining relation, the mesonic moduli space can be expressed as follows,
\beal{es140na80}
\mathcal{M}^{mes}_{\text{BFT}[L_{n-1.n.n-1}]} 
&=&
\text{Spec}~  \mathbb{C}[A_+, A_-,B_{+}, B_{-}] /  \langle A_{+}^{n-1} A_{-}^{n} - B_{+} B_{-} \rangle 
\nn\\
&=& L_{n-1.n.n-1}
~,~
\eea
where we note that the mesonic moduli space is a toric Calabi-Yau 3-fold known as $L_{n-1.n.n-1}$ \cite{Benvenuti:2005ja,Butti:2005sw,Franco:2005sm}.
\\

\subsection{$\text{BFT}[\text{CM}_5]$ Model \label{sec_e160}}

\begin{figure} [h]
\centering
\includegraphics{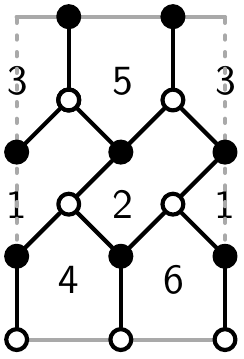}
\caption{
The bipartite graph for the $\text{BFT}[\text{CM}_5]$ Model.
\label{fig_160}
}
\end{figure}

\begin{figure} [h]
\centering
\includegraphics{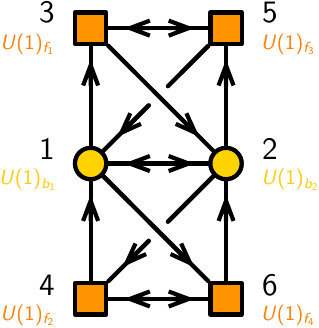}
\caption{
The quiver for the $\text{BFT}[\text{CM}_5]$ Model.
\label{fig_161}
}
\end{figure}

With the following example, we study by calculating the Hilbert series the mesonic moduli space of a BFT on a cylinder whose dimension is higher than 3. 
The following BFT is defined by the bipartite graph on a cylinder as shown in \fref{fig_160}.
The corresponding quiver diagram is given in \fref{fig_161} and the superpotential takes the form, 
\beal{es160a10}
W = 
&
X_{12} X_{24} X_{41} + 
 X_{21} X_{16} X_{62} + 
 X_{35} X_{51} X_{13} + 
 X_{53} X_{32} X_{25} 
 &
 \nn\\
 &
 - 
 X_{12} X_{25} X_{51} - 
 X_{21} X_{13} X_{32} - 
 X_{24} X_{46} X_{62} - X_{16} X_{64} X_{41}
 &
~,~
\eea
where the gauge and flavor symmetry charges on the chiral fields $X_{ij}$ are summarized in \tref{tab_160}.
We note that the global symmetry of the BFT based on the quiver diagram in \fref{fig_161} takes the form $U(1)_{f_1} \times U(1)_{f_2} \times U(1)_{f_3} \times U(1)_{f_4} \times U(1)_R$.
However, given that $B=2$ for this BFT, we have an additional $U(1)_h$ symmetry factor coming from the boundaries of the bipartite graph on the cylinder, as shown in \fref{fig_160}. 
Accordingly, we can rewrite the global symmetry by replacing one of the $U(1)_f$ flavor symmetry factors with $U(1)_h$.
For now, we keep the global symmetry as $U(1)_{f_1} \times U(1)_{f_2} \times U(1)_{f_3} \times U(1)_{f_4} \times U(1)_R$.
The $U(1)_R$ charges on the chiral fields as shown in \tref{tab_160} are obtained using $a$-maximization. 

\begin{table}[ht!]
\centering
\begin{tabular}{|c|c|c|c|c|c|c|c|c|c|c|}
\hline
\; & $U(1)_{b_1}$ & $U(1)_{b_2}$ & $U(1)_{f_1}$ & $U(1)_{f_2}$ & $U(1)_{f_3}$ & $U(1)_{f_4}$ & $U(1)_R$ & fugacity \\
\hline\hline
$X_{12}$ & $-1$ & $+1$ & 0 & 0 & 0 & 0 & 2/3 &  $b_1^{-1} b_2 \bar{t}$ \\
$X_{21}$ & $+1$ & $-1$ & 0 & 0 & 0 & 0 & 2/3 &  $b_1 b_2^{-1} \bar{t}$ \\
\hline
$X_{13}$ & $-1$ & 0 & $+1$ & 0 & 0 & 0 & 2/3 &   $b_1^{-1} f_1 \bar{t}$ \\
$X_{41}$ & $+1$ & 0 & 0 & $-1$ & 0 & 0 & 2/3 &   $b_1 f_2^{-1} \bar{t}$\\
$X_{16}$ & $-1$ & 0 & 0 & 0 & 0 & $+1$ & 2/3 &   $b_1^{-1} f_4 \bar{t}$\\
$X_{51}$ & $+1$ & 0 & 0 & 0 & $-1$ & 0 & 2/3 &   $b_1 f_3^{-1} \bar{t}$\\
$X_{24}$ & 0 & $-1$ & 0 & $+1$ & 0 & 0 & 2/3 &   $b_2^{-1} f_2 \bar{t}$\\
$X_{32}$ & 0 & $+1$ & $-1$ & 0 & 0 & 0 & 2/3 &   $b_2 f_1^{-1} \bar{t}$\\
$X_{25}$ & 0 & $-1$ & 0 & 0 & $+1$ & 0 & 2/3 &   $b_2^{-1} f_3 \bar{t}$\\
$X_{62}$ & 0 & $+1$ & 0 & 0 & 0 & $-1$ & 2/3 &   $b_2 f_4^{-1} \bar{t}$\\
\hline
$X_{35}$ & 0 & 0 & $-1$ & 0 & $+1$ & 0 & 2/3 &   $f_1^{-1} f_3 \bar{t}$\\
$X_{53}$ & 0 & 0 & $+1$ & 0 & $-1$ & 0 & 2/3 &   $f_1 f_3^{-1} \bar{t}$\\
$X_{46}$ & 0 & 0 & 0 & $-1$ & 0 & $+1$ & 2/3 &   $f_2^{-1} f_4 \bar{t}$\\
$X_{64}$ & 0 & 0 & 0 & $+1$ & 0 & $-1$ & 2/3 &   $f_2 f_4^{-1} \bar{t}$\\
\hline
\end{tabular}
\caption{Gauge and global symmetry charges on the chiral fields of the $\text{BFT}[\text{CM}_5]$ Model.}
\label{tab_160}
\end{table}

\paragraph{Forward Algorithm.}
The superpotential of the  $\text{BFT}[\text{CM}_5]$ Model gives us the following $F$-terms, 
\beal{es160a11}
&
\partial_{X_{12}} W =
X_{24} X_{41} -X_{25} X_{51} ~,~   
\partial_{X_{13}} W =
 X_{35} X_{51} -X_{32} X_{21} ~,~   
 &
 \nn\\
 &
  \partial_{X_{16}} W =
 X_{62} X_{21} -X_{64} X_{41} ~,~   
 \partial_{X_{21}}W =
 X_{16} X_{62} -X_{13} X_{32}~,~  
 &
 \nn\\
 &
 \partial_{X_{24}} W =
 X_{41} X_{12} -X_{46} X_{62} ~,~   
 \partial_{X_{25}}W =
 X_{53} X_{32} -X_{51} X_{12}~,~  
 &
 \nn\\
 &
 \partial_{X_{32}}W =
 X_{25} X_{53} -X_{21} X_{13}~,~  
\partial_{X_{41}} W = 
X_{12} X_{24} -X_{16} X_{64} ~,~   
&
\nn\\
&
\partial_{X_{51}} W =
X_{13} X_{35} -X_{12} X_{25} ~,~  
\partial_{X_{62}} W =
X_{21} X_{16} -X_{24} X_{46} ~.~
&
\eea
which allow us to identify the following independent chiral fields,
\beal{es160a12}
&
v_1 = X_{12} ~,~
v_2 = X_{16} ~,~
v_3 = X_{21} ~,~
v_4 = X_{24} ~,~
&
\nn\\
&
v_5 = X_{25} ~,~
v_6 = X_{32} ~,~
v_7 = X_{41} ~.~
&
\eea
These can be used to express the remaining chiral fields as follows,
\beal{es160a13}
&
X_{13} = 
\frac{v_1 v_4 v_7}{v_3 v_6} ~,~
X_{35} = 
\frac{v_3 v_5 v_6}{v_4 v_7} ~,~
X_{46} =
\frac{v_2 v_3}{v_4} ~,~
X_{51} = 
\frac{v_4 v_7}{v_5} ~,~
X_{53} =
\frac{v_1 v_4 v_7}{v_5 v_6} ~,~
&
\nn\\
&
X_{62} =
\frac{v_1 v_4 v_7}{v_2 v_3} ~,~
X_{64} =
\frac{v_1 v_4}{v_2} ~.~
&
\eea
The above relations can be summarized in the following $K$-matrix
\beal{es160a60}
K=\left(
\begin{array}{c|ccccccc}
\; & v_1 & v_2 & v_3 & v_4 & v_5 & v_6 & v_7\\
\hline
X_{12} & 1 & 0 & 0 & 0 & 0 & 0 & 0 \\
X_{21} & 0 & 0 & 1 & 0 & 0 & 0 & 0 \\
X_{13} & 1 & 0 & -1 & 1 & 0 & -1 & 1 \\
X_{41} & 0 & 0 & 0 & 0 & 0 & 0 & 1 \\
X_{16} & 0 & 1 & 0 & 0 & 0 & 0 & 0 \\
X_{51} & 0 & 0 & 0 & 1 & -1 & 0 & 1 \\
X_{24} & 0 & 0 & 0 & 1 & 0 & 0 & 0 \\
X_{32} & 0 & 0 & 0 & 0 & 0 & 1 & 0 \\
X_{25} & 0 & 0 & 0 & 0 & 1 & 0 & 0 \\
X_{62} & 1 & -1 & -1 & 1 & 0 & 0 & 1 \\
X_{35} & 0 & 0 & 1 & -1 & 1 & 1 & -1 \\
X_{53} & 1 & 0 & 0 & 1 & -1 & -1 & 1 \\
X_{46} & 0 & 1 & 1 & -1 & 0 & 0 & 0 \\
X_{64} & 1 & -1 & 0 & 1 & 0 & 0 & 0 \\
\end{array}
\right)
~.~
 \eea
 
The $K$-matrix under the forward algorithm allows us to construct the perfect matchings which correspond to GLSM fields in the BFT. 
They are summarized in the following $P$-matrix,
 \beal{es160a61}
P=\left(
\begin{array}{c|ccccccc|cc|cc|cc}
\; & p_1 & p_2 & p_3 & p_4 & p_5 & p_6 & p_7 & q_1 & q_2 & s_1 & s_2 & u_1 & u_2 \\
\hline
X_{12} & 1 & 1 & 0 & 0 & 0 & 0 & 0 & 1 & 1 & 0 & 0 & 0 & 1 \\
X_{21} & 0 & 0 & 0 & 1 & 1 & 1 & 0 & 1 & 0 & 0 & 0 & 1 & 0 \\
X_{13} & 0 & 1 & 1 & 0 & 0 & 0 & 0 & 0 & 1 & 0 & 1 & 0 & 0 \\
X_{41} & 0 & 0 & 1 & 1 & 1 & 0 & 0 & 0 & 0 & 1 & 0 & 0 & 0 \\
X_{16} & 1 & 0 & 0 & 0 & 0 & 0 & 1 & 0 & 1 & 0 & 1 & 0 & 0 \\
X_{51} & 0 & 0 & 0 & 1 & 0 & 1 & 1 & 0 & 0 & 1 & 0 & 0 & 0 \\
X_{24} & 0 & 0 & 0 & 0 & 0 & 1 & 1 & 0 & 0 & 0 & 1 & 1 & 0 \\
X_{32} & 1 & 0 & 0 & 0 & 0 & 0 & 1 & 0 & 0 & 1 & 0 & 0 & 1 \\
X_{25} & 0 & 0 & 1 & 0 & 1 & 0 & 0 & 0 & 0 & 0 & 1 & 1 & 0 \\
X_{62} & 0 & 1 & 1 & 0 & 0 & 0 & 0 & 0 & 0 & 1 & 0 & 0 & 1 \\
X_{35} & 1 & 0 & 0 & 0 & 1 & 0 & 0 & 1 & 0 & 0 & 0 & 1 & 1 \\
X_{53} & 0 & 1 & 0 & 1 & 0 & 1 & 0 & 1 & 1 & 0 & 0 & 0 & 0 \\
X_{46} & 1 & 0 & 0 & 1 & 1 & 0 & 0 & 1 & 1 & 0 & 0 & 0 & 0 \\
X_{64} & 0 & 1 & 0 & 0 & 0 & 1 & 0 & 1 & 0 & 0 & 0 & 1 & 1 \\
 \end{array}
\right)~.~
\eea
The chiral fields of the BFT can be expressed as products of the GLSM fields as shown below,
\beal{es160a61b}
&
X_{12}   =   p_1 p_2 q_1 q_2 u_2~,~     
X_{21}   =   p_3 p_4 p_5 s_1~,~     
X_{13}   =   p_4 p_5 p_6 q_1 u_1~,~     
X_{41}   =   p_3 p_5 s_2 u_1~,~     
&
\nn\\
&
X_{16}   =   p_2 p_3 q_2 s_2~,~     
X_{51}   =   p_1 p_5 q_1 u_1 u_2~,~     
X_{24}   =   p_1 p_7 q_2 s_2~,~   
X_{32}   =   p_6 p_7 s_2 u_1~,~     
&
\nn\\
&
X_{25}   =   p_4 p_6 p_7 s_1~,~     
X_{62}   =   p_1 p_4 p_5 q_1 q_2~,~     
X_{35}   =   p_1 p_7 s_1 u_2~,~     
X_{53}   =   p_2 p_4 p_6 q_1 q_2~,~     
&
\nn\\
&
X_{46}   =   p_2 p_3 s_1 u_2~,~     
X_{64}   =   p_2 p_6 q_1 u_1 u_2
~.~
&
\eea
The $F$-terms of the BFT can be expressed as $U(1)$ charges on the GLSM fields as summarized in the charge matrix below,
\beal{es160a62}
Q_{F} =\left(
\begin{array}{ccccccc|cc|cc|cc}
 p_1 & p_2 & p_3 & p_4 & p_5 & p_6 & p_7 & q_1 & q_2 & s_1 & s_2 & u_1 & u_2 \\
\hline
 1 & 0 & 0 & 0 & 0 & 1 & -1 & 0 & -1 & 0 & 1 & -1 & 0 \\
 0 & 1 & 0 & 0 & 0 & -1 & 1 & 0 & 0 & 0 & -1 & 1 & -1 \\
 0 & 0 & 1 & 0 & 0 & 0 & 1 & 0 & 0 & -1 & -1 & 0 & 0 \\
 0 & 0 & 0 & 1 & 0 & 0 & 0 & 0 & -1 & -1 & 1 & -1 & 1 \\
 0 & 0 & 0 & 0 & 1 & 1 & 0 & 0 & -1 & -1 & 1 & -2 & 1 \\
 0 & 0 & 0 & 0 & 0 & 0 & 0 & 1 & -1 & 0 & 1 & -1 & 0 \\
\end{array}
\right)~.~
\eea
We also have a charge matrix for the $D$-terms, which takes the following form,
\beal{es160a63}
Q_{D} =\left(
\begin{array}{ccccccc|cc|cc|cc}
 p_1 & p_2 & p_3 & p_4 & p_5 & p_6 & p_7 & q_1 & q_2 & s_1 & s_2 & u_1 & u_2 \\
\hline
 0 & 0 & 1 & 0 & 0 & 0 & 1 & 0 & 0 & -2 & 0 & 0 & 0 \\
 0 & 0 & 0 & 1 & -1 & -1 & 0 & 0 & 0 & 0 & 0 & 0 & 1 \\
\end{array}
\right)~.~
\eea

The toric diagram of the $\text{BFT}[\text{CM}_5]$ Model can be obtained from the $Q_F$ and $Q_D$-matrices.
In summary, the toric diagram is given by,
\beal{es160a65}
G_{t} =\left(
\begin{array}{ccccccc|cc|cc|cc}
 p_1 & p_2 & p_3 & p_4 & p_5 & p_6 & p_7 & q_1 & q_2 & s_1 & s_2 & u_1 & u_2 \\
\hline
 1 & 0 & 0 & 1 & 1 & 0 & 0 & 1 & 1 & 0 & 0 & 0 & 0 \\
 0 & 1 & 0 & 1 & 0 & 1 & 0 & 1 & 1 & 0 & 0 & 0 & 0 \\
 0 & 0 & 1 & 0 & 1 & -1 & -1 & 0 & 0 & 0 & 0 & 0 & 0 \\
 0 & 0 & 0 & 1 & 0 & 0 & 0 & 0 & 1 & 0 & 0 & -1 & -1 \\
 1 & 1 & 1 & 1 & 1 & 1 & 1 & 1 & 1 & 1 & 1 & 1 & 1 \\
\end{array}
\right)~,~
\eea
where we note that the mesonic moduli space of the $\text{BFT}[\text{CM}_5]$ Model is a toric Calabi-Yau 5-fold. 
\\

\paragraph{Hilbert Series.}
Using the $Q_F$ and $Q_D$-matrices summarizing charges on GLSM fields, we can express the mesonic moduli space as the following symplectic quotient,
\beal{es160a66}
\mathcal{M}^{mes}_{\text{BFT}[\text{CM}_5]}
= \text{Spec}
\left(\mathbb{C}[p_1,\dots,p_7, q_1,q_2,s_1,s_2,u_1,u_2] // Q_F \right) // Q_D
~.~
\eea
Using the symplectic quotient description of the mesonic moduli space of the $\text{BFT}[\text{CM}_5]$ Model, we can calculate the refined Hilbert series as shown below,
\beal{es160a70}
&&
g(t_1,\dots, t_7, y_{q_1},y_{q_2}, y_s, y_u;\mathcal{M}^{mes}_{\text{BFT}[\text{CM}_5]})
=
\frac{P(t_1,\dots, t_7, y_{q_1},y_{q_2}, y_s, y_u)}{
(1 -  t_1 t_4 t_5 y_{q_1} y_{q_2}) (1 -  t_2 t_4 t_6 y_{q_1} y_{q_2})
}
\nn\\
&&
\hspace{1cm}
\times
\frac{1}{
 (1 -     t_2 t_3^2 t_4 t_5 y_{q_2} y_{s}) 
 (1 -  t_1 t_3 t_4 t_5 t_7 y_{q_2} y_{s}) 
 (1 -     t_2 t_3 t_4 t_6 t_7 y_{q_2} y_{s}) 
 (1 -  t_1 t_4 t_6 t_7^2 y_{q_2} y_{s})
}
\nn\\
&&
\hspace{1cm}
\times
\frac{1}{
  (1 -     t_1 t_5 y_{q_1} y_{u}) 
  (1 -  t_2 t_6 y_{q_1} y_{u}) 
  (1 -     t_2 t_3^2 t_5 y_{s} y_{u}) 
  (1 -  t_1 t_3 t_5 t_7 y_{s} y_{u}) 
}
\nn\\
&&
\hspace{1cm}
\times
\frac{1}{
(1 -     t_2 t_3 t_6 t_7 y_{s} y_{u}) (1 -  t_1 t_6 t_7^2 y_{s} y_{u})
}
~,~
\eea
where the fugacities $t_1, \dots, t_7$ correspond to GLSM fields $p_1, \dots, p_7$, the fugacities $y_{q_1}$ and $y_{q_2}$ correspond to GLSM fields $q_1$ and $q_2$, and the fugacities $y_s$ and $y_u$ correspond to the products of GLSM fields $s=s_1 s_2$ and $u=u_1 u_2$, respectively. 
Here, the numerator $P(t_1,\dots, t_7, y_{q_1},y_{q_2}, y_s, y_u)$ is a palindromic polynomial which we choose not to present here due to its length. 
The plethystic logarithm of the Hilbert series takes the following form,
\beal{es160a71}
&&
PL[g(t_1,\dots, t_7, y_{q_1},y_{q_2}, y_s, y_u;\mathcal{M}^{mes}_{\text{BFT}[\text{CM}_5]})]
= 
(t_1 t_4 t_5 y_{q_1} y_{q_2} + t_2 t_4 t_6 y_{q_1} y_{q_2} + t_1 t_5 y_{q_1} y_{u}
\nn\\
&& 
\hspace{1cm}
 +     t_2 t_6 y_{q_1} y_{u}) 
 + (t_2 t_3^2 t_4 t_5 y_{q_2} y_{s} +     t_1 t_3 t_4 t_5 t_7 y_{q_2} y_{s} + t_2 t_3 t_4 t_6 t_7 y_{q_2} y_{s} +     t_1 t_4 t_6 t_7^2 y_{q_2} y_{s} 
 \nn\\
 &&
 \hspace{1cm}
 + t_2 t_3^2 t_5 y_{s} y_{u} +     t_1 t_3 t_5 t_7 y_{s} y_{u} + t_2 t_3 t_6 t_7 y_{s} y_{u} +     t_1 t_6 t_7^2 y_{s} y_{u}) 
  -  t_1 t_2 t_4 t_5 t_6 y_{q_1}^2 y_{q_2} y_{u}  
  \nn\\
 &&
 \hspace{1cm}
  - (t_1 t_2 t_3 t_4^2 t_5 t_6 t_7 y_{q_1} y_{q_2}^2 y_{s} + t_1 t_2 t_3^2 t_4 t_5^2 y_{q_1} y_{q_2} y_{s} y_{u} +     t_2^2 t_3^2 t_4 t_5 t_6 y_{q_1} y_{q_2} y_{s} y_{u} 
    \nn\\
  &&
  \hspace{1cm}
  +     t_1^2 t_3 t_4 t_5^2 t_7 y_{q_1} y_{q_2} y_{s} y_{u} +     3 t_1 t_2 t_3 t_4 t_5 t_6 t_7 y_{q_1} y_{q_2} y_{s} y_{u} +    t_2^2 t_3 t_4 t_6^2 t_7 y_{q_1} y_{q_2} y_{s} y_{u} 
    \nn\\
  &&
  \hspace{1cm}
  +    t_1^2 t_4 t_5 t_6 t_7^2 y_{q_1} y_{q_2} y_{s} y_{u} +     t_1 t_2 t_4 t_6^2 t_7^2 y_{q_1} y_{q_2} y_{s} y_{u} +     t_1 t_2 t_3 t_5 t_6 t_7 y_{q_1} y_{s} y_{u}^2) 
    \nn\\
  &&
  \hspace{1cm}
  - (t_1 t_2 t_3^2 t_4^2 t_5 t_6 t_7^2 y_{q_2}^2 y_{s}^2 + t_1 t_2 t_3^3 t_4 t_5^2 t_7 y_{q_2} y_{s}^2 y_{u} +     t_2^2 t_3^3 t_4 t_5 t_6 t_7 y_{q_2} y_{s}^2 y_{u} 
  \nn\\
  &&
  \hspace{1cm}
  +     3 t_1 t_2 t_3^2 t_4 t_5 t_6 t_7^2 y_{q_2} y_{s}^2 y_{u} +     t_1^2 t_3 t_4 t_5 t_6 t_7^3 y_{q_2} y_{s}^2 y_{u} +     t_1 t_2 t_3 t_4 t_6^2 t_7^3 y_{q_2} y_{s}^2 y_{u} 
    \nn\\
  &&
  \hspace{1cm}
  +     t_1 t_2 t_3^2 t_5 t_6 t_7^2 y_{s}^2 y_{u}^2) 
  + \dots
~,~
\eea
where we can see that the infinite expansion of the plethystic logarithm implies that the mesonic moduli space is a non-complete intersection.

\begin{table}[ht!]
\centering
\begin{tabular}{|l|c|c|c|c|c|c|c|}
\hline
\; & $U(1)_{\bar{f}_1}$ & $U(1)_{\bar{f}_2}$ & $U(1)_{\bar{f}_3}$ & $U(1)_{\bar{f}_4}$ & $U(1)_{R}$ & fugacity\\
\hline\hline
$p_1$ & $+1$ & $0$ & $0$ & $0$ & $1/3$ & $t_1 = \bar{f}_1 t^{16}$\\
$p_2$ & $-1$ & $0$ & $0$ & $0$ & $1/6$ & $t_2 = \bar{f}_1^{-1}t^{8}$\\
$p_3$ & $0$ & $+2$ & $0$ & $0$ & $5/16$ & $t_3 = \bar{f}_2^2 t^{15}$\\
$p_4$ & $0$ & $-1$ & $+1$ & $0$ & $5/48$ & $t_4 = \bar{f}_2^{-1} \bar{f}_3 t^{5}$\\
$p_5$ & $0$ & $-1$ & $-1$ & $0$ & $5/48$ & $t_5 = \bar{f}_2^{-1} \bar{f}_3^{-1} t^{5}$\\
$p_6$ & $0$ & $0$ & $0$ & $0$ & $13/48$ & $t_6 =  t^{13}$\\
$p_7$ & $0$ & $0$ & $0$ & $0$ & $7/48$ & $t_7 =  t^{7}$\\
\hline
$q_1$ & $0$ & $0$ & $0$ & $+1$ & $1/12$ & $y_{q_1} = \bar{f}_4 t^{4}$\\
$q_2$ & $0$ & $0$ & $0$ & $-1$ & $1/24$ & $y_{q_2} = \bar{f}_4^{-1} t^{2}$\\
\hline
$s_1$ & $0$ & $0$ & $0$ & $0$ & $7/48$ & $y_{s_1} = t^{7}$\\
$s_2$ & $0$ & $0$ & $0$ & $0$ & $7/48$ & $y_{s_2} = t^{7}$\\
\hline
$u_1$ & $0$ & $0$ & $0$ & $0$ & $5/48$ & $y_{u_1} = t^{5}$\\
$u_2$ & $0$ & $0$ & $0$ & $0$ & $1/24$ & $y_{u_2} = t^{2}$\\
\hline
\end{tabular}
\caption{Global symmetry charges according to $U(1)_{\bar{f}_1} \times U(1)_{\bar{f}_2} \times U(1)_{\bar{f}_3} \times U(1)_{\bar{f}_4}\times U(1)_R$ on perfect matchings of the $\text{BFT}[\text{CM}_5]$ Model.}
\label{tab160_2}
\end{table}

Based on the global symmetry charge assignment on the GLSM fields, as summarized in \tref{tab160_2}, 
we can use the following fugacity map, 
\beal{es160a75}
&
\bar{f}_1 = 
\frac{t_1^{1/3}}{t_2^{2/3}}
~,~
\bar{f}_2 = 
\frac{t_3^{1/5}}{t_4^{3/10} t_5^{3/10}}
~,~
\bar{f}_3 = 
\frac{t_4^{1/2}}{t_5^{1/2}}
~,~
\bar{f}_4 = 
\frac{y_{q_1}^{1/3}}{y_{q_2}^{2/3}}
~,~
&
\nn\\
&
t=
t_1^{1/96} t_2^{1/96} t_3^{1/96} t_4^{1/96} t_5^{1/96} t_6^{1/96} y_{q_1}^{1/96} y_{q_2}^{1/96} y_{s_1}^{1/96} y_{s_2}^{1/96} y_{u_1}^{1/96} y_{u_2}^{1/96}
~,~
&
\eea
in order to express the Hilbert series of the mesonic moduli space in terms of global symmetry fugacities as shown below, 
\beal{es160a76}
&&
g(t, \bar{f}_i ;\mathcal{M}^{mes}_{\text{BFT}[\text{CM}_5]}))=
\frac{
P(t,\bar{f}_i)
}
{
(1 -  \bar{f}_1 \bar{f}_2^{-2} t^{32})
(1 -  \bar{f}_1^{-1} \bar{f}_2^{-1} \bar{f}_3 t^{32})
(1 -  \bar{f}_1^{-1} \bar{f}_4 t^{32})
}
\nn\\
&&
\hspace{1cm}
\times
\frac{1}{
(1 -  \bar{f}_1 \bar{f}_2^{-1} \bar{f}_3^{-1} \bar{f}_4 t^{32})
(1 -  \bar{f}_1 t^{64}) 
(1 -  \bar{f}_1^{-1} \bar{f}_2^2 t^{64})
(1 -  \bar{f}_1 \bar{f}_2 \bar{f}_3^{-1} t^{64})
}
\nn\\
&&
\hspace{1cm}
\times
\frac{1}{
(1 -  \bar{f}_1^{-1} \bar{f}_2^3 \bar{f}_3^{-1} t^{64})
(1 -  \bar{f}_1 \bar{f}_4^{-1} t^{64})
(1 -  \bar{f}_1^{-1} \bar{f}_2^2 \bar{f}_4^{-1} t^{64})
(1 -  \bar{f}_1 \bar{f}_2^{-1} \bar{f}_3 \bar{f}_4^{-1} t^{64})
}
\nn\\
&&
\hspace{1cm}
\times
\frac{1}{
(1 -  \bar{f}_1^{-1} \bar{f}_2 \bar{f}_3 \bar{f}_4^{-1} t^{64})
}
 ~,~
\eea
where the corresponding plethystic logarithm is given by, 
\beal{es160a76}
&&
PL[
g(t, \bar{f}_i ;\mathcal{M}^{mes}_{\text{BFT}[\text{CM}_5]}))
]
= 
\Big(\frac{\bar{f}_{1}}{\bar{f}_{2}^2}+\frac{\bar{f}_{1} \bar{f}_{4}}{\bar{f}_{2} \bar{f}_{3}}+\frac{\bar{f}_{3}}{\bar{f}_{1} \bar{f}_{2}}+\frac{\bar{f}_{4}}{\bar{f}_{1}}\Big)t^{32} 
+\Big(\frac{\bar{f}_{2}^3}{\bar{f}_{1} \bar{f}_{3}}+\frac{\bar{f}_{2}^2}{\bar{f}_{1} \bar{f}_{4}}+\frac{\bar{f}_{2}^2}{\bar{f}_{1}}
\nn\\
&&
\hspace{1cm}
+\frac{\bar{f}_{2} \bar{f}_{3}}{\bar{f}_{1} \bar{f}_{4}}+\frac{\bar{f}_{1} \bar{f}_{3}}{\bar{f}_{2} \bar{f}_{4}}+\frac{\bar{f}_{1} \bar{f}_{2}}{\bar{f}_{3}}+\frac{\bar{f}_{1}}{\bar{f}_{4}}+\bar{f}_{1}\Big)t^{64} 
-\frac{\bar{f}_{4} }{\bar{f}_{2}^2} t^{64}
-\Big(\frac{\bar{f}_{1}^2}{\bar{f}_{2}^2}+\frac{\bar{f}_{2}^2}{\bar{f}_{1}^2}+\frac{\bar{f}_{1} ^2}{\bar{f}_{2} \bar{f}_{3}}
\nn\\
&&
\hspace{1cm}
+\frac{\bar{f}_{2} \bar{f}_{3}}{\bar{f}_{1}^2}+\frac{\bar{f}_{2} \bar{f}_{4}}{\bar{f}_{3}}+\frac{\bar{f}_{3}}{\bar{f}_{2} \bar{f}_{4}}+\frac{\bar{f}_{3}}{\bar{f}_{2}}+\frac{\bar{f}_{2}}{\bar{f}_{3}}+3\Big)
t^{96} 
- \Big(\frac{\bar{f}_{2}^4}{\bar{f}_{1}^2 \bar{f}_{4}}+\frac{\bar{f}_{1}^2}{\bar{f}_{4}}+\frac{\bar{f}_{2}^3}{\bar{f}_{3} \bar{f}_{4}}+\frac{\bar{f}_{2}^3}{\bar{f}_{3}}
\nn\\
&&
\hspace{1cm}
+\frac{3 \bar{f}_{2}^2}{\bar{f}_{4}}+\frac{\bar{f}_{2} \bar{f}_{3}}{\bar{f}_{4}^2}+\frac{\bar{f}_{2} \bar{f}_{3}}{\bar{f}_{4}}\Big) t^{128}
+ \dots
~.~
\eea
Here, $P(t,\bar{f}_i)$ is the palindromic numerator of the refined Hilbert series.
When we unrefine the Hilbert series by setting the flavor symmetry fugacities to $\bar{f}_i = 1$, we obtain,
\beal{es160a77}
&&
g(t, \bar{f}_i=1 ;\mathcal{M}^{mes}_{\text{BFT}[\text{CM}_5]}))=
\frac{
1 + 3 t^{32} + 9 t^{64} + 8 t^{96} + 9 t^{128} + 3 t^{160} + t^{192}
}{
(1-t^{32})(1-t^{64})^4
}
~,~
\eea
where we clearly see how the numerator of the Hilbert series is palindromic. 
This implies that the mesonic moduli space of the $\text{BFT}[\text{CM}_5]$ Model is Calabi-Yau \cite{Stanley}.

\begin{table}[ht!]
\centering
\begin{tabular}{|l|l|c|c|c|c|c|c|}
\hline
generator & GLSM fields & $U(1)_{\bar{f}_1}$ & $U(1)_{\bar{f}_2}$ & $U(1)_{\bar{f}_3}$ & $U(1)_{\bar{f}_4}$ & $U(1)_R$ & fugacities \\
\hline\hline
$A_1 = X_{46}$ & $p_1 p_4 p_5 q_1 q_2$ & $+1$ &  $-2$ &  $0$ &  $0$ &  $2/3$ & $\bar{f}_1 \bar{f}_2^{-2} t^{32}$
\\
$A_2 = X_{64}$ & $p_2 p_6 q_1 u_1 u_2$ & $-1$ &  $0$ &  $0$ &  $+1$ &  $2/3$ & $\bar{f}_1^{-1} \bar{f}_4 t^{32}$
\\
$A_3 = X_{35}$ & $p_1 p_5 q_1 u_1 u_2$ & $+1$ &  $-1$ &  $-1$ &  $+1$ &  $2/3$ & $\bar{f}_1 \bar{f}_2^{-1} \bar{f}_3^{-1} \bar{f}_4 t^{32}$
\\
$A_4 = X_{53}$ & $p_2 p_4 p_6 q_1 q_2$ & $-1$ &  $-1$ &  $+1$ &  $0$ &  $2/3$ & $\bar{f}_1^{-1} \bar{f}_2^{-1} \bar{f}_3 t^{32}$
\\
$B_1 = X_{41} X_{13}$ & $p_2 p_3^2 p_4 p_5 q_2 s_1 s_2$ & $-1$ &  $+2$ &  $0$ &  $-1$ &  $4/3$ & $\bar{f}_1^{-1} \bar{f}_2^{2} \bar{f}_4^{-1} t^{64}$
\\
$B_2 = X_{41} X_{16}$ & $p_1 p_3 p_4 p_5 p_7 q_2 s_1 s_2$ & $+1$ &  $0$ &  $0$ &  $-1$ &  $4/3$ & $\bar{f}_1 \bar{f}_4^{-1} t^{64}$
\\
$B_3 = X_{51} X_{13}$ & $p_2 p_3 p_4 p_6 p_7 q_2 s_1 s_2$ & $-1$ &  $+1$ &  $+1$ &  $-1$ &  $4/3$ & $\bar{f}_1^{-1} \bar{f}_2 \bar{f}_3 \bar{f}_4^{-1} t^{64}$
\\
$B_4 = X_{51} X_{16}$ & $p_1 p_4 p_6 p_7^2 q_2 s_1 s_2$ & $+1$ &  $-1$ &  $+1$ &  $-1$ &  $4/3$ & $\bar{f}_1 \bar{f}_2^{-1} \bar{f}_3 \bar{f}_4^{-1} t^{64}$
\\
$B_5 = X_{32} X_{24}$ & $p_1 p_6 p_7^2 s_1 s_2 u_1 u_2$ & $+1$ &  $0$ &  $0$ &  $0$ &  $4/3$ & $\bar{f}_1 t^{64}$
\\
$B_6 = X_{32} X_{25}$ & $p_1 p_3 p_5 p_7 s_1 s_2 u_1 u_2$ & $+1$ &  $+1$ &  $-1$ &  $0$ &  $4/3$ & $\bar{f}_1 \bar{f}_2 \bar{f}_3^{-1} t^{64}$
\\
$B_7 = X_{62} X_{24}$ & $p_2 p_3 p_6 p_7 s_1 s_2 u_1 u_2$ & $-1$ &  $+2$ &  $0$ &  $0$ &  $4/3$ & $\bar{f}_1^{-1} \bar{f}_2^{2} t^{64}$
\\
$B_8 = X_{62} X_{25}$ & $p_2 p_3^2 p_5 s_1 s_2 u_1 u_2$ & $-1$ &  $+3$ &  $-1$ &  $0$ &  $4/3$ & $\bar{f}_1^{-1} \bar{f}_2^{3} \bar{f}_3^{-1} t^{64}$
\\
\hline
\end{tabular}
\caption{Generators of the mesonic moduli space of the $\text{BFT}[\text{CM}_5]$ Model.}
\label{tab160_3}
\end{table}

The plethystic logarithm of the Hilbert series for the mesonic moduli space of the $\text{BFT}[\text{CM}_5]$ Model indicates that there are in total 12 generators that satisfy defining quadratic relations, which are summarized below,
\beal{es160a80}
&&
\mathcal{M}^{mes}_{[\text{CM}_5]}
=
\text{Spec}~  \mathbb{C}[A_i, B_i] /  \langle 
\nn\\
&&
\hspace{1cm}
A_4 A_3 - A_1 A_2 ,~
\nn\\
&&
\hspace{1cm}
A_2 B_6-B_7 A_3 ,~
A_4 B_6-B_7 A_1 ,~
A_4 B_2-B_3 A_1 ,~
A_4 B_5-B_4 A_2 ,~
A_4 B_7-B_3 A_2 ,~
\nn\\
&&
\hspace{1cm}
A_1 B_7-B_2 A_2 ,~
A_4 B_8-B_1 A_2 ,~
A_1 B_5-B_4 A_3 ,~
A_1 B_7-B_3 A_3 ,~
A_1 B_6-B_2 A_3 ,~
\nn\\
&&
\hspace{1cm}
A_1 B_8-B_1 A_3 ,~
\nn\\
&&
\hspace{1cm}
B_6 B_7-B_8 B_5 ,~
B_4 B_7-B_3 B_5 ,~
B_2 B_7-B_1 B_5 ,~
B_4 B_6-B_2 B_5 ,~
B_3 B_6-B_1 B_5 ,~
\nn\\
&&
\hspace{1cm}
B_4 B_8-B_1 B_5 ,~
B_3 B_8-B_1 B_7 ,~
B_2 B_8-B_1 B_6 ,~
B_2 B_3-B_1 B_4 
\nn\\
&&
\hspace{1cm}
 \rangle 
~.~
\eea
Combined with the information about the toric diagram of the mesonic moduli space, as computed in \eref{es160a65}, we identify the mesonic moduli space of the $\text{BFT}[\text{CM}_5]$ Model as a toric Calabi-Yau 5-folds with 12 generators that form a non-complete intersection. 
We give the name $\text{CM}_5$ for this toric Calabi-Yau 5-fold. 
\\

By expressing the mesonic moduli space in terms of the binomial ideal formed by the $F$-terms of the BFT, as shown below, 
\beal{es160a85}
\mathcal{M}^{mes}_{[\text{CM}_5]}
= \text{Spec}~
(\mathbb{C}[X_{ij}] / \mathcal{I}^{\text{Irr}}_{\partial W} ) // (U(1)_{b_1} \times U(1)_{b_2})
~,~
\eea
where the coherent component of the ideal takes the form, 
\beal{es160a86}
&&
\mathcal{I}^{\text{Irr}}_{\partial W}=
\langle
~
X_{21}  X_{62} -X_{41}  X_{64} ~,~
X_{35}  X_{53} -X_{46}  X_{64} ~,~
X_{12}  X_{51} -X_{32}  X_{53} ~,~
\nn
\\
&&
\hspace{1.6cm}
X_{24}  X_{41} -X_{25}  X_{51} ~,~
X_{12}  X_{41} -X_{46}  X_{62} ~,~
X_{21}  X_{32} -X_{35}  X_{51} ~,~
\nn
\\
&&
\hspace{1.6cm}
X_{13}  X_{32} -X_{16}  X_{62} ~,~
X_{12}  X_{25} -X_{13}  X_{35} ~,~
X_{12}  X_{24} -X_{16}  X_{64} ~,~
\nn
\\
&&
\hspace{1.6cm}
X_{16}  X_{21} -X_{24}  X_{46} ~,~
X_{13}  X_{21} -X_{25}  X_{53} ~,~
X_{12}  X_{21} -X_{46}  X_{64} ~,~
\nn
\\
&&
\hspace{1.6cm}
X_{24}  X_{53}  X_{62} -X_{13}  X_{51}  X_{64} ~,~
X_{24}  X_{35}  X_{62} -X_{25}  X_{32}  X_{64} ~,~
\nn
\\
&&
\hspace{1.6cm}
X_{13}  X_{46}  X_{51} -X_{16}  X_{41}  X_{53} ~,~
X_{16}  X_{35}  X_{41} -X_{25}  X_{32}  X_{46} 
~
\rangle 
~,~
\eea
we can calculate the Hilbert series of the mesonic moduli space in terms of fugacities associated to the global symmetry charge on chiral fields as summarized in \tref{tab_160}.
The resulting Hilbert series takes the following form, 
\beal{es160a87}
&&
g(\bar{t}, f_i; \mathcal{M}^{mes}_{\text{BFT}[\text{CM}_5]})
=
\frac{P(\bar{t},f_i)}{
(1 - f_1 f_3^{-1} \bar{t}) 
(1 - f_2 f_4^{-1} \bar{t}) 
(1 - f_1^{-1} f_3 \bar{t}) 
(1 - f_2^{-1} f_4 \bar{t}) 
}
\nn\\
&&
\hspace{1cm}
\times
\frac{1}{
(1 - f_1 f_2^{-1} \bar{t}^2) 
(1 - f_1 f_3^{-1} \bar{t}^2) 
(1 - f_1^{-1} f_2 \bar{t}^2) 
(1 - f_2 f_4^{-1} \bar{t}^2) 
}
\nn\\
&&
\hspace{1cm}
\times
\frac{1}{
(1 - f_1^{-1} f_3 \bar{t}^2) 
(1 - f_3 f_4^{-1} \bar{t}^2) 
(1 - f_2^{-1} f_4 \bar{t}^2) 
(1 - f_3^{-1} f_4 \bar{t}^2)
}
~,~
\eea
where the $P(\bar{t},f_i)$ is the refined palindromic numerator of the Hilbert series.
The corresponding plethystic logarithm takes the form, 
\beal{es160a88}
&&
PL[
g(\bar{t}, f_i; \mathcal{M}^{mes}_{\text{BFT}[\text{CM}_5]})
]
=
\Big(\frac{f_{1}}{f_{3}}+\frac{f_{3}}{f_{1}}+\frac{f_{4}}{f_{2}}+\frac{f_{2}}{f_{4}}\Big)\bar{t}
+
\Big(\frac{f_{1}}{f_{2}}+\frac{f_{2}}{f_{1}}+\frac{f_{1}}{f_{3}}+\frac{f_{3}}{f_{1}} +\frac{f_{4}}{f_{2}}
\nn\\
&&
\hspace{1cm}
+\frac{f_{2}}{f_{4}}+\frac{f_{4}}{f_{3}}+\frac{f_{3}}{f_{4}}\Big) \bar{t}^2
- \bar{t}^2
- \Big(\frac{f_{1} f_{4}}{f_{2} f_{3}}+\frac{f_{1}f_{2}}{f_{3} f_{4}}+\frac{f_{3} f_{4}}{f_{1}f_{2}}+\frac{f_{2} f_{3}}{f_{1}f_{4}}
+\frac{f_{1}}{f_{4}}+\frac{f_{4}}{f_{1}}
\nn\\
&&
\hspace{1cm}
+\frac{f_{3}}{f_{2}}+\frac{f_{2}}{f_{3}}+3\Big) \bar{t}^3
- \Big( \frac{f_{1} f_{4}}{f_{2} f_{3}}+\frac{f_{2} f_{3}}{f_{1}f_{4}}+\frac{f_{1}}{f_{4}}+\frac{f_{4}}{f_{1}}+\frac{f_{3}}{f_{2}}
+\frac{f_{2}}{f_{3}}+3 \Big) \bar{t}^4
+ \dots
~.~
\nn\\
\eea
By setting the flavor symmetry fugacities to $f_i = 1$, we obtain the unrefined Hilbert series of the mesonic moduli space, 
\beal{es160a88}
&&
g(\bar{t}, f_i=1; \mathcal{M}^{mes}_{\text{BFT}[\text{CM}_5]})
=
\frac{
1 + 3 \bar{t} + 9 \bar{t}^2 + 8 \bar{t}^3 + 9 \bar{t}^4 + 3 \bar{t}^5 + \bar{t}^6
}{
(1 - \bar{t}) (1 - \bar{t}^2)^4
}
~,~
\eea
which we identify to be identical to the corresponding Hilbert series in \eref{es160a77} up to the fugacity map $t^{32} = \bar{t}$.
We conclude that the Hilbert series both in terms of fugacities corresponding to global symmetry charges on chiral fields and GLSM fields are equivalent, implying that they describe the same non-complete intersection as the mesonic moduli space of the $\text{BFT}[\text{CM}_5]$ Model.
\\

\section{Conclusion and Discussions}

In this work, we study the algebraic structure of mesonic moduli spaces of bipartite field theories (BFTs) by computing the corresponding Hilbert series. 
In particular, we concentrate on BFTs that are defined on disks and cylinders. 
By calculating the Hilbert series, we are able to identify whether the mesonic moduli spaces of the BFTs are complete intersections.
Moreover, the refined Hilbert series whose fugacities count charges under the global symmetry of the BFT allows us to express the Hilbert series expansion in terms of characters of irreducible representations of any enhanced non-abelian factors of the global symmetry. 
An example is the $\text{BFT}[Q^{1,1,1}]$ Model, where the global symmetry based on the quiver diagram of the BFT is of the form $U(1)_{f_1} \times U(1)_{f_2} \times U(1)_{f_3} \times U(1)_R$.
By closer inspection of the Hilbert series of the mesonic moduli space, we show that under a fugacity map, the Hilbert series can be expressed in terms of characters of representations of an enhanced global symmetry of the form $SU(2)_x \times SU(2)_y \times SU(2)_z \times U(1)_R$. This enhanced global symmetry also matches the isometry of the mesonic moduli space of the BFT, which is identified to be the Calabi-Yau cone over $Q^{1,1,1}$. 

Our work also identifies two infinite one-parameter families of BFTs defined on cylinders whose mesonic moduli spaces are complete intersection toric Calabi-Yau 3-folds.
The computation of the Hilbert series allows us to identify the generators of the mesonic moduli spaces explicitly by the use of plethystics.
These generators of the mesonic moduli spaces can be expressed in terms of chiral fields of the BFT.
The charges that these generators carry under the global symmetry of the BFT can also be identified through the Hilbert series. 
We identify the defining relations formed amongst the generators in order to characterize the algebraic structure of the mesonic moduli spaces of these BFTs.

We plan to expand our analysis of the algebraic structure of BFT mesonic moduli spaces to a wider set of BFTs defined on more general Riemann surfaces with arbitrary number of boundaries. 
Here, we note that some of these BFTs are related to class $\mathcal{S}_k$ theories as studied in \cite{Gaiotto:2015usa,Franco:2015jna,Hanany:2015pfa,Coman:2015bqq,Razamat:2016dpl,Razamat:2018zus,Bourton:2020rfo}.
Furthermore, while we focused on the computation of Hilbert series of the mesonic moduli space of BFTs in this work, the Hilbert series can also be calculated for the master space of such BFTs.
We expect to discover more intricate algebro-geometric structures for the moduli spaces of BFTs and plan to report on these results in the near future.
\\

\section*{Acknowledgements}

R.-K. S. would like to thank Jiakang Bao, Sebastian Franco, Georgios P. Goulas, Dongwook Ghim, Amihay Hanany, Yang-Hui He, Alessandro Pini and Masahito Yamazaki
for discussions and collaborations on related topics. 
He would also like to thank 
the Simons Center for Geometry and Physics at Stony Brook University,
the Merkin Center for Pure and Applied Mathematics at the California Institute of Technology,
the Kavli Institute for the Physics and Mathematics of the University at the University of Tokyo,
as well as the Aspen Center for Physics 
for hospitality during various stages of this work.
R.-K. S. is supported by a Basic Research Grant of the National Research Foundation of Korea (NRF-2022R1F1A1073128).
He is also supported by a Start-up Research Grant for new faculty at UNIST (1.210139.01) and a UNIST AI Incubator Grant (1.230038.01).
He is also partly supported by the BK21 Program (``Next Generation Education Program for Mathematical Sciences'', 4299990414089) funded by the Ministry of Education in Korea and the National Research Foundation of Korea (NRF).

\bibliographystyle{JHEP}
\bibliography{mybib}

\end{document}